%% file: main.tex
\documentclass[lettersize,journal]{IEEEtran}
\usepackage{amsmath,amsfonts}
\usepackage{algorithmic}
\usepackage{algorithm}
\usepackage{array}
\usepackage[caption=false,font=footnotesize,labelfont=sf,textfont=sf]{subfig}
\usepackage{textcomp}
\usepackage{stfloats}
\usepackage{url}
\usepackage{verbatim}
\usepackage{graphicx}
\usepackage{cite}
\usepackage{caption2}
\usepackage{amsmath}
\usepackage{multirow}
\usepackage[table,xcdraw]{xcolor}
\begin{document}

\title{Textured mesh Quality Assessment using Geometry and Color Field Similarity}

\author{Kaifa Yang,
        Qi Yang,~\IEEEmembership{Member,~IEEE},
        Zhu Li,~\IEEEmembership{Senior Member,~IEEE},
        Yiling Xu,~\IEEEmembership{Member,~IEEE}
        \IEEEcompsocitemizethanks{\IEEEcompsocthanksitem K. Yang, Y. Xu are from Shanghai Jiao Tong University, Shanghai, China (e-mail: \{sekiroyyy, yl.xu\}@sjtu.edu.cn). Q. Yang, Z. Li are from University of Missouri–Kansas City, Kansas, USA (e-mail: qiyang@umkc.edu, lizhu@umkc.edu).}
        \thanks{This paper is supported in part by National Key R\&D Program of China (2024YFB2907204), National Natural Science Foundation of China (62371290), the Fundamental Research Funds for the Central Universities of China, and STCSM under Grant (22DZ2229005). The corresponding author is Yiling Xu(e-mail: yl.xu@sjtu.edu.cn).}
}

\markboth{Journal of \LaTeX\ Class Files,~Vol.~14, No.~8, August~2021}%
{Shell \MakeLowercase{\textit{et al.}}: A Sample Article Using IEEEtran.cls for IEEE Journals}

\maketitle

\begin{abstract}
    Textured mesh quality assessment (TMQA) is critical for various 3D mesh applications. However, existing TMQA methods often struggle to provide accurate and robust evaluations. Motivated by the effectiveness of fields in representing both 3D geometry and color information, we propose a novel point-based TMQA method called field mesh quality metric (FMQM). FMQM utilizes signed distance fields and a newly proposed color field named nearest surface point color field to realize effective mesh feature description. Four features related to visual perception are extracted from the geometry and color fields: geometry similarity, geometry gradient similarity, space color distribution similarity, and space color gradient similarity. Experimental results on three benchmark datasets demonstrate that FMQM outperforms state-of-the-art (SOTA) TMQA metrics. Furthermore, FMQM exhibits low computational complexity, making it a practical and efficient solution for real-world applications in 3D graphics and visualization. Our code is publicly available at: \url{https://github.com/yyyykf/FMQM}.
\end{abstract}


\section{Introduction}
\IEEEPARstart{M}{eshs} consist of serial polygonal faces, which collectively describe the 3D shape of an object or scene. Additional attributes such as color, normal, and UV coordinates are commonly associated with faces to satisfy different use cases. Generally, meshes can be categorized into three types, i.e., non-colored, vertex-colored, and texture-colored, based on the form in which the color attributes exist. Among these, texture-colored meshes are the most prevalent, as texture mapping and editing techniques enable artists and developers to create visually appealing and realistic models without increasing geometric complexity. The famous international standardization group such as MPEG WG7 \cite{mpegCfP} and AOMedia Volumetric Visual Media (VVM) Working Group \cite{vvmCfP} also prioritize the compression of texture-colored meshes.
Therefore, this paper mainly concentrates on texture-colored meshes, with the term \textquotedblleft meshes\textquotedblright\ referring specifically to such texture-colored representations throughout the subsequent paper.

Meshes have extensive applications in fields such as game development, virtual reality, medical image processing, and architectural modeling.
However, algorithms\cite{meshCompression1, meshlod} utilized in these applications, such as lossy compression and simplification, may introduce distortions that affect perceptual quality.
Therefore, effective textured mesh quality assessment (TMQA) metrics are imperative to measure perceptual distortions. Referring to the study experience in the quality assessment community,
TMQA can generally be classified into subjective and objective categories.
Subjective TMQA involves human participants scoring distorted meshes. Although it offers accurate evaluations, it is costly, time-consuming, and unsuitable for real-time applications. 
Objective TMQA metrics aim to replicate the perceptual judgments of the Human Visual System (HVS) to predict quality scores. Consequently, an accurate objective metric can guide mesh processing algorithms effectively, which has significant research value for practical mesh application and forms the primary focus of this paper.

Based on the availability of the original reference samples, TMQA metrics can be categorized into three types: full-reference (FR) metrics, which require access to the original mesh and are commonly used for benchmarking mesh processing algorithms; reduced-reference metrics, which utilize partial features of the reference mesh and are particularly useful in scenarios with limited bandwidth or computational resources; and no-reference metrics, which do not rely on any reference and are typically applied in evaluating the perceptual quality of 3D content generation. In this work, we focus solely on FR objective metrics, as our proposed method is designed within this setting. 

Current FR objective TMQA metrics are commonly categorized into three types: projection-based, model-based, and point-based.
Projection-based metrics utilize predetermined viewpoints to render the mesh and apply established image quality assessment (IQA) or video quality assessment (VQA) metrics to evaluate its perceived quality. However, their performance can be affected by the choice of viewpoint position\cite{yana2021}, the number of viewpoints\cite{pcqa1}, the pooling strategy \cite{lavo_viewpoint}, the resolution of rendered images, and the presence of background information.
Model-based metrics evaluate mesh quality by directly comparing features extracted from vertices and faces, avoiding distortions from format conversion (e.g., mesh-to-image). However, they often emphasize geometry while struggling to capture texture information, leading to underestimation of texture-related quality. Additionally, model-based methods suffer from correspondence issues due to topological differences and feature collapse in non-manifold regions \cite{geodesicpsim}.
Point-based metrics sample points on mesh surfaces and evaluate quality using established point cloud quality assessment (PCQA) methods \cite{graphsim, pcqm}. Over the years, numerous efficient PCQA metrics have been developed, many of which incorporate texture information by associating color with each sampled point.
By avoiding reliance on irregular mesh connectivity, this approach simplifies TMQA compared to model-based metrics.
However, its effectiveness is highly dependent on sampling density. Higher sampling rates help reduce correspondence ambiguities inherent in both point clouds and meshes, and are also essential for preserving geometric details in the absence of explicit topology. Consequently, both the accuracy and reliability of point-based metrics are fundamentally constrained by sampling strategy \cite{TDMD}.

To address the aforementioned challenges and achieve more accurate mesh quality evaluation, we propose a novel point-based metric, termed the field mesh quality metric (FMQM). This metric is specifically designed for textured meshes and is tailored to fully account for their unique characteristics.
FMQM generalizes traditional point-based methods through a field-based formulation, which assigns each point in 3D space a mesh-related meaning by establishing correspondences with specific locations on the mesh surface
While both FMQM and traditional point-based metrics avoid issues from mesh topology, point clouds typically lose structural information and thus require dense sampling. In contrast, FMQM preserves the link between samples and the original mesh via field definitions, enabling accurate structure representation with far fewer points (see Section \ref{sec:robutness}).
Furthermore, FMQM directly compares field values at the same spatial locations, avoiding nearest-neighbor matching and effectively resolving the correspondence problem.

To facilitate the extraction of features relevant to mesh quality, FMQM incorporates two key field components: the signed distance field (SDF) and a newly proposed color field, termed the nearest surface point color field (NCF).
The SDF represents the minimum distance from a point in space to the mesh surface, with the sign indicating whether the point lies inside or outside the surface. Due to its robust capability in capturing geometric structure, the SDF has been widely adopted in reconstruction \cite{meshSDF1} and generation \cite{deepsdf} tasks. In this work, we leverage its strength to quantify geometric distortions in textured meshes.
However, relying solely on geometric information may not be sufficient for TMQA, as texture maps can significantly affect visual perception and potentially mask geometric errors. To address this, we introduce NCF to jointly evaluate color and geometric distortions, which encodes both the color value and its spatial distribution by assigning to each point the color of its nearest surface location on the mesh, enabling a more comprehensive assessment of perceived quality.

The proposed FMQM consists of four key steps:
First, local surface patches are constructed based solely on the topological information of the reference mesh.
Second, sample point locations are selected near these local surfaces, and the SDF and NCF values are computed for both the reference and distorted meshes according to the field definition.
Third, four 3D field features related to visual perception are extracted and compared from the geometry and color fields: geometry similarity, geometry gradient similarity, spatial color distribution similarity, and spatial color gradient similarity.
Finally, these features are combined to derive a final quality score for the mesh.

In summary, our contributions are as follows:
\begin{itemize}
    \item Introduction of a FR point-based TMQA method specifically designed for textured meshes, which effectively addresses the challenges of incorporating both topological and texture information.
    \item Development of a novel field-based color representation for TMQA, leveraging field theory to enable the simultaneous analysis of both geometric and texture distribution differences, thereby providing a more effective way to extract features for TMQA.
    \item Presentation of experimental results on three benchmark databases, demonstrating that the proposed model outperforms current state-of-the-art (SOTA) FR methods.
\end{itemize}

The remainder of the paper is structured as follows:
Section~\ref{sec:relatedwork} provides an overview of current TMQA metrics, highlighting the strengths and weaknesses of projection-based, model-based, and point-based approaches.
Section~\ref{sec:preliminary} presents the mathematical formulation of the TMQA problem, including the definition of the SDF.
Section~\ref{sec:ncf} introduces the definition and properties of our proposed color field, NCF, along with two key feature extraction units essential for mesh quality assessment.
Section~\ref{sec:method} outlines the overall framework and details the sub-steps of the FMQM methodology.
Section~\ref{sec:experiment} compares the performance of FMQM with SOTA metrics on three benchmark TMQA datasets.
Section~\ref{sec:conclusion} draws the conclusion.

\section{Related Work}
\label{sec:relatedwork}
In this section, we provide an overview of the evolution of quality assessment methods in the current research on TMQA.
\subsection{Projection-based Methods}
\label{sec:projection}
Projection-based methods involve rendering meshes from specific viewpoints, thereby transforming the TMQA problem into a well-established IQA or VQA task. This approach aligns with the real-world application scenarios of textured meshes by converting unstructured 3D geometry into structured 2D representations, and avoids the challenges associated with complex 3D topologies. A representative example is the method proposed by MPEG \cite{mmetric}, which selects 16 uniformly distributed viewports using the Fibonacci sphere lattice and renders both color and depth images at a resolution of 2048×2048. Quality scores are then calculated based on PSNR in the RGB and YUV color spaces.

\begin{figure}[htbp]
    \centering
    \subfloat[Inside view of drawingRoom\label{projection:sub1}]{
        \includegraphics[width=0.22\textwidth]{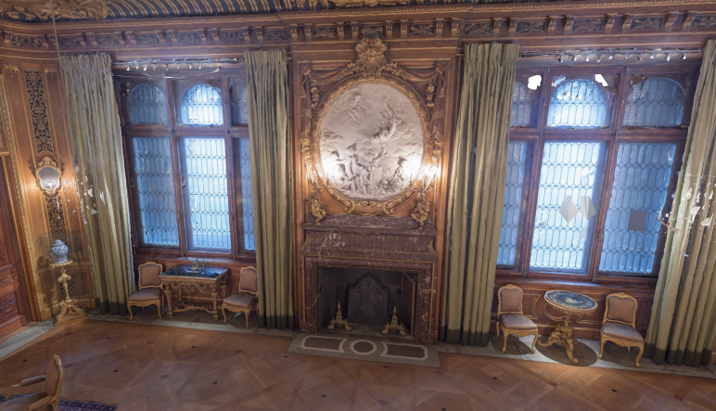}
    }
    \hfill
    \subfloat[Outside view of drawingRoom\label{projection:sub2}]{
        \includegraphics[width=0.22\textwidth]{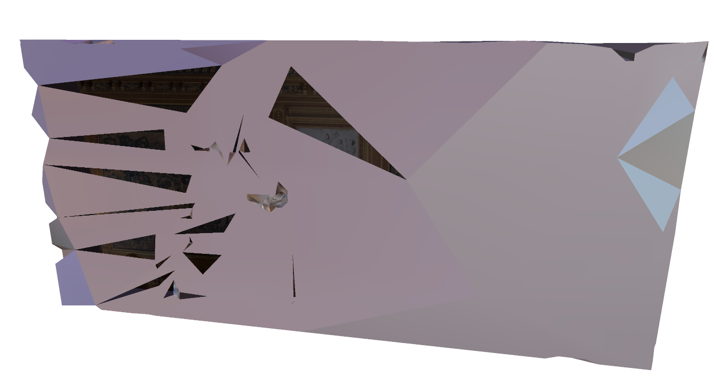}
    }
    \par\bigskip
    \subfloat[Upper view of bennetTomb\label{projection:sub3}]{
        \includegraphics[width=0.22\textwidth]{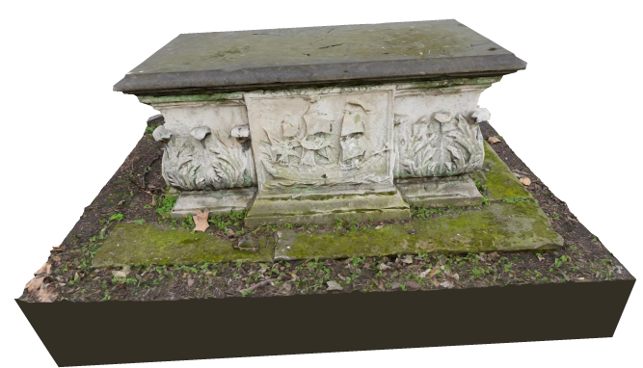}
    }
    \hfill
    \subfloat[Bottom view of bennetTomb\label{projection:sub4}]{
        \includegraphics[width=0.22\textwidth]{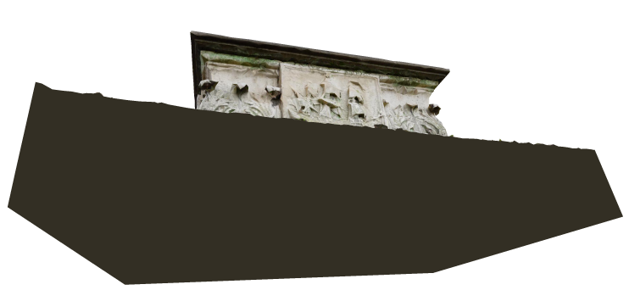} 
    }
    \caption{Projection Views of drawingRoom from TSMD\cite{tsmd} and bennetTomb from SJTU-TMQA\cite{sjtumqa}}
    \label{fig:projection}
\end{figure}

However, the performance of these methods is highly sensitive to viewpoint selection. Nehmé et al. \cite{yana2023} evaluated several perceptual metrics using one main and four auxiliary views, and found that perceptual pooling across views is notably non-uniform.
In addition, projection-based methods also face intrinsic limitations. As shown in Fig.~\ref{fig:projection}, while most meshes are best viewed from the outside, certain types like indoor scenes are only meaningful from interior views (e.g., Fig.~\ref{projection:sub1}), with exterior views offering little value (e.g., Fig.~\ref{projection:sub2}). Similarly, some meshes have uninformative regions—such as the dark bottom view of BennettTomb (Fig.~\ref{projection:sub4})—while only specific angles, like the top view (Fig.~\ref{projection:sub3}), reveal useful content. 
Furthermore, most IQA/VQA metrics treat background pixels equally with foreground pixels, which dilutes the contribution of actual texture or geometric artifacts. Liu et al. \cite{wpc} noted that the prevalence of high-scoring background regions often masks visible degradations in the foreground.

To address these issues, Graphics-LPIPS \cite{yana2023} computes LPIPS on overlapping 64×64 patches from a manually selected main view, discarding patches with over 35\% background to reduce irrelevant influence. This patch-based strategy enhances local sensitivity and improves metric stability. Similarly, 3D-PSSIM \cite{3dpssim} employs a multi-view, multi-layer projection framework that incorporates depth images to better capture geometric distortion, and applies perceptual weighting to emphasize high-curvature regions, improving sensitivity to subtle artifacts.

While these methods enhance robustness, they still fundamentally depend on 2D projections, which inevitably discard crucial 3D information such as spatial continuity, surface topology, and occlusion relationships. The rendering process thus becomes a bottleneck, as it undermines the structural integrity of the original mesh. In addition, there is no established guideline for how many viewpoints are sufficient or which ones are most representative. Without a principled view selection strategy, projection-based approaches often struggle to achieve consistent and comprehensive results, especially for meshes with complex or self-occluded geometry.

\subsection{Model-based Methods}
\label{sec:model-based} 
Model-based TMQA methods operate directly on mesh topology and attributes, avoiding the need for media-type conversion and preserving full 3D structural information.
Early approaches focused solely on geometric distortion. Metro \cite{metro} and MESH \cite{mesh} quantified distortion via vertex-to-vertex or vertex-to-face distances, offering computational efficiency and robustness to non-manifold geometry. However, they neglect surface appearance, resulting in weak correlation with the HVS.

Subsequent methods introduced HVS-aligned features at the vertex or face level. GL \cite{gl} and GL2 \cite{gl2} use geometric Laplacians to characterize local surface roughness. 3DWPM \cite{3dwpm} captures dihedral angles between adjacent faces and aggregates them via area-weighted averaging. TPDM \cite{tpdm} employs curvature tensors and introduces perceptual weighting to simulate visual masking. Zhang et al. \cite{zhang2022} extract a diverse set of geometric and color features and compare their statistical distributions between reference and distorted meshes. While these methods improve perceptual relevance, most rely on point-to-point correspondence or similar mesh densities, which are often disrupted by real-world distortions. Zhang et al. address this by comparing global feature distributions, but at the cost of discarding spatial layout information.

To better reflect local perceptual quality, recent methods define surface patches for localized feature extraction. MSDM2 \cite{msdm2} uses spherical neighborhoods with Gaussian-weighted curvature statistics. GeodesicPSIM \cite{geodesicpsim} defines one-ring patches and compares curvature and color features, using cropping to normalize patch size. These methods exploit topology for patch construction, but still face challenges: patch correspondence may be unstable, Gaussian weighting may misrepresent irregular structures, and color comparison ignores spatial distribution in 3D space.
HybridMQA~\cite{hybridmqa} employs a hybrid framework that combines model-based and projection-based features. It encodes geometry via 2D projections aligned with texture maps, refines them using a graph neural network, and fuses geometric and texture features through cross-attention in 2D. Despite its joint reasoning capability, the method remains dependent on projections, making it sensitive to viewpoint bias and prone to incomplete information integration.

In summary, model-based TMQA methods benefit from direct geometric access and perceptually motivated features, allowing finer-grained assessment without relying on rendering. However, they remain challenged by correspondence assumptions, irregular sampling, and the difficulty of preserving spatial relationships within global or patch-level comparisons. Moreover, effectively leveraging texture values and their distribution across the surface remains an open problem—most methods either ignore texture, treat it locally without spatial context, or struggle to integrate it with geometry in a unified perceptual framework.

\subsection{Point-based Methods}
\label{sec:point}
Point-based TMQA methods assess mesh quality by first converting meshes into point clouds and then applying established PCQA metrics. A representative example is the method adopted by MPEG \cite{mmetric}, which uses grid-based sampling to extract colored points from mesh surfaces. This approach directly enables the use of PCQA metrics such as D1, D2 \cite{d1d2}, pointwise color PSNR, and $\rm PCQM_{psnr}$ \cite{pcqm}.

Point-based methods naturally incorporate mesh color information and are particularly effective at capturing the color value distribution over the surface. When the sampling is dense enough, these methods can produce perceptually reliable results and alleviate the correspondence problem, as nearest-neighbor matching becomes more stable compared to sparse vertex-level comparisons.
However, point-based approaches discard explicit mesh connectivity. While dense sampling can approximate geometric structure, recovering topological information requires significantly higher resolution. Moreover, performance remains sensitive to the choice of sampling strategy. For instance, the grid-based method adopted by MPEG offers limited control over point count and may inadequately capture complex geometry. This highlights the open question of how many points are necessary to faithfully represent the original mesh for reliable quality assessment.

\subsection{Summary}
In summary, TMQA methods face persistent challenges arising from topology mismatches, vertex density differences, and the difficulty of leveraging texture information in a perceptually meaningful way. While media-type conversions—such as projecting to 2D or converting to point clouds—can alleviate certain issues, they often introduce new limitations, such as view selection bias or loss of structural context.

To address these challenges, this paper introduces visual-related geometry and color fields, which retain the strengths of point-based methods—such as uniform spatial density and effective modeling of color distribution—while extending feature representation from the mesh surface to the entire 3D space. Each point in space is explicitly linked to its corresponding face information on the original mesh, enabling feature extraction that is both geometrically meaningful and spatially consistent. This formulation allows for reliable inference of topological structure even under significantly sparser sampling compared to conventional point-based approaches, providing a more efficient and robust foundation for mesh quality assessment.

\section{Preliminary}
\label{sec:preliminary}
In this section, we present the mathematical formulation of the TMQA problem and introduce the definition of the spatial fields used in our method.

\subsection{TMQA Problem Formulation}
A textured mesh \( M := \{f_1, f_2, \ldots, f_N\} \) is composed of a set of triangular faces, where each face \( f \) is defined by three 3D vertices \( f := (\mathbf{v}_0, \mathbf{v}_1, \mathbf{v}_2) \). Each vertex \( \mathbf{v}_i \) is associated with a 2D UV coordinate \( \mathbf{v}_t = (u, v) \), which maps the vertex to a position on a texture image \( \mathbf{I} \in \mathbb{R}^{W \times H \times 3} \), thereby enabling color information to be assigned to the mesh surface.

The goal of a TMQA metric \( D \) is to quantify the perceptual difference between a reference mesh \( M_r \) and a distorted mesh \( M_d \). The predicted quality score should align with human perception and exhibit a strong correlation with the corresponding subjective score \( s \). This relationship can be expressed as:
\begin{equation}
s = \phi(D(M_r, M_d)),
\end{equation}
where \( \phi \) denotes a monotonically non-linear mapping function, typically used to align the objective score range with subjective evaluations.

\subsection{Field Description}
\label{sec:sdf}
Fields can be broadly classified into scalar fields and vector fields. A $D$-dimensional scalar field is a function that maps each point in $\mathbb{R}^D$ to a scalar value:
\begin{equation}
    f(\mathbf{p}) = s, \quad \mathbf{p} \in \mathbb{R}^D, \quad s \in \mathbb{R}
\end{equation}
Vector fields can be viewed as a composition of multiple scalar fields, where each component represents a scalar field in one dimension.

In computer graphics, unsigned distance fields (UDF) and SDF are widely used for mesh representation and processing, as they provide continuous and spatially coherent descriptions of geometry.
Given a mesh $M$ and a query point $\mathbf{p} \in \mathbb{R}^3$, the UDF is defined as:
\begin{equation}
    f_{UDF}(\mathbf{p}) = \min_{\mathbf{q} \in M} \left\| \mathbf{p} - \mathbf{q} \right\|
\end{equation}
That is, $f_{UDF}(\mathbf{p})$ returns the shortest Euclidean distance from $\mathbf{p}$ to any point $\mathbf{q}$ on the surface of the mesh $M$.  
In practice, the nearest surface point \( \mathbf{q} \) can be efficiently estimated by evaluating triangle-wise distances, often accelerated using voxel-based spatial partitioning to limit the search region.

Based on the definition of the UDF, the SDF is further defined as:
\begin{equation}
\label{equ:sdfDef}
f_{SDF}(\mathbf{p}) = \left\{
\begin{array}{ll}
f_{UDF}(\mathbf{p}) & \text{if } \mathbf{p} \text{ is outside } M \\
- f_{UDF}(\mathbf{p}) & \text{if } \mathbf{p} \text{ is inside } M
\end{array}
\right.
\end{equation}
The sign of the SDF relies on distinguishing whether \( \mathbf{p} \) lies inside or outside the mesh, which is well-defined only when the mesh is watertight. In cases where watertightness is not guaranteed, a common heuristic involves ray casting: a ray is cast from \( \mathbf{p} \) toward its nearest surface point \( \mathbf{q} \), and the number of mesh-face intersections is counted. An odd number implies the point lies inside the mesh, while an even number indicates it is outside.

\section{NCF: A New Color Field Description for TMQA}
\label{sec:ncf}
In this section, we propose a new color field representation, termed the NCF, specifically designed for the TMQA task. We introduce its formal definition, key properties, and two quality-related feature extraction units.

\subsection{Motivation and Definition}
\label{sec:ncfDefinition}
\begin{figure}[htbp]
    \centering
    \subfloat[\label{maskeffect:sub1}]{
        \centering
        \includegraphics[width=0.1\textwidth]{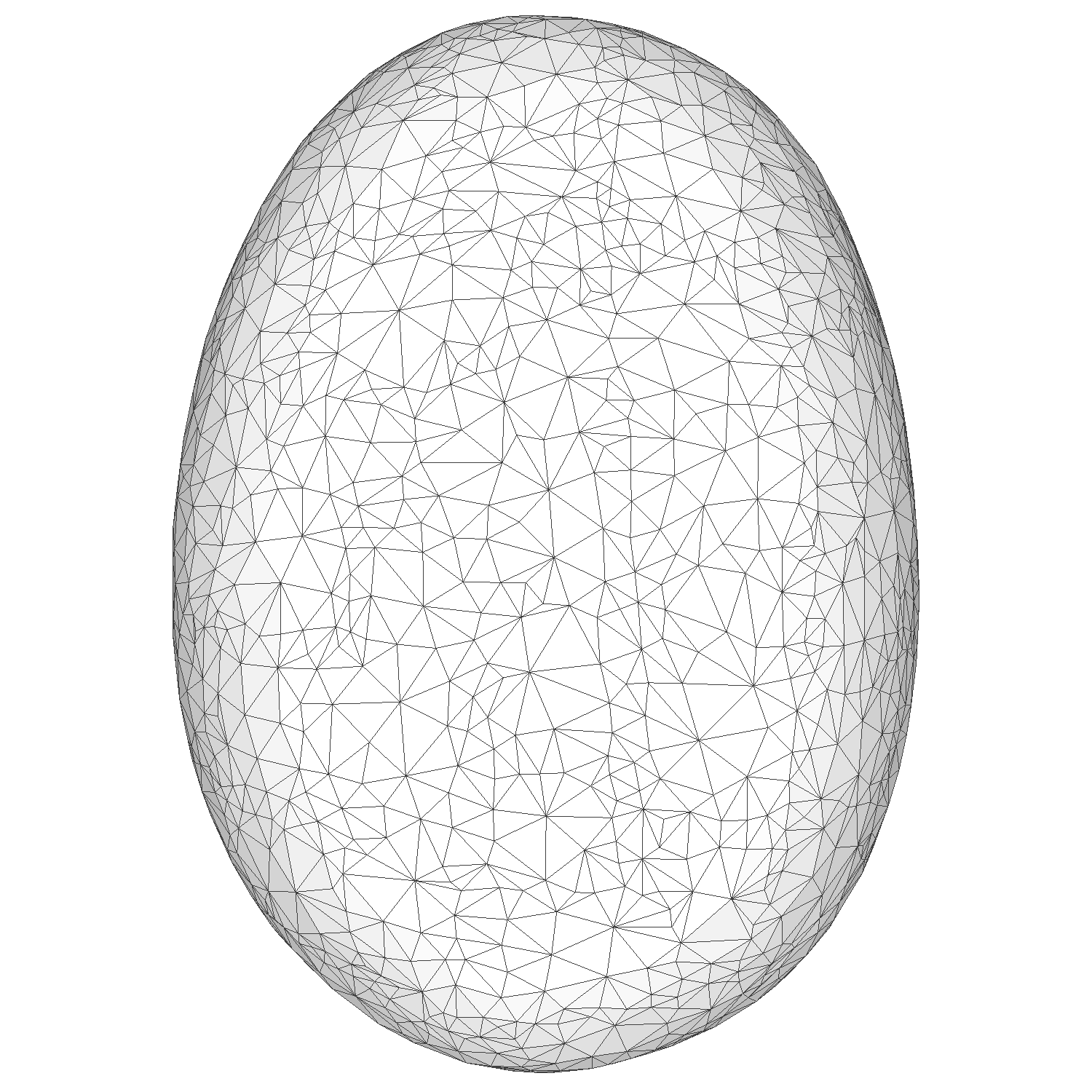}
    }
    \hfill
    \subfloat[\label{maskeffect:sub2}]{
        \centering
        \includegraphics[width=0.1\textwidth]{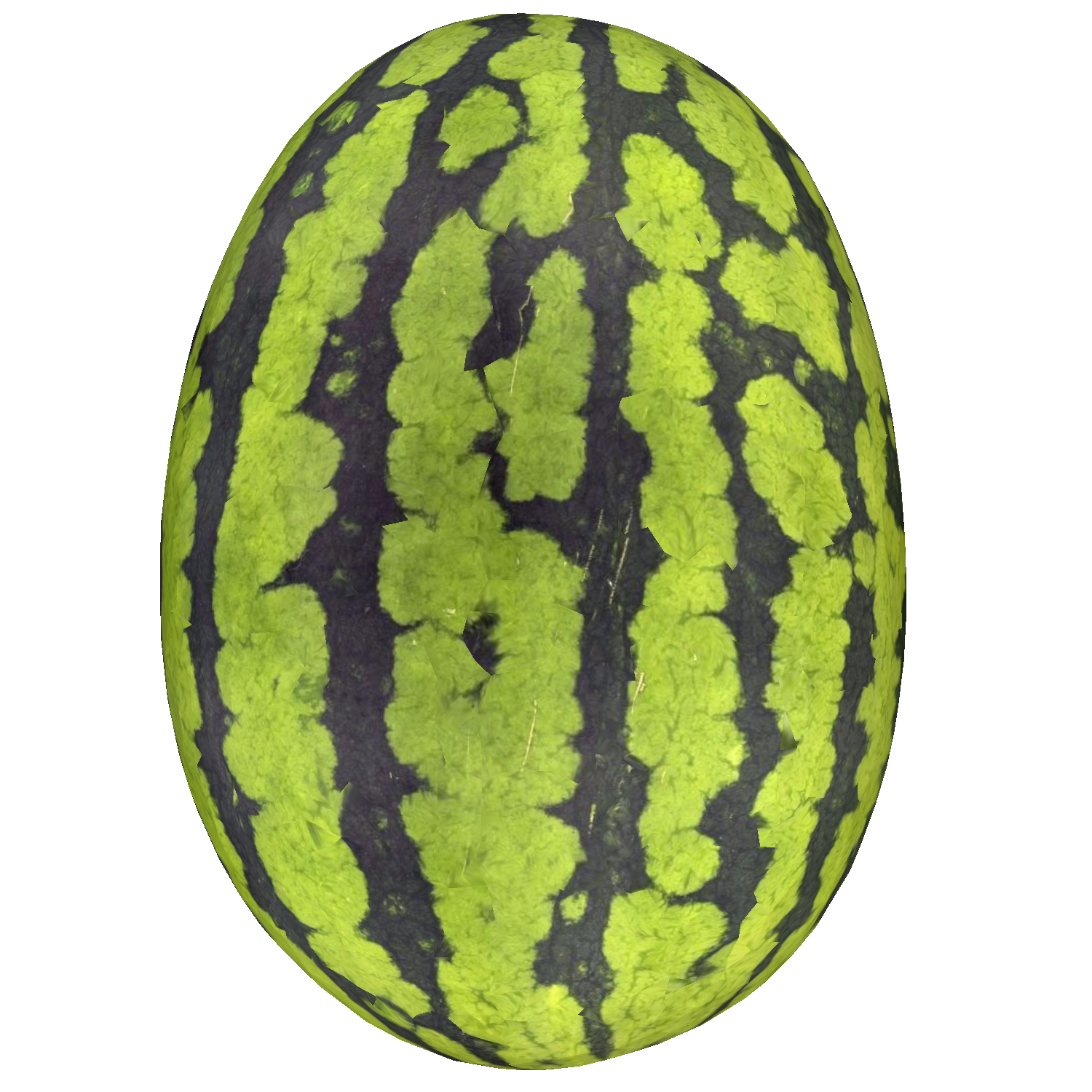}
    }
    \hfill
    \subfloat[\label{maskeffect:sub3}]{
        \centering
        \includegraphics[width=0.1\textwidth]{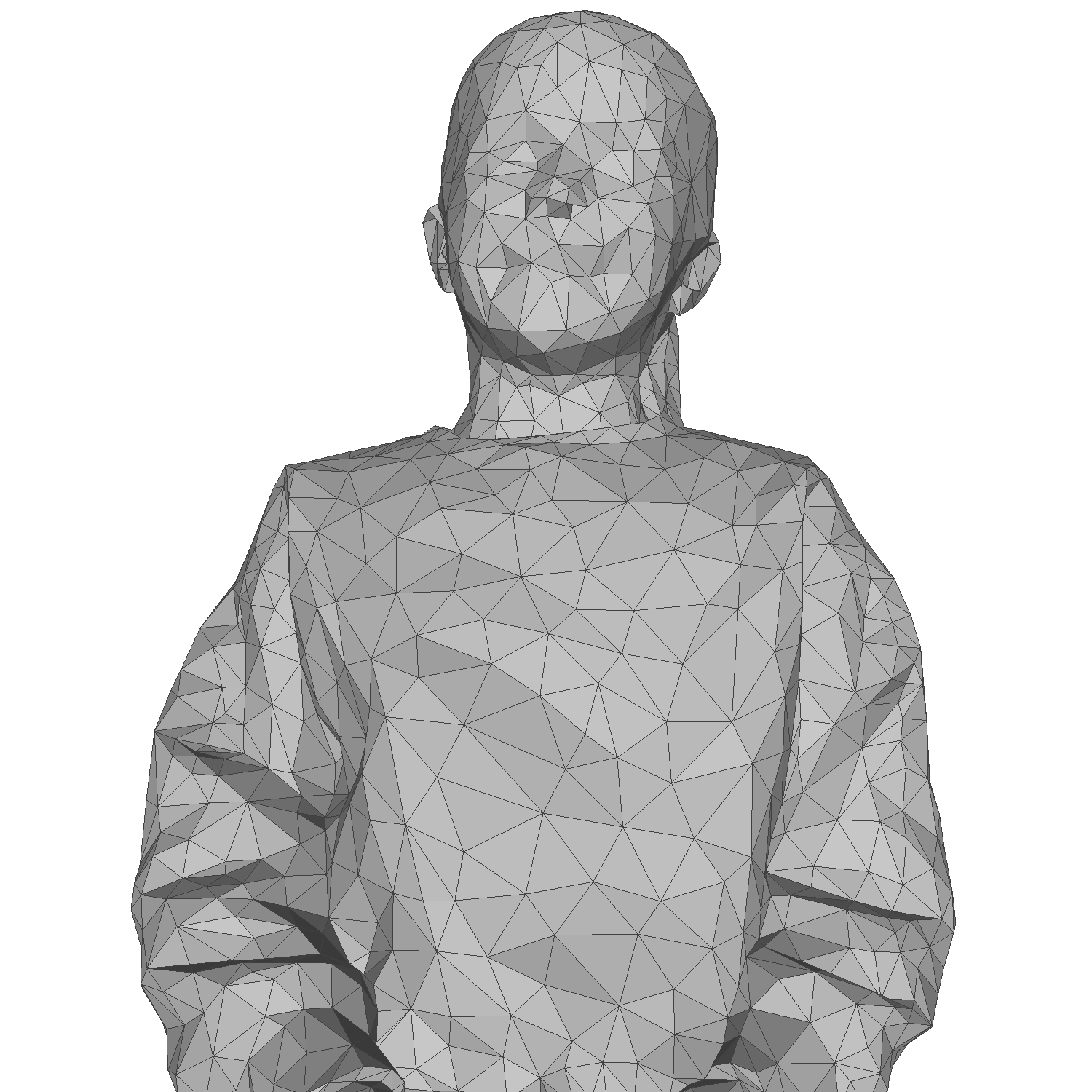}
    }
    \hfill
    \subfloat[\label{maskeffect:sub4}]{
        \centering
        \includegraphics[width=0.1\textwidth]{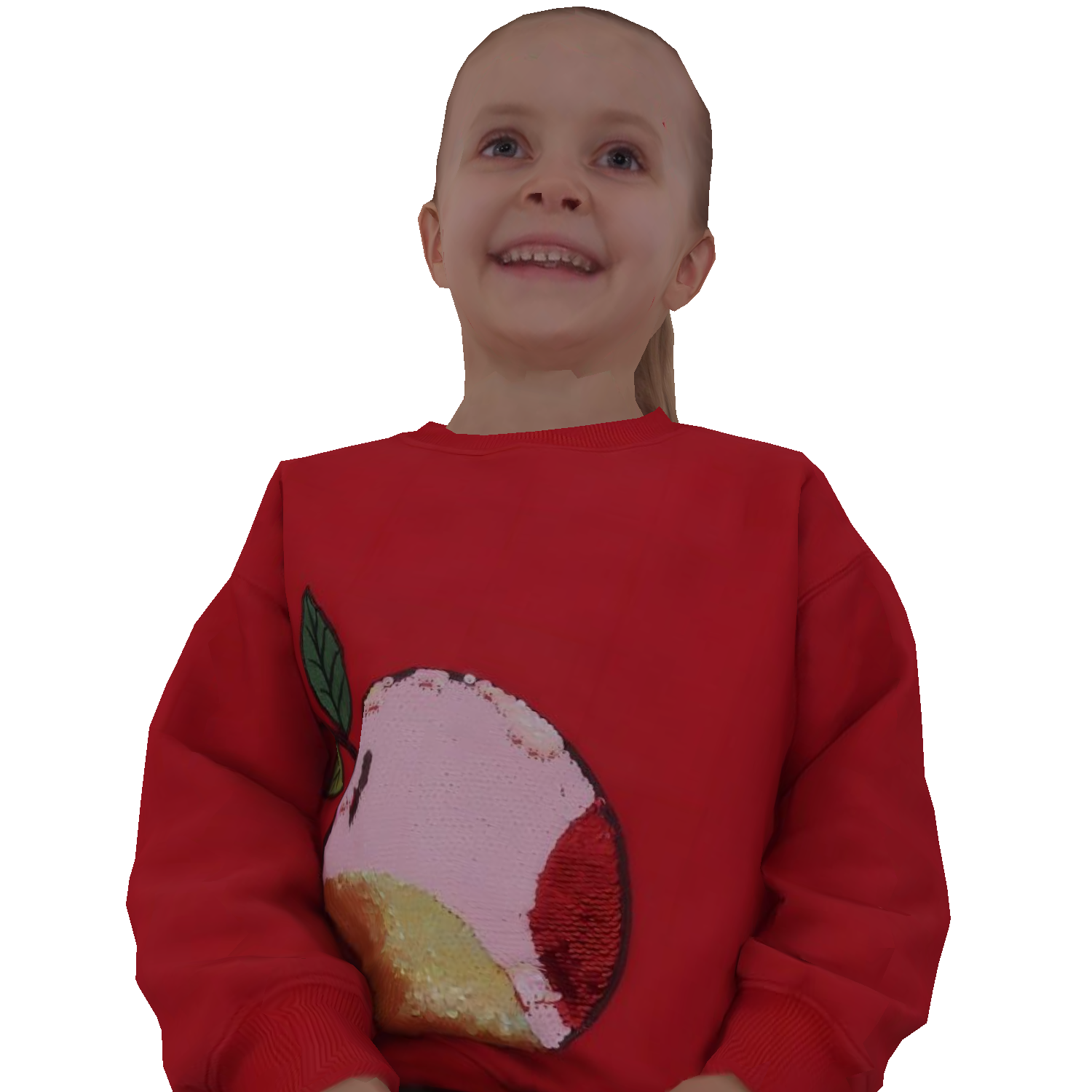} 
    }
    \caption{Masking Effect Caused by Mesh Texture}
    \label{fig:maskeffect}
\end{figure}

The texture of a 3D mesh not only enhances visual realism but can also perceptually mask geometric artifacts. For instance, complex textures can introduce visual richness to otherwise smooth surfaces (e.g., Fig. \ref{maskeffect:sub1} and \ref{maskeffect:sub2}), while even a coarse mesh geometry may appear smoother when textured with a homogeneous color (e.g., Fig. \ref{maskeffect:sub3} and \ref{maskeffect:sub4}).
Given this perceptual significance, a comprehensive TMQA metric should incorporate a faithful description of surface color. However, existing TMQA methods lack a dedicated formulation that captures the spatial distribution of texture in 3D space. To this end, we introduce the NCF, which provides a continuous and spatially coherent representation of color anchored to the underlying geometry.

Formally, we define the value of the NCF at a query point \( \mathbf{p} \in \mathbb{R}^3 \) as the RGB color associated with its nearest point \( \mathbf{p}_n \in M \) on the mesh surface, as determined during the SDF computation in Section~\ref{sec:sdf}. Assume that \( \mathbf{p}_n \) lies on a triangle formed by the vertices \( \{\mathbf{v}_0, \mathbf{v}_1, \mathbf{v}_2\} \), with corresponding UV coordinates \( \{(u_0, v_0), (u_1, v_1), (u_2, v_2)\} \). The UV coordinates of \( \mathbf{p}_n \) can then be interpolated using its barycentric coordinates \( \{\lambda_0, \lambda_1, \lambda_2\} \) as follows:
\begin{equation}
    (u_n, v_n) = \lambda_0 (u_0, v_0) + \lambda_1 (u_1, v_1) + \lambda_2 (u_2, v_2).
\end{equation}
These UV coordinates are mapped to pixel locations \( (i, j) \) on the texture image \( \mathbf{I} \in \mathbb{R}^{W \times H \times 3} \) using the standard convention:
\begin{equation}
    i = u_n \cdot W - 0.5, \quad j = (1 - v_n) \cdot H - 0.5.
\end{equation}
Finally, the NCF value at \( \mathbf{p} \) is defined as the sampled color at pixel location \( (i, j) \):
\begin{equation}
\label{equ:ncfDef}
    \mathbf{f}_{\mathrm{NCF}}(\mathbf{p}) = g(\mathbf{I}, i, j),
\end{equation}
where \( g(\cdot) \) denotes a texture sampling function. In our implementation, we employ bilinear interpolation over the four nearest neighboring pixels, in accordance with common practices in computer graphics~\cite{fundCG}.

Fig.~\ref{ncf:sub1} illustrates this process. Each sampling point \( P^i \) in the 3D space is assigned a color by tracing to its nearest surface point \( P_n^i \), and subsequently evaluating Equation~\eqref{equ:ncfDef}. Intuitively, the resulting NCF values exhibit a smooth spatial extension of surface texture into the surrounding volume, effectively resembling a diffusion of surface color along local normal directions. This property provides a rich and geometry-aware representation of texture, which we leverage in our quality assessment framework as discussed in the following sections.

\subsection{NCF Property}
\label{sec:ncfProperty} 
In this section, we discuss the core properties of the proposed NCF, with a focus on its alignment with the perceptual results of mesh rendering. Given that textured mesh applications are highly dependent on rendering pipelines, we examine how the spatial distribution of NCF values corresponds to rendered images and highlight its benefits for mesh quality analysis.

\begin{figure*}[htbp]
    \centering
    \subfloat[3D polygon object with NCF values\label{ncf:sub1}]{
        \includegraphics[width=0.3\textwidth]{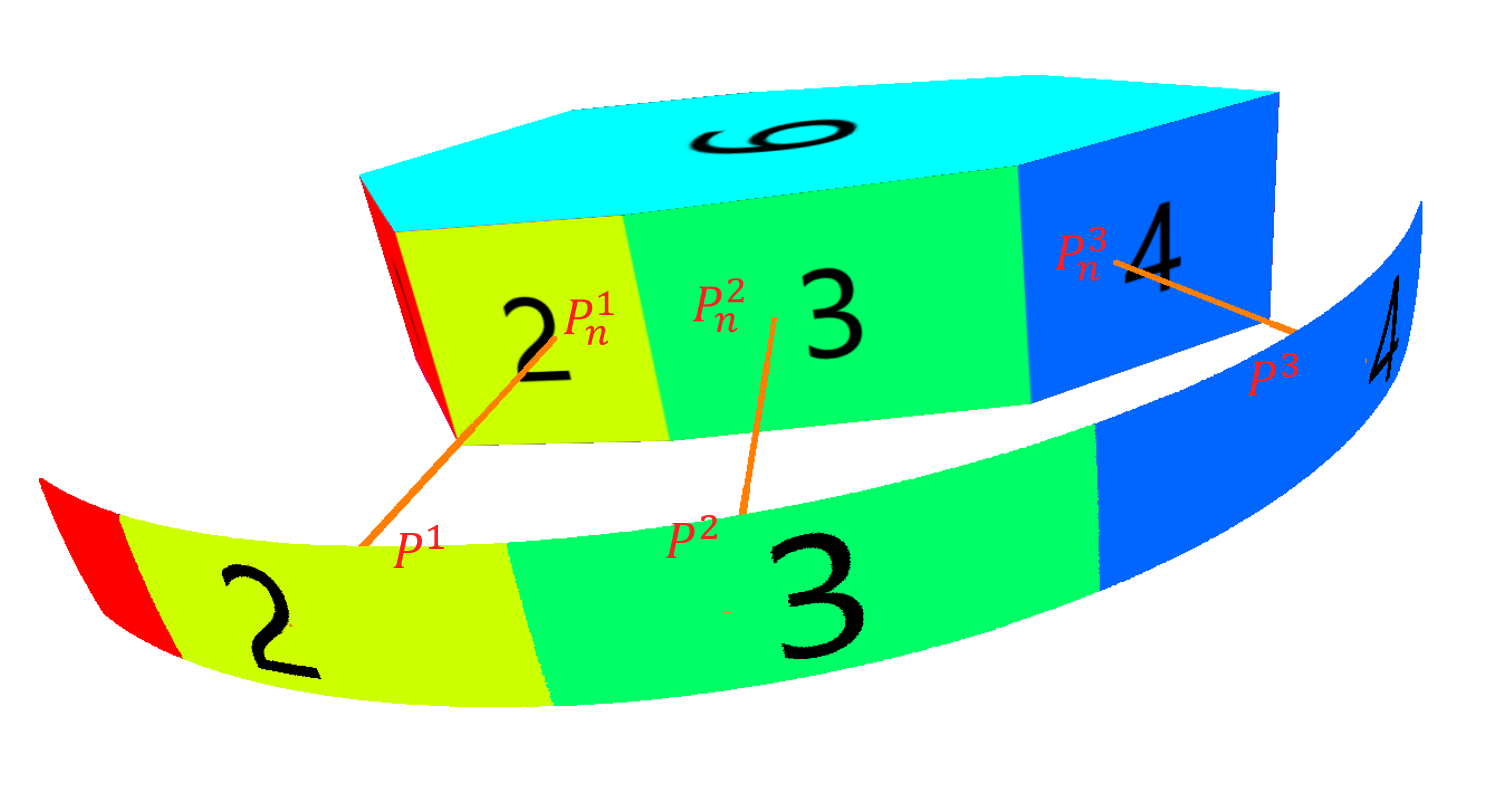}
    }
    \hfill
    \subfloat[Traditional rendering result from viewpoint $P$\label{ncf:sub2}]{
        \includegraphics[width=0.32\textwidth]{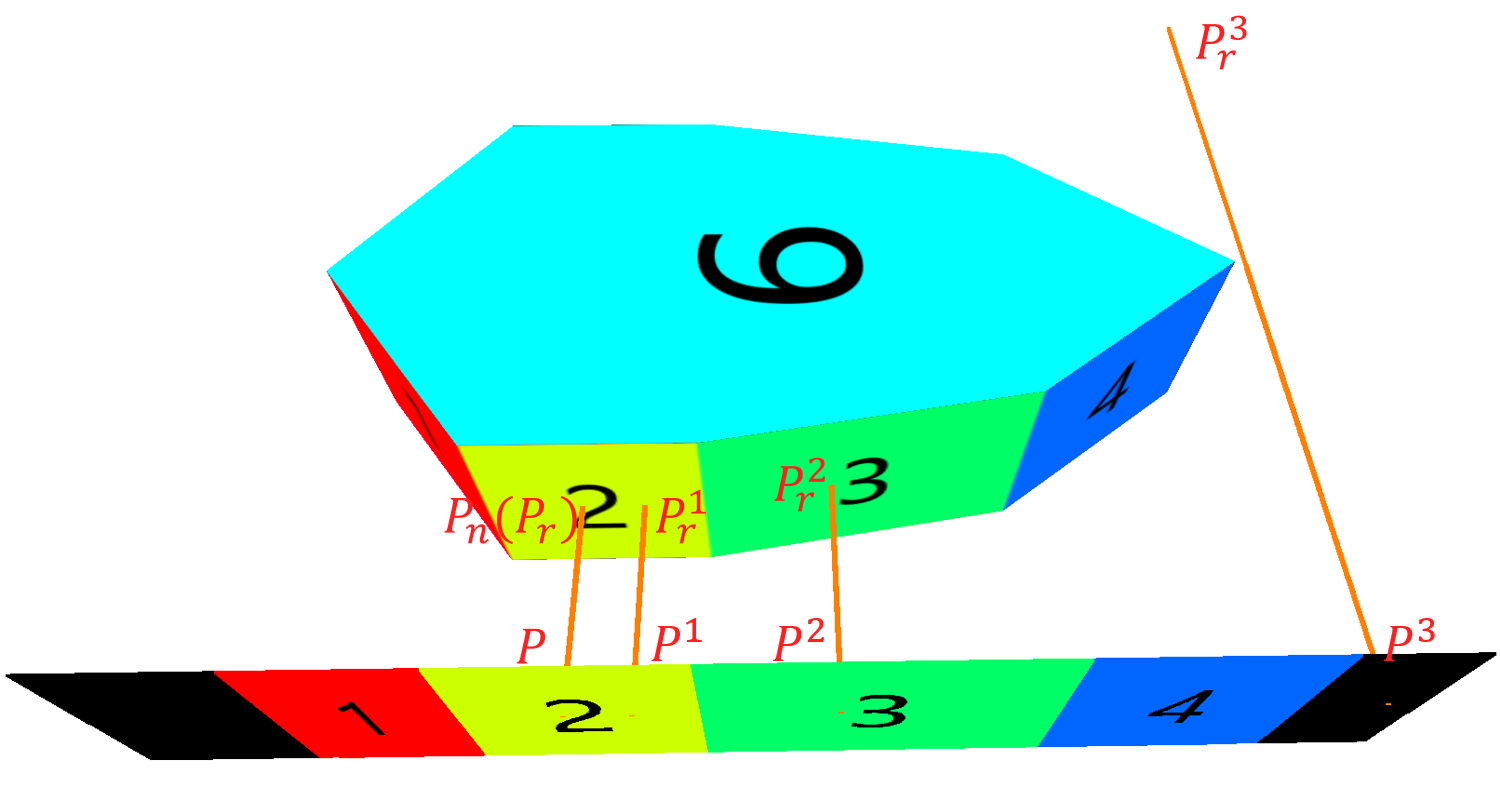}
    }
    \hfill
    \subfloat[NCF values on the same image plane\label{ncf:sub3}]{
        \includegraphics[width=0.3\textwidth]{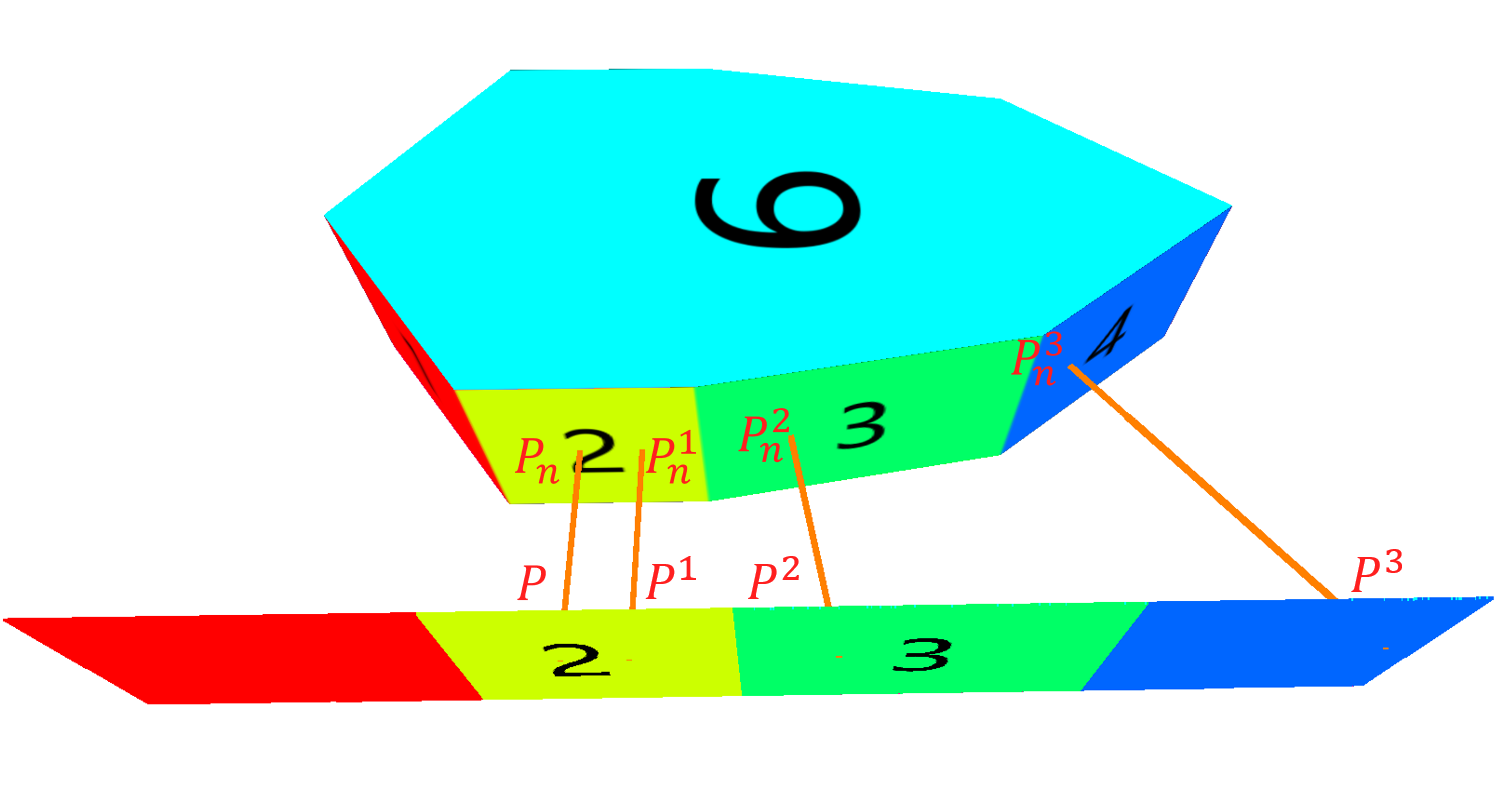}
    }
    \caption{Definition and Properties of NCF}
    \label{fig:ncf}
\end{figure*}

Consider a viewpoint \( P \) with view direction \( \overrightarrow{PP^n} \), where \( P^n \) denotes the nearest surface point of \( P \). A traditional rendering result at pixel \( P^i \) is obtained via ray casting: a ray is emitted from \( P^i \) along direction \( \overrightarrow{PP^n} \), and the color is retrieved from the first intersection point \( P^i_r \) between the ray and the mesh surface. The rendering result is illustrated in Fig.~\ref{ncf:sub2}, where some pixels, such as \( P^3 \), remain unshaded due to the absence of any ray-surface intersection (i.e., background color).

In contrast, the corresponding NCF map (Fig.~\ref{ncf:sub3}) assigns a color to each 3D position based on the closest surface point, regardless of visibility. Let \( \mathbf{I}_{P} \) represent the rendered pixel from viewpoint \( P \). Based on visual and geometric analysis, we summarize several key properties of NCF:

\textbf{Property~(1): normal-aligned diffusion of surface texture.}  
For any 3D point \( P^i \), whose closest surface point is \( P^i_n \), the vector \( \overrightarrow{P^iP^i_n} \) is parallel to the surface normal \( \mathbf{n}(P^i_n) \). Consequently, all points lying along this normal direction share the same NCF value:
\begin{equation}
\forall Q_i \in \overline{P^iP^i_n},\quad \mathbf{f}_{\mathrm{NCF}}(Q_i) = \mathbf{f}_{\mathrm{NCF}}(P^i).
\end{equation}
This property reflects that NCF extends surface texture information outward along the surface normals in a perceptually meaningful manner.

\textbf{Property~(2): view-consistent behavior under normal-aligned viewing.}
When the viewing direction aligns with the surface normal at the intersection point \( P^j_r \) (i.e., \( \overrightarrow{PP^n} \parallel \mathbf{n}(P^j_r) \)), the rendered value and the NCF value at that spatial location are identical:
\begin{equation}
\mathbf{f}_{\mathrm{NCF}}(P^j) = \mathbf{I}_{P^j}.
\end{equation}
This demonstrates that NCF can accurately reproduce the surface appearance under front-facing views, as exemplified by \( P^1 \) on face 2 in Fig.~\ref{ncf:sub3}.

\textbf{Property~(3): structural coherence under normal-deviated viewing.}
When the surface normal at the intersection point deviates from the viewing direction (i.e., \( \overrightarrow{PP^n} \not\parallel \mathbf{n}(P^k_r) \)), NCF may yield different values compared to rendering. However, the spatial structure remains highly consistent, preserving local geometry-aware color patterns. This behavior is observed on faces 1, 3, and 4 in Fig.~\ref{ncf:sub3}: for instance, the number ``3'' on the green face remains recognizable as a ``3'', but appears slightly stretched due to the angular difference.

\textbf{Property~(4): NCF as a virtual multi-camera projection.}
One of the most distinctive features of NCF is its ability to emulate local, surface-aligned rendering behavior across the entire space. Since NCF assigns values based on surface-normal directions, sampling NCF values near the mesh effectively simulates a collection of virtual cameras, each locally aligned to the nearest surface patch. This leads to spatially coherent and view-independent color fields that resemble per-surface renderings—without requiring explicit viewpoint specification.

Such behavior is particularly advantageous for full-field quality assessment: it eliminates the limitations of fixed-view rendering and supports region-wise comparison under consistent local context. Fig.~\ref{ncf:sub1} illustrates this property, where the sampled NCF values surrounding the mesh produce a continuous and perceptually aligned visualization of the original texture. Additional NCF visualizations on various mesh samples are provided in the supplementary materials.

These properties reveal that NCF serves as a rendering-aware volumetric representation that faithfully encodes surface appearance throughout space. By decoupling the representation from explicit viewpoints, NCF facilitates objective and comprehensive texture quality analysis, making it particularly well-suited for the TMQA task.

\subsection{NCF Feature Extraction Unit}
\label{sec:ncfFeature}
To quantitatively assess mesh quality via the proposed NCF, we introduce two feature extraction units: a normal-direction color patch unit and a tangent-direction color patch unit, that are naturally derived from the spatial and perceptual properties of the field. These units enable the extraction of color features that are both visually meaningful and structurally aligned with the surface geometry and texture layout.

Motivated by \textbf{Property~(1)} in Sec.~\ref{sec:ncfProperty}, which states that NCF values remain constant along the surface normal direction, we define the normal-direction color patch unit as the set of points lying along a straight line toward the nearest surface point:
\begin{equation}
\scalebox{0.95}{$
    P_{\text{normal}} = \left\{ P' \;:\; \left| \frac{\overrightarrow{PP_n} \cdot \overrightarrow{PP'}}{|\overrightarrow{PP_n}| \cdot |\overrightarrow{PP'}|} \right| > \text{norThre},\;\; |\overrightarrow{PP'}| < r \right\}
    $}
\end{equation}
Here, \( P \) is the query point, \( P_n \) its closest surface point, and \( \text{norThre} \approx 1 \) enforces alignment with the surface normal. The radius \( r \) confines the patch to a perceptually local region, mitigating interference from distant or unrelated geometry.

This unit captures distortion along a single, view-consistent sampling path, analogous to a ray in the rendering process. If texture distortion occurs, the NCF value at \( P_n \) is altered, propagating inconsistencies along \( \overrightarrow{PP_n} \). In the presence of geometric distortion, the spatial correspondence along this path is disrupted—points along \( \overrightarrow{PP_n} \) no longer share the same nearest surface point as \( P \), leading to NCF value deviations even when the texture remains unchanged.

While the normal-direction unit captures distortion along a view-consistent ray, the tangent-direction unit evaluates the visual coherence of the surface region surrounding \( P_n \), corresponding to a local image patch in rendered views. It aggregates nearby NCF values whose directions are approximately tangent to the surface:
\begin{equation}
\scalebox{0.95}{$
    P_{\text{tangent}} = \left\{ P' \;:\; \left| \frac{\overrightarrow{PP_n} \cdot \overrightarrow{PP'}}{|\overrightarrow{PP_n}| \cdot |\overrightarrow{PP'}|} \right| < \text{tanThre},\;\; |\overrightarrow{PP'}| < r \right\}
    $}
\end{equation}
The threshold \( \text{tanThre} \approx 0 \) selects directions orthogonal to the surface normal, forming a planar neighborhood around \( P \) that approximates a rendered pixel patch, as supported by \textbf{Property~(2)} and \textbf{(3)}.
The radius \( r \) ensures perceptual locality and is critical for preserving the structural integrity of the patch. When \( r \) is too large, the patch may span regions with significant normal variation, leading to the surface stretching effects described in \textbf{Property~(3)}. This introduces projection ambiguity and local distortion in the NCF representation. By constraining the patch size, such artifacts are mitigated, maintaining fidelity to the underlying geometry and texture.

In summary, the normal and tangent patch units provide complementary perspectives for evaluating texture fidelity: the former captures distortion along individual view rays, while the latter captures appearance coherence across local surface regions. Both units are defined in a view-independent manner and avoid issues associated with background pixels or viewpoint selection—common limitations in projection-based TMQA methods.
For more intuitive understanding, visualizations of the two patch units are included in the supplementary materials.

\section{Method}
\label{sec:method}
In this section, we present the overall framework of FMQM, along with its key components and feature extraction procedures.
\subsection{Framework}
\label{sec:framework}
\begin{figure*}[htbp]  
    \centering  
    \includegraphics[width=\linewidth]{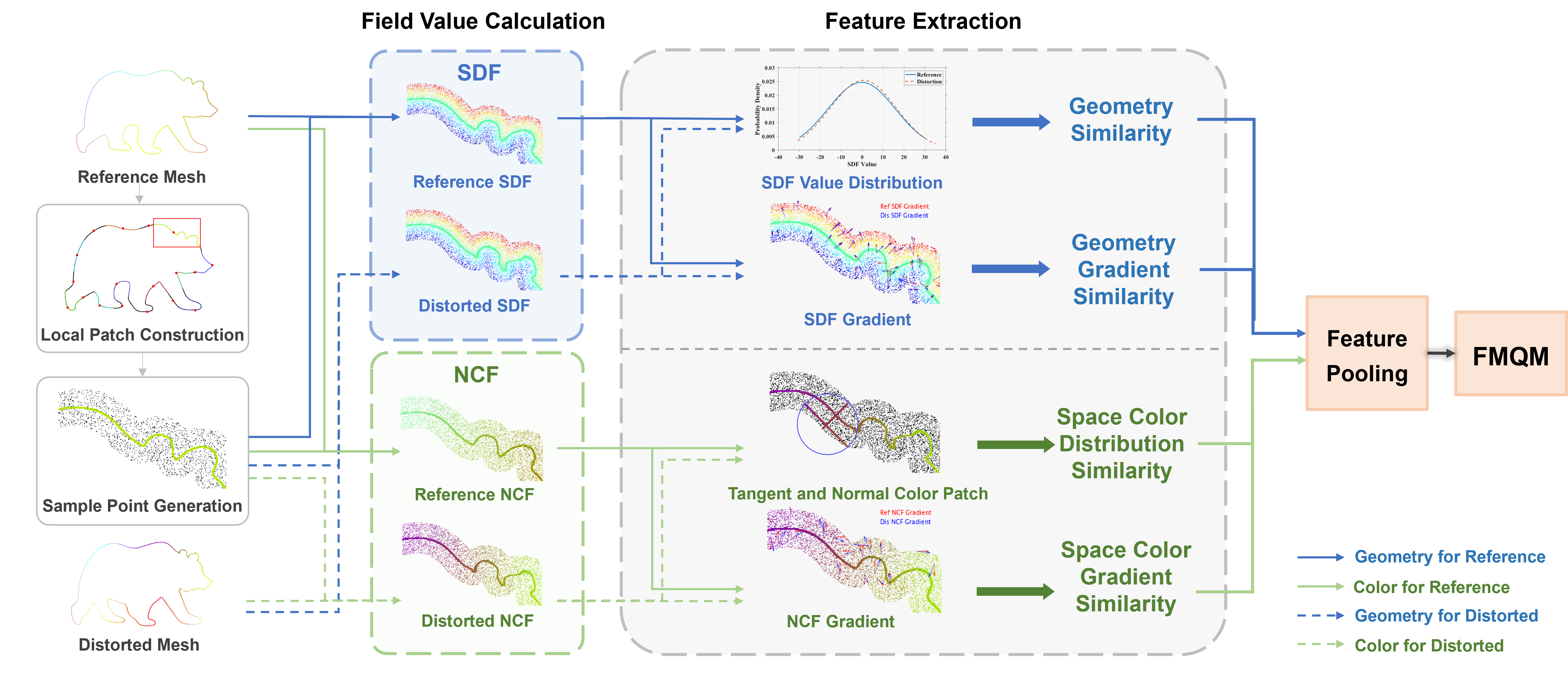}  
    \caption{Framework of FMQM (2D illustration; 3D visualization provided in supplementary material)}  
    \label{fig:framework}  
\end{figure*}

The FMQM pipeline is illustrated in Fig.~\ref{fig:framework}.
Given a reference textured mesh and its distorted version, FMQM evaluates their perceptual difference by operating on spatial fields defined over local surface regions.
First, we construct local surface patches using the topology of the reference mesh.  
Then, we uniformly generate discrete sampling locations within each patch, based on its geometric size and shape.  
At each sampled location, we compute the SDF and NCF from both reference and distorted meshes, following the definitions introduced in Section~\ref{sec:preliminary} and Section~\ref{sec:ncf}.
Subsequently, we extract and compare multiple perceptual features from the two fields, including:
(i) geometry similarity,  
(ii) geometry gradient similarity,  
(iii) spatial color distribution similarity, and  
(iv) spatial color gradient similarity.
Finally, a quality score is obtained by aggregating the feature-wise similarities across all patches.

\subsection{Local Patch Construction}
\label{sec:localpatch}
Considering that different regions of a mesh may exhibit varying levels of geometric and texture complexity, as well as differing sensitivity to distortions, all subsequent processing in FMQM is based on segmented local surface patches.

Assume that a total of \( N_{\text{patch}} \) local patches are to be constructed.
We first select \( N_{\text{patch}} \) patch centers from the vertices of the reference mesh using either farthest point sampling (FPS) or random sampling (RS).  
For each patch center \( v_{\text{center}} \), we dynamically construct a local surface patch \( M_{\text{local}} \) composed of a set of vertices \( V_{\text{local}} \) and faces \( F_{\text{local}} \). 
Initially, \( V_{\text{local}} \) and \( F_{\text{local}} \) include the one-ring vertex and face neighbors of \( v_{\text{center}} \), denoted as \( \{\mathbf{v}_1, \mathbf{v}_2, \dots, \mathbf{v}_n\} \) and \( \{f_1, f_2, \dots, f_m\} \), respectively.
To ensure that each patch is large enough to extract useful features, we compare the total area of the faces in \( F_{\text{local}} \), denoted \( S_{\text{local}} \), with the total surface area of the entire mesh, denoted \( S_{\text{global}} \). 
If the ratio \( \frac{S_{\text{local}}}{S_{\text{global}}} < \frac{1}{N_{\text{patch}}} \), additional one-ring neighbors of the current vertices in \( V_{\text{local}} \) are iteratively added until the area ratio exceeds \( \frac{1}{N_{\text{patch}}} \), or no further neighbors are available.

Different choices of \( N_{\text{patch}} \) would result in different sizes of the local patch according to the construction process described above, thereby implicitly adjusting the perceptual domain size at which the mesh is locally analyzed. To illustrate the robustness of our method, we use a unique predefined \( N_{\text{patch}} \) for all mesh samples. The impact of \( N_{\text{patch}} \) will be studied in Section~\ref{sec:robutness}.

\subsection{Sample Point Generation}
\label{sec:sample}
To evaluate mesh quality, FMQM relies on discretized samples of the underlying geometry and color fields. Since these fields are defined in continuous 3D space, it is neither practical nor necessary to sample the entire domain. Instead, we restrict sampling to regions near the surface, where the local field behavior sufficiently captures the essential geometric and textural characteristics. Sampling beyond this vicinity may introduce ambiguity due to interference from unrelated geometry. Focusing on near-surface regions thus ensures both representational fidelity and perceptual relevance.

Suppose the total target number of samples is \( N_{\text{target}} \). After constructing a local surface patch, the number of samples allocated to this patch is computed as:
\(
N_{\text{local}} = \frac{S_{\text{local}}}{S_{\text{global}}} \cdot N_{\text{target}}
\),
where \( S_{\text{local}} \) and \( S_{\text{global}} \) denote the areas of the local patch and the entire mesh, respectively.

To ensure uniform sampling near the surface, we introduce a normal-offset strategy. The offset distance from the surface, \( \Delta d \), is drawn from a zero-mean Gaussian distribution:
\(
\Delta d \sim \mathcal{N}(0, \sigma_{\text{local}}^2)
\).
The variance \( \sigma_{\text{local}} \) adapts to the local patch scale:
\begin{equation}
\sigma_{\text{local}} = \sigma_{\text{base}} \cdot L_{\text{diagonal}}
\end{equation}
where \( L_{\text{diagonal}} \) is the diagonal length of the local patch bounding box and \( \sigma_{\text{base}} \) is a global scaling parameter.

For each triangle face \( f \in F_{\text{local}} \), with vertices \( (\mathbf{v}_0, \mathbf{v}_1, \mathbf{v}_2) \) and surface normal \( \mathbf{n}_f \), the allocated sample count is given by:
\(
N_f = \frac{S_f}{S_{\text{local}}} \cdot N_{\text{local}}
\).
For each face, \( N_f \) barycentric coordinate triples \( (\omega_0, \omega_1, \omega_2) \) are randomly generated, alongside \( N_f \) offset distances \( \Delta d \sim \mathcal{N}(0, \sigma_{\text{local}}^2) \). The 3D field sample positions are then computed as:
\begin{equation}
\mathbf{p} = \omega_0 \mathbf{v}_0 + \omega_1 \mathbf{v}_1 + \omega_2 \mathbf{v}_2 + \Delta d \cdot \mathbf{n}_f
\end{equation}

Importantly, these sample positions are determined solely based on the reference mesh, and are consistently used for both the reference and distorted fields, which indicates that the sampling inconsistency observed in other point-based metrics does not arise in our method. By comparing field values at identical spatial locations, FMQM avoids the need for explicit vertex-to-vertex correspondence. This enables robust comparison under varying mesh vertex densities and topologies, effectively resolving the correspondence ambiguity discussed in Section~\ref{sec:model-based}.

\subsection{Field Value Calculation}
\label{sec:field value}

For each sampled location \( \mathbf{p} \), FMQM computes the corresponding field vectors from both the reference and distorted meshes. These vectors are composed of geometric and color field attributes derived from the SDF and NCF, defined as:
\begin{equation}
\scalebox{0.95}{$
\begin{aligned}
F^r &= (x_p, y_p, z_p, x_{p_n}^r, y_{p_n}^r, z_{p_n}^r, f_{\text{SDF}}^r, f_{\text{NCF1}}^r, f_{\text{NCF2}}^r, f_{\text{NCF3}}^r), \\
F^d &= (x_p, y_p, z_p, x_{p_n}^d, y_{p_n}^d, z_{p_n}^d, f_{\text{SDF}}^d, f_{\text{NCF1}}^d, f_{\text{NCF2}}^d, f_{\text{NCF3}}^d)
\end{aligned}
$}
\end{equation}
Each vector encodes the sample position, its nearest surface point, and corresponding SDF and NCF responses. These descriptors provide a unified representation of geometry and texture at each sampling location, serving as the foundation for perceptual feature extraction in the subsequent stage.

\subsection{Field Feature Extraction and Comparison}
\label{sec:feature}
\subsubsection{Geometry Similarity}
\label{sec:geossim}
as described in Section~\ref{sec:sample}, the field samples are uniformly distributed in the vicinity of the reference surface. In this context, the SDF values effectively characterize the spatial alignment between the sampling points and the mesh surface.

To capture the global offset of the surface, we calculate the mean of the SDF values across all local samples, denoted as \( \mathrm{geoMean} \). As a first-order descriptor, \( \mathrm{geoMean} \) reflects the average displacement between the surface and the sample distribution, offering a quantitative measure of depth-related distortion.

In addition, local surface curvature plays a key role in visual phenomena such as bumpiness and shading. To assess consistency in this aspect, we extract second-order statistics from the SDF values within each patch, including the standard deviation \( \mathrm{geoStd} \) and the covariance \( \mathrm{geoCov} \). These descriptors reflect the spread and directional correlation of distances between the sampling points and the surface.  
Distortions such as noise, compression, or quantization may perturb vertex positions in a spatially non-uniform manner. Such deformations alter the curvature or smoothness of the surface, which in turn changes the local distribution of SDF values—resulting in variations in both \( \mathrm{geoStd} \) and \( \mathrm{geoCov} \). For instance, surface roughening due to quantization may increase variance, while directional artifacts can disrupt the natural correlation structure captured by the covariance.

Following the formulation in SSIM~\cite{ssim}, we evaluate geometry similarity from three statistical perspectives:

\begin{equation}
    \mathrm{geoMeanSSIM} = \frac{2 \cdot \mathrm{geoMean}^r \cdot \mathrm{geoMean}^d}{(\mathrm{geoMean}^r)^2 + (\mathrm{geoMean}^d)^2}
\end{equation}

\begin{equation}
    \mathrm{geoStdSSIM} = \frac{2 \cdot \mathrm{geoStd}^r \cdot \mathrm{geoStd}^d}{(\mathrm{geoStd}^r)^2 + (\mathrm{geoStd}^d)^2}
\end{equation}

\begin{equation}
    \mathrm{geoCovSSIM} = \frac{\mathrm{geoCov}}{\mathrm{geoStd}^r \cdot \mathrm{geoStd}^d}
\end{equation}
where the superscripts \( r \) and \( d \) refer to the reference and distorted meshes, respectively.
The final geometry similarity score is computed as the geometric mean of the three components:
\begin{equation}
\scalebox{0.9}{$
    \mathrm{geoSSIM} = \sqrt[3]{\mathrm{geoMeanSSIM} \cdot \mathrm{geoStdSSIM} \cdot \mathrm{geoCovSSIM}}
$}
\end{equation}

This formulation enables a holistic and perceptually relevant assessment of geometric similarity between the reference and distorted meshes.

\subsubsection{Geometry Gradient Similarity}
\label{sec:geoGssim}
The gradient of the SDF indicates the direction of the steepest change in distance values. Given that the gradient of a scalar field is always perpendicular to its level sets, the gradient near the surface naturally aligns with the normal direction of the nearest point \( p_n \). Therefore, discrepancies in SDF gradients effectively reflect differences in surface bending or curvature direction between reference and distorted meshes.

Since the mesh surface is piecewise smooth, the SDF is differentiable in most regions and its gradient satisfies the Eikonal equation~\cite{levelset}, i.e.,
\begin{equation}
|\nabla f_{\text{SDF}}| = 1
\end{equation}
Leveraging this property, we estimate the SDF gradient as
\begin{equation}
\nabla f_{\text{SDF}} = \text{sign}(f_{\text{SDF}}) \cdot \frac{\overrightarrow{p_n p}}{|\overrightarrow{p_n p}|}
\end{equation}

Because the gradient magnitude is always 1, we focus on the directional consistency between the gradients. The geometry gradient similarity is defined as the root mean square of the cosine similarities between reference and distorted gradients:
\begin{equation}
\mathrm{geoGraSSIM} = \sqrt{ \frac{1}{N_{\mathrm{local}}} \sum_{i=1}^{N_{\mathrm{local}}} \left( \nabla f^r_{\mathrm{SDF}_i} \cdot \nabla f^d_{\mathrm{SDF}_i} \right)^2 }
\label{eq:geoGraSSIM}
\end{equation}

This formulation captures directional bending consistency and complements scalar statistics like mean and variance by introducing gradient-level sensitivity to surface orientation, thereby enriching geometric analysis from a differential perspective.

\subsubsection{Space Color Distribution Similarity}
\label{sec:colorssim}
as introduced in Section~\ref{sec:ncfFeature}, the normal-direction and tangent-direction color patch units offer structurally aligned and perceptually grounded contexts for evaluating local variations in the color field. These units respectively characterize two complementary aspects of texture fidelity: the directional consistency of color propagation along surface normals and the local coherence of appearance within tangent-plane neighborhoods. Both are essential for modeling the visual impact of surface and texture distortions as perceived by the HVS.

Building upon these insights, we define a perceptual similarity measure that captures color field consistency in a spatially localized and channel-wise manner. Following the formulation of SSIM, the proposed metric evaluates statistical agreement between reference and distorted patches via three components—mean similarity, standard deviation similarity, and cross-covariance similarity—computed over the RGB color channels. Specifically, for the normal-direction patch unit, let \( \mu_j^r, \mu_j^d, \sigma_j^r, \sigma_j^d, \sigma_j^{rd} \) denote the mean, standard deviation, and cross-covariance of the color samples in the reference (\( r \)) and distorted (\( d \)) meshes for each channel \( j \in \{r, g, b\} \). The similarity terms are then defined as:

\begin{equation}
\mathrm{MeanSim}_j = \frac{2\mu_j^r \mu_j^d}{(\mu_j^r)^2 + (\mu_j^d)^2}
\end{equation}

\begin{equation}
\mathrm{StdSim}_j = \frac{2\sigma_j^r \sigma_j^d}{(\sigma_j^r)^2 + (\sigma_j^d)^2}
\end{equation}

\begin{equation}
\mathrm{CovSim}_j = \frac{\sigma_j^{rd}}{\sigma_j^r \sigma_j^d}
\end{equation}

\begin{equation}
\scalebox{0.95}{$
\mathrm{ColorNorSSIM}_j = \sqrt[3]{\mathrm{MeanSim}_j \cdot \mathrm{StdSim}_j \cdot \mathrm{CovSim}_j}
$}
\end{equation}

\begin{equation}
\mathrm{ColorNorSSIM} = \frac{1}{3} \sum_{j \in \{r,g,b\}} \mathrm{ColorNorSSIM}_j
\end{equation}
A similar computation is performed for the tangent-direction color patch unit, yielding \( \mathrm{ColorTanSSIM} \). Finally, the space color distribution similarity is computed as the geometric mean of the two components:
\begin{equation}
\mathrm{ColorSSIM} = \sqrt{\mathrm{ColorNorSSIM} \cdot \mathrm{ColorTanSSIM}}
\end{equation}

To reduce redundancy and improve computational efficiency, color patch units are only extracted from a randomly selected subset of \( N_{\mathrm{color\_patch}} \) sample points, since neighboring units yield nearly identical statistics. The final \( \mathrm{ColorSSIM} \) score is averaged over all selected units, where \( N_{\mathrm{color\_patch}} \ll N_{\mathrm{local}} \).

\subsubsection{Space Color Gradient Similarity}
\label{sec:colorGssim}
In traditional IQA, spatial gradients play a critical role in capturing salient structural information such as edges, contours, and texture transitions—all of which are highly perceptible to the HVS. Extending this insight to 3D textured meshes, we interpret the gradient of the NCF as an indicator of color and structure transitions over the surface, allowing us to quantify distortions that may not be captured by pixel-wise color differences alone.

Direct computation of the NCF gradient is intractable in the continuous domain. However, as established in Section~\ref{sec:ncfProperty}, NCF values remain nearly invariant along the surface normal and exhibit meaningful variation primarily along the tangent direction. This observation suggests that perceptually relevant gradients reside within the local tangent plane. We therefore estimate the NCF gradient using color differences between sampled points within the tangent-direction patch unit defined in Section~\ref{sec:ncfFeature}, which structurally resembles a rendered image patch and preserves local surface coherence.

For a query point \( P \), and for each \( P' \in P_{\mathrm{tangent}} \), we approximate the gradient magnitude of the \( \mathrm{NCF} \) for color channel \( j \in \{r, g, b\} \) as the maximum directional color difference:
\begin{equation}
    \left| \nabla f_{\mathrm{NCF}_j}(P) \right| = \max_{P' \in P_{\mathrm{tangent}}} \left| \frac{f_{\mathrm{NCF}_j}(P') - f_{\mathrm{NCF}_j}(P)}{|\overrightarrow{P'P}|} \right|
\end{equation}
The corresponding gradient direction is estimated using the unit vector pointing toward the direction of maximum change:
\begin{equation}
\scalebox{0.85}{$
    \mathrm{dir}(\nabla f_{\mathrm{NCF}_j}(P)) = \mathrm{sign}(f_{\mathrm{NCF}_j}(P_{\max}) - f_{\mathrm{NCF}_j}(P)) \cdot \frac{\overrightarrow{PP_{\max}}}{|\overrightarrow{PP_{\max}}|}
$}
\end{equation}

To evaluate structural fidelity between the reference and distorted color fields, we compare both the angular and magnitude components of the estimated gradients. Let \( \nabla f_{\mathrm{NCF}_j}^r(P_i) \) and \( \nabla f_{\mathrm{NCF}_j}^d(P_i) \) denote the estimated gradients at sample \( P_i \) from the reference and distorted fields, respectively. Then the angular similarity is computed as:
\begin{equation}
\scalebox{0.80}{$
    \mathrm{colorGraASSIM}_j = \sqrt{ \frac{1}{N_{\mathrm{local}}} \sum_{i=1}^{N_{\mathrm{local}}} \left( \frac{ \nabla f_{\mathrm{NCF}_j}^r(P_i) }{ |\nabla f_{\mathrm{NCF}_j}^r(P_i)| } \cdot \frac{ \nabla f_{\mathrm{NCF}_j}^d(P_i) }{ |\nabla f_{\mathrm{NCF}_j}^d(P_i)| } \right)^2 }
$}
\end{equation}
and the magnitude similarity is computed as:
\begin{equation}
\scalebox{0.80}{$
    \mathrm{colorGraLSSIM}_j = 1 - \sqrt{ \frac{1}{N_{\mathrm{local}}} \sum_{i=1}^{N_{\mathrm{local}}} \frac{ \left| |\nabla f_{\mathrm{NCF}_j}^r(P_i)|^2 - |\nabla f_{\mathrm{NCF}_j}^d(P_i)|^2 \right| }{ |\nabla f_{\mathrm{NCF}_j}^r(P_i)|^2 + |\nabla f_{\mathrm{NCF}_j}^d(P_i)|^2 + \epsilon } }
$}
\end{equation}
where \( \epsilon \) is a small constant for numerical stability.

Averaging across RGB channels and combining the two components multiplicatively, we define the final space color gradient similarity as:
\begin{equation}
\scalebox{0.95}{$
    \mathrm{colorGraSSIM} = \sqrt{ \mathrm{colorGraASSIM} \cdot \mathrm{colorGraLSSIM} }
    $}
\end{equation}

The estimated NCF gradient characterizes local transitions in color and structure over the mesh surface. By comparing its direction and magnitude between reference and distorted fields, our gradient-based similarity effectively highlights subtle perceptual degradations such as edge blurring or misaligned texture contours. As a complement to color-based patch statistics, this metric enhances the sensitivity to structural distortions that are critical for assessing texture fidelity in 3D mesh representations.

\subsubsection{Feature Pooling}
Each proposed feature is inherently defined within the range \([0, 1]\), where \(0\) indicates severe distortion and \(1\) denotes perfect quality. To derive an overall mesh quality score, we aggregate the geometric and color-based components using the geometric mean, which balances their contributions while preserving scale consistency:
\begin{equation}
\scalebox{0.8}{$
\mathrm{FMQM} = \sqrt[4]{\mathrm{geoSSIM} \cdot \mathrm{geoGraSSIM} \cdot \mathrm{colorSSIM} \cdot \mathrm{colorGraSSIM}}
$}
\end{equation}
Additional results with Lab color space and average pooling are included in the supplementary to verify robustness.

\section{Experiments and Results}
\label{sec:experiment}
In this section, we comprehensively evaluate the performance of FMQM against SOTA TMQA metrics across three textured mesh datasets.  
We begin by describing the datasets, including their scale and the types of distortions they contain.  
Next, we detail the SOTA baseline methods and their parameter configurations.  
We then present comparisons from three perspectives: overall performance, distortion-specific performance, and per-sample prediction accuracy.  
Additionally, we analyze the robustness and efficiency of FMQM.  
Finally, we conduct an ablation study to assess the individual contributions of each proposed feature.

\subsection{Datasets}
To ensure the generalizability and robustness of the proposed metric, evaluations are conducted on three publicly available TMQA datasets: TSMD\cite{tsmd}, SJTU-TMQA\cite{sjtumqa}, and YANA\cite{yana2023}.

The TSMD dataset contains 42 reference meshes and 210 distorted meshes generated using the lossy compression pipeline developed by the AOMedia VVM Working Group~\cite{vvmCfP}.  
The distortion process includes a sequence of operations applied to mesh geometry—dequantization, triangulation, decimation, and Draco compression~\cite{draco}—as well as texture processing steps such as downscaling and AV1 (libaom) compression.  
Each reference mesh is associated with five rate-distortion levels, covering a wide range of perceptual qualities.

The SJTU-TMQA dataset consists of 945 distorted samples derived from 21 reference meshes.  
Each reference mesh is subjected to eight types of distortions:  
(i) Geometry distortions: 5 levels of Vertex Gaussian Noise (GN), 5 levels of Quantization Position (QP), 4 levels of Simplification without Texture (SOT), and 5 levels of Simplification with Texture (SWT);  
(ii) Texture distortions: 5 levels of Downsampling (DS) and 6 levels of Texture Map Compression (TMC);  
(iii) Mixed distortions: 6 levels of Mixed Quantization (MQ), and 9 levels of combined Geometry and Texture Map Compression (GTC).

The YANA dataset includes 55 reference meshes and 343,750 distorted samples, among which 3,000 are annotated with subjective quality scores.  
Each reference mesh is degraded using five types of mixed distortions: 10 levels of LoD simplification, 5 levels of geometry quantization, 5 levels of UV coordinate quantization, 5 levels of texture downsampling, and 5 levels of JPEG compression.  
Consistent with the evaluation protocol adopted by the authors of the YANA dataset, we restrict our experiments to the 3,000 distorted samples with subjective annotations.

\subsection{SOTA Metrics}
Three types of SOTA metrics mentioned in Section~\ref{sec:relatedwork} are used for comparison. All these methods are non-learning-based and require no dataset-specific training, enabling consistent evaluation across datasets under a unified protocol. 
\begin{itemize}
    \item \textbf{Projection-based metrics}: These are further divided into image-based and video-based metrics. 
    For image-based metrics, $16$ evenly distributed viewpoints are selected as suggested by MPEG~\cite{mmetric}, and depth/color images of resolution 2048\texttimes2048 are rendered.
    Then, depth image PSNR ($\rm Geo_{psnr}$) and color image PSNR ($\rm RGB_{psnr}$ and $\rm YUV_{psnr}$) are computed.
    For video-based metrics, PSNR, SSIM~\cite{ssim}, MS-SSIM~\cite{msssim}, 3-SSIM~\cite{3ssim}, VQM~\cite{vqm}, and VMAF~\cite{vmaf} are calculated using the MSU Video Quality Measurement Tool~\cite{vqmt} based on the rendered videos used in subjective experiments. 
    The video resolutions for TSMD, SJTU-TMQA, and YANA are 1920\texttimes1080, 1920\texttimes1080, and 650\texttimes550, respectively. All videos have a framerate of 30 fps and durations of 18s, 16.5s, and 8s.

    \item \textbf{Model-based metrics}: These include MESH~\cite{mesh} based on vertex-to-face distance, GL~\cite{gl} and TPDM~\cite{tpdm} based on per-vertex geometry features, MSDM2~\cite{msdm2} capturing geometry structure, and GeodesicPSIM~\cite{geodesicpsim} integrating both geometry and texture.

    \item \textbf{Point-based metrics}: Point clouds are sampled using grid sampling with a resolution of $1024$ recommended by MPEG~\cite{mmetric}.  
    The number of sampled points varies among different samples but typically ranges from $1,000,000$ to $2,000,000$.
    Then, five full-reference PCQA metrics are applied: D1, D2~\cite{d1d2}, point color PSNR ($\rm YUV_{psnr}$), $\rm PCQM_{psnr}$~\cite{pcqm}, and GraphSIM~\cite{graphsim}.
\end{itemize}

\subsection{Experimental Setup}
The hyperparameters required for FMQM are configured as follows:
\begin{itemize}
    \item Target number of sampled points: $N_\text{target} = 200{,}000$
    \item Patch selection: $N_\text{patch} = 250$ local surface patches are selected using FPS
    \item Sampling variance: the offset variance for spatial sampling is set to $\sigma_\text{base} = 0.05$
    \item Neighbor selection in color field: neighbor points are selected based on the criteria $\text{norThre} = 0.95$, $\text{tanThre} = 0.05$, and a radius $r = 0.125 \times L_\text{diagonal}$
    \item Color field feature extraction: for each surface patch, $N_\text{color\_patch} = 100$ points are used to extract NCF features
\end{itemize}

\subsection{Performance Evaluation}
Following the recommendation of the Video Quality Experts Group~\cite{logestic}, a nonlinear regression function is applied to map the raw scores produced by each objective metric onto the same scale as the subjective quality scores.
After regression, three commonly used correlation metrics are computed to evaluate the performance of the objective TMQA metrics:
Pearson’s Linear Correlation Coefficient (PLCC), Spearman’s Rank Order Correlation Coefficient (SROCC), and Root Mean Square Error (RMSE).
\subsubsection{Overall performance}
\begin{table*}[]
\centering
\caption{Metric performance on TSMD, SJTU-TMQA, and YANA dataset. The {\color[HTML]{FE0000} best}, {\color[HTML]{32CB00} second} and {\color[HTML]{3166FF} third} results are bolded in {\color[HTML]{FE0000} red}, {\color[HTML]{32CB00} green} and {\color[HTML]{3166FF} blue}.}
\label{tab:overall}
\resizebox{\textwidth}{!}{%
\begin{tabular}{|l|l|ccc|ccc|ccc|}
\hline
\textbf{}                                   & \textbf{}                  & \multicolumn{3}{c|}{\textbf{TSMD}}                                                                                                                              & \multicolumn{3}{c|}{\textbf{SJTU-TMQA}}                                                                                                                         & \multicolumn{3}{c|}{\textbf{YANA}}                                                                                                                              \\ \hline
\textbf{}                                   & \textbf{}                  & \multicolumn{1}{c|}{\textbf{PLCC}}                         & \multicolumn{1}{c|}{\textbf{SROCC}}                        & \textbf{RMSE}                         & \multicolumn{1}{c|}{\textbf{PLCC}}                         & \multicolumn{1}{c|}{\textbf{SROCC}}                        & \textbf{RMSE}                         & \multicolumn{1}{c|}{\textbf{PLCC}}                         & \multicolumn{1}{c|}{\textbf{SROCC}}                        & \textbf{RMSE}                         \\ \hline
                                            & \textbf{Geo\_PSNR}         & \multicolumn{1}{c|}{\textbf{0.730}}                        & \multicolumn{1}{c|}{\textbf{0.731}}                        & \textbf{0.796}                        & \multicolumn{1}{c|}{\textbf{0.329}}                        & \multicolumn{1}{c|}{\textbf{0.091}}                        & \textbf{2.258}                        & \multicolumn{1}{c|}{\textbf{0.314}}                        & \multicolumn{1}{c|}{\textbf{0.262}}                        & \textbf{0.927}                        \\ \cline{2-11} 
                                            & \textbf{RGB\_PSNR}         & \multicolumn{1}{c|}{\textbf{0.688}}                        & \multicolumn{1}{c|}{\textbf{0.674}}                        & \textbf{0.845}                        & \multicolumn{1}{c|}{\textbf{0.561}}                        & \multicolumn{1}{c|}{\textbf{0.585}}                        & \textbf{1.980}                        & \multicolumn{1}{c|}{\textbf{0.391}}                        & \multicolumn{1}{c|}{\textbf{0.393}}                        & \textbf{0.898}                        \\ \cline{2-11} 
                                            & \textbf{YUV\_PSNR}         & \multicolumn{1}{c|}{\textbf{0.684}}                        & \multicolumn{1}{c|}{\textbf{0.677}}                        & \textbf{0.849}                        & \multicolumn{1}{c|}{\textbf{0.523}}                        & \multicolumn{1}{c|}{\textbf{0.536}}                        & \textbf{2.038}                        & \multicolumn{1}{c|}{\textbf{0.402}}                        & \multicolumn{1}{c|}{\textbf{0.402}}                        & \textbf{0.894}                        \\ \cline{2-11} 
                                            & \textbf{PSNR}              & \multicolumn{1}{c|}{\textbf{0.527}}                        & \multicolumn{1}{c|}{\textbf{0.544}}                        & \textbf{0.989}                        & \multicolumn{1}{c|}{\textbf{0.406}}                        & \multicolumn{1}{c|}{\textbf{0.442}}                        & \textbf{2.185}                        & \multicolumn{1}{c|}{\textbf{0.490}}                        & \multicolumn{1}{c|}{\textbf{0.475}}                        & \textbf{0.851}                        \\ \cline{2-11} 
                                            & \textbf{SSIM}              & \multicolumn{1}{c|}{\textbf{0.512}}                        & \multicolumn{1}{c|}{\textbf{0.506}}                        & \textbf{1.000}                        & \multicolumn{1}{c|}{\textbf{0.462}}                        & \multicolumn{1}{c|}{\textbf{0.484}}                        & \textbf{2.121}                        & \multicolumn{1}{c|}{\textbf{0.415}}                        & \multicolumn{1}{c|}{\textbf{0.385}}                        & \textbf{0.888}                        \\ \cline{2-11} 
                                            & \textbf{MS-SSIM}           & \multicolumn{1}{c|}{\textbf{0.661}}                        & \multicolumn{1}{c|}{\textbf{0.654}}                        & \textbf{0.873}                        & \multicolumn{1}{c|}{\textbf{0.518}}                        & \multicolumn{1}{c|}{\textbf{0.553}}                        & \textbf{2.046}                        & \multicolumn{1}{c|}{\textbf{0.495}}                        & \multicolumn{1}{c|}{\textbf{0.473}}                        & \textbf{0.848}                        \\ \cline{2-11} 
                                            & \textbf{3-SSIM}            & \multicolumn{1}{c|}{\textbf{0.683}}                        & \multicolumn{1}{c|}{\textbf{0.674}}                        & \textbf{0.850}                        & \multicolumn{1}{c|}{\textbf{0.474}}                        & \multicolumn{1}{c|}{\textbf{0.518}}                        & \textbf{2.105}                        & \multicolumn{1}{c|}{{\color[HTML]{32CB00} \textbf{0.640}}} & \multicolumn{1}{c|}{{\color[HTML]{32CB00} \textbf{0.631}}} & {\color[HTML]{32CB00} \textbf{0.750}} \\ \cline{2-11} 
                                            & \textbf{VQM}               & \multicolumn{1}{c|}{\textbf{0.671}}                        & \multicolumn{1}{c|}{\textbf{0.670}}                        & \textbf{0.862}                        & \multicolumn{1}{c|}{\textbf{0.431}}                        & \multicolumn{1}{c|}{\textbf{0.457}}                        & \textbf{2.157}                        & \multicolumn{1}{c|}{\textbf{0.520}}                        & \multicolumn{1}{c|}{\textbf{0.511}}                        & \textbf{0.834}                        \\ \cline{2-11} 
\multirow{-9}{*}{\textbf{Projection-based}} & \textbf{VMAF}              & \multicolumn{1}{c|}{\textbf{0.704}}                        & \multicolumn{1}{c|}{\textbf{0.700}}                        & \textbf{0.827}                        & \multicolumn{1}{c|}{\textbf{0.493}}                        & \multicolumn{1}{c|}{\textbf{0.527}}                        & \textbf{2.080}                        & \multicolumn{1}{c|}{{\color[HTML]{3166FF} \textbf{0.607}}} & \multicolumn{1}{c|}{{\color[HTML]{3166FF} \textbf{0.592}}} & {\color[HTML]{3166FF} \textbf{0.775}} \\ \hline
                                            & \textbf{MESH}              & \multicolumn{1}{c|}{\textbf{0.281}}                        & \multicolumn{1}{c|}{\textbf{0.390}}                        & \textbf{1.117}                        & \multicolumn{1}{c|}{\textbf{0.148}}                        & \multicolumn{1}{c|}{\textbf{0.058}}                        & \textbf{2.365}                        & \multicolumn{1}{c|}{\textbf{0.142}}                        & \multicolumn{1}{c|}{\textbf{0.173}}                        & \textbf{0.966}                        \\ \cline{2-11} 
                                            & \textbf{GL}                & \multicolumn{1}{c|}{\textbf{0.202}}                        & \multicolumn{1}{c|}{\textbf{0.323}}                        & \textbf{1.140}                        & \multicolumn{1}{c|}{\textbf{0.106}}                        & \multicolumn{1}{c|}{\textbf{0.078}}                        & \textbf{2.378}                        & \multicolumn{1}{c|}{\textbf{0.139}}                        & \multicolumn{1}{c|}{\textbf{0.119}}                        & \textbf{0.967}                        \\ \cline{2-11} 
                                            & \textbf{TPDM}              & \multicolumn{1}{c|}{\textbf{0.180}}                        & \multicolumn{1}{c|}{\textbf{0.064}}                        & \textbf{1.149}                        & \multicolumn{1}{c|}{\textbf{0.344}}                        & \multicolumn{1}{c|}{\textbf{0.101}}                        & \textbf{2.245}                        & \multicolumn{1}{c|}{\textbf{0.438}}                        & \multicolumn{1}{c|}{\textbf{0.404}}                        & \textbf{0.878}                        \\ \cline{2-11} 
                                            & \textbf{MSDM2}             & \multicolumn{1}{c|}{\textbf{0.000}}                        & \multicolumn{1}{c|}{\textbf{0.000}}                        & \textbf{0.000}                        & \multicolumn{1}{c|}{\textbf{0.393}}                        & \multicolumn{1}{c|}{\textbf{0.053}}                        & \textbf{2.198}                        & \multicolumn{1}{c|}{\textbf{0.357}}                        & \multicolumn{1}{c|}{\textbf{0.335}}                        & \textbf{0.912}                        \\ \cline{2-11} 
\multirow{-5}{*}{\textbf{Model-based}}      & \textbf{GeodesicPSIM(FPS)} & \multicolumn{1}{c|}{{\color[HTML]{32CB00} \textbf{0.806}}} & \multicolumn{1}{c|}{{\color[HTML]{32CB00} \textbf{0.807}}} & {\color[HTML]{32CB00} \textbf{0.688}} & \multicolumn{1}{c|}{{\color[HTML]{32CB00} \textbf{0.725}}} & \multicolumn{1}{c|}{{\color[HTML]{32CB00} \textbf{0.757}}} & {\color[HTML]{32CB00} \textbf{1.647}} & \multicolumn{1}{c|}{\textbf{0.444}}                        & \multicolumn{1}{c|}{\textbf{0.416}}                        & \textbf{0.874}                        \\ \hline
                                            & \textbf{D1}                & \multicolumn{1}{c|}{\textbf{0.725}}                        & \multicolumn{1}{c|}{\textbf{0.650}}                        & \textbf{0.802}                        & \multicolumn{1}{c|}{\textbf{0.367}}                        & \multicolumn{1}{c|}{\textbf{0.132}}                        & \textbf{2.224}                        & \multicolumn{1}{c|}{\textbf{0.378}}                        & \multicolumn{1}{c|}{\textbf{0.391}}                        & \textbf{0.904}                        \\ \cline{2-11} 
                                            & \textbf{D2}                & \multicolumn{1}{c|}{\textbf{0.541}}                        & \multicolumn{1}{c|}{\textbf{0.468}}                        & \textbf{0.978}                        & \multicolumn{1}{c|}{\textbf{0.360}}                        & \multicolumn{1}{c|}{\textbf{0.129}}                        & \textbf{2.231}                        & \multicolumn{1}{c|}{\textbf{0.214}}                        & \multicolumn{1}{c|}{\textbf{0.217}}                        & \textbf{0.953}                        \\ \cline{2-11} 
                                            & \textbf{YUV\_PSNR}         & \multicolumn{1}{c|}{\textbf{0.619}}                        & \multicolumn{1}{c|}{\textbf{0.604}}                        & \textbf{0.914}                        & \multicolumn{1}{c|}{{\color[HTML]{333333} \textbf{0.635}}} & \multicolumn{1}{c|}{{\color[HTML]{333333} \textbf{0.671}}} & {\color[HTML]{333333} \textbf{1.848}} & \multicolumn{1}{c|}{\textbf{0.181}}                        & \multicolumn{1}{c|}{\textbf{0.200}}                        & \textbf{0.960}                        \\ \cline{2-11} 
                                            & \textbf{PCQM\_PSNR}        & \multicolumn{1}{c|}{\textbf{0.757}}                        & \multicolumn{1}{c|}{\textbf{0.747}}                        & \textbf{0.761}                        & \multicolumn{1}{c|}{\textbf{0.540}}                        & \multicolumn{1}{c|}{\textbf{0.576}}                        & \textbf{2.012}                        & \multicolumn{1}{c|}{\textbf{0.384}}                        & \multicolumn{1}{c|}{\textbf{0.384}}                        & \textbf{0.901}                        \\ \cline{2-11} 
                                            & \textbf{GraphSIM}          & \multicolumn{1}{c|}{{\color[HTML]{3166FF} \textbf{0.765}}} & \multicolumn{1}{c|}{{\color[HTML]{3166FF} \textbf{0.752}}} & {\color[HTML]{3166FF} \textbf{0.750}} & \multicolumn{1}{c|}{{\color[HTML]{3166FF} \textbf{0.711}}} & \multicolumn{1}{c|}{{\color[HTML]{3166FF} \textbf{0.755}}} & {\color[HTML]{3166FF} \textbf{1.681}} & \multicolumn{1}{c|}{\textbf{0.562}}                        & \multicolumn{1}{c|}{\textbf{0.557}}                        & \textbf{0.807}                        \\ \cline{2-11} 
\multirow{-6}{*}{\textbf{Point-based}}      & \textbf{FMQM}              & \multicolumn{1}{c|}{{\color[HTML]{FE0000} \textbf{0.810}}} & \multicolumn{1}{c|}{{\color[HTML]{FE0000} \textbf{0.811}}} & {\color[HTML]{FE0000} \textbf{0.683}} & \multicolumn{1}{c|}{{\color[HTML]{FE0000} \textbf{0.728}}} & \multicolumn{1}{c|}{{\color[HTML]{FE0000} \textbf{0.760}}} & {\color[HTML]{FE0000} \textbf{1.639}} & \multicolumn{1}{c|}{{\color[HTML]{FE0000} \textbf{0.674}}} & \multicolumn{1}{c|}{{\color[HTML]{FE0000} \textbf{0.666}}} & {\color[HTML]{FE0000} \textbf{0.721}} \\ \hline
\end{tabular}
}
\end{table*}
Table~\ref{tab:overall} summarizes the performance of all evaluated metrics on the TSMD, SJTU-TMQA, and YANA datasets. Note that MSDM2 was excluded from TSMD due to non-manifold geometry issues encountered in multiple samples.
Several important observations can be drawn from the results:

i) FMQM consistently outperforms all other metrics across the three datasets.  
It ranks first in both PLCC and SROCC on all datasets (e.g., SROCC: 0.811 on TSMD, 0.760 on SJTU-TMQA, 0.666 on YANA), and achieves the lowest RMSE in each case. These results underscore its robustness and superior generalization ability across diverse mesh types and distortion categories.

ii) Projection-based metrics perform well on TSMD but may degrade significantly on YANA.
For example, $\rm Geo_{psnr}$ achieves a relatively high PLCC of 0.730 on TSMD, but drops to 0.314 on YANA. This performance gap illustrates the limited generalizability of projection-based methods when handling complex 3D distortions beyond what 2D projections can effectively capture.

iii) Among model-based metrics, only GeodesicPSIM demonstrates stable and competitive performance.
It achieves the second-highest PLCC scores on TSMD (0.806) and SJTU-TMQA (0.725), markedly outperforming alternative model-based approaches such as GL and TPDM (e.g., GL PLCC: 0.202 on TSMD, 0.139 on YANA).
This superior performance is attributable to its integration of both geometric and texture information, in contrast to other methods in this category, which primarily focus on geometric structure alone and therefore exhibit limited perceptual sensitivity.

iv) Point-based metrics vary in performance, with GraphSIM showing the most consistency.
D1 and $\rm PCQM_{psnr}$ perform well on TSMD (PLCC: 0.725 and 0.757) but decline on YANA (PLCC: 0.378 and 0.384). In contrast, GraphSIM maintains relatively stable performance across all datasets (PLCC: 0.765, 0.711, 0.562),  likely due to its perceptually motivated design that incorporates the spatial distribution and gradient of color through three-moment color features.

v) SJTU-TMQA and YANA present greater challenges than TSMD.
The correlation scores on these two datasets are generally lower across all metrics, reflecting the increased distortion diversity and content complexity. This further emphasizes the need for TMQA methods that are both perceptually grounded and broadly generalizable.

\subsubsection{Per-distortion performance}
To further investigate metric behavior under different types of distortion, we perform a fine-grained analysis on the SJTU-TMQA dataset. The results are summarized in Table~\ref{tab:per-dis}, from which several insights can be drawn:

i) Projection-based metrics are particularly effective for quantization (QP, MQ) and downsampling (DS) distortions. These degradations largely preserve the mesh’s global structure and maintain approximate pixel-wise correspondence in the rendered images. For instance, quantization mainly involves rounding vertex attributes, which introduces limited displacement and preserves the overall shape in projections. In contrast, distortions such as Gaussian noise (GN) or mesh simplification (SOT, SWT) can cause significant vertex displacements or texture misalignments, leading to noticeable structural inconsistencies in rendered views and a sharp drop in correlation.

ii) Model-based metrics such as MESH, GL, and MSDM2 generally underperform across most distortions, highlighting their limited capacity to capture perceptually relevant texture degradation. TPDM, which utilizes curvature-based geometric descriptors, shows relatively strong performance for GN (SROCC: 0.766) and SWT (SROCC: 0.801), suggesting that curvature cues are helpful in modeling structure-sensitive distortions. GeodesicPSIM achieves top performance on DS and TMC, suggesting that incorporating geodesic distance with color information is effective for modeling perceptual quality degradation caused by texture impairments.

iii) Point-based metrics indicate that D1 and D2 are particularly suited to quantization artifacts, achieving strong correlations on QP and MQ distortions. GraphSIM shows the most consistent performance across different distortion types (except GN), achieving top performance in SOT and GTC. This highlights the robustness of its graph- and gradient-based feature representations for perceptual quality assessment in 3D content.

iv) The proposed FMQM exhibits stable performance across all distortion categories, ranking first in three cases, second in three, and third in another three. This consistency across diverse distortion types underscores the robustness and generalizability of the proposed features.
\begin{table*}[]
\centering
\caption{Metric SROCC on difference types of distortion on SJTU-TMQA dataset.
}
\label{tab:per-dis}
\resizebox{\textwidth}{!}{%
\begin{tabular}{|l|l|cccc|cc|cc|c|}
\hline
\textbf{}                                   & \textbf{}             & \multicolumn{4}{c|}{\textbf{Geometry}}                                                                                                                                                                                       & \multicolumn{2}{c|}{\textbf{Texture}}                                                              & \multicolumn{2}{c|}{\textbf{Mix}}                                                                  & \textbf{Overall}                      \\ \hline
\textbf{}                                   & \textbf{}             & \multicolumn{1}{c|}{\textbf{GN}}                           & \multicolumn{1}{c|}{\textbf{QP}}                           & \multicolumn{1}{c|}{\textbf{SOT}}                          & \textbf{SWT}                          & \multicolumn{1}{c|}{\textbf{DS}}                           & \textbf{TMC}                          & \multicolumn{1}{c|}{\textbf{MQ}}                           & \textbf{GTC}                          & \textbf{Overall}                      \\ \hline
                                            & \textbf{GEO\_PSNR}    & \multicolumn{1}{c|}{{\color[HTML]{333333} \textbf{0.412}}} & \multicolumn{1}{c|}{{\color[HTML]{333333} \textbf{0.621}}} & \multicolumn{1}{c|}{{\color[HTML]{333333} \textbf{0.300}}} & {\color[HTML]{333333} \textbf{0.463}} & \multicolumn{1}{c|}{{\color[HTML]{333333} \textbf{-}}}     & {\color[HTML]{333333} \textbf{-}}     & \multicolumn{1}{c|}{{\color[HTML]{333333} \textbf{0.687}}} & {\color[HTML]{333333} \textbf{0.196}} & {\color[HTML]{333333} \textbf{0.091}} \\ \cline{2-11} 
                                            & \textbf{RGB\_PSNR}    & \multicolumn{1}{c|}{{\color[HTML]{333333} \textbf{0.354}}} & \multicolumn{1}{c|}{{\color[HTML]{333333} \textbf{0.637}}} & \multicolumn{1}{c|}{{\color[HTML]{333333} \textbf{0.629}}} & {\color[HTML]{333333} \textbf{0.505}} & \multicolumn{1}{c|}{{\color[HTML]{333333} \textbf{0.738}}} & {\color[HTML]{333333} \textbf{0.619}} & \multicolumn{1}{c|}{{\color[HTML]{333333} \textbf{0.672}}} & {\color[HTML]{333333} \textbf{0.566}} & {\color[HTML]{333333} \textbf{0.585}} \\ \cline{2-11} 
                                            & \textbf{YUV\_PSNR}    & \multicolumn{1}{c|}{{\color[HTML]{333333} \textbf{0.336}}} & \multicolumn{1}{c|}{{\color[HTML]{333333} \textbf{0.621}}} & \multicolumn{1}{c|}{{\color[HTML]{333333} \textbf{0.634}}} & {\color[HTML]{333333} \textbf{0.478}} & \multicolumn{1}{c|}{{\color[HTML]{333333} \textbf{0.739}}} & {\color[HTML]{333333} \textbf{0.591}} & \multicolumn{1}{c|}{{\color[HTML]{333333} \textbf{0.657}}} & {\color[HTML]{333333} \textbf{0.487}} & {\color[HTML]{333333} \textbf{0.536}} \\ \cline{2-11} 
                                            & \textbf{PSNR}         & \multicolumn{1}{c|}{{\color[HTML]{333333} \textbf{0.379}}} & \multicolumn{1}{c|}{{\color[HTML]{333333} \textbf{0.659}}} & \multicolumn{1}{c|}{{\color[HTML]{333333} \textbf{0.477}}} & {\color[HTML]{333333} \textbf{0.514}} & \multicolumn{1}{c|}{{\color[HTML]{333333} \textbf{0.729}}} & {\color[HTML]{333333} \textbf{0.460}} & \multicolumn{1}{c|}{{\color[HTML]{333333} \textbf{0.678}}} & {\color[HTML]{333333} \textbf{0.053}} & {\color[HTML]{333333} \textbf{0.442}} \\ \cline{2-11} 
                                            & \textbf{SSIM}         & \multicolumn{1}{c|}{{\color[HTML]{333333} \textbf{0.029}}} & \multicolumn{1}{c|}{{\color[HTML]{333333} \textbf{0.606}}} & \multicolumn{1}{c|}{{\color[HTML]{333333} \textbf{0.469}}} & {\color[HTML]{333333} \textbf{0.396}} & \multicolumn{1}{c|}{{\color[HTML]{333333} \textbf{0.692}}} & {\color[HTML]{333333} \textbf{0.498}} & \multicolumn{1}{c|}{{\color[HTML]{333333} \textbf{0.674}}} & {\color[HTML]{333333} \textbf{0.013}} & {\color[HTML]{333333} \textbf{0.484}} \\ \cline{2-11} 
                                            & \textbf{MS-SSIM}      & \multicolumn{1}{c|}{{\color[HTML]{333333} \textbf{0.226}}} & \multicolumn{1}{c|}{{\color[HTML]{333333} \textbf{0.676}}} & \multicolumn{1}{c|}{{\color[HTML]{333333} \textbf{0.502}}} & {\color[HTML]{333333} \textbf{0.522}} & \multicolumn{1}{c|}{{\color[HTML]{333333} \textbf{0.803}}} & {\color[HTML]{333333} \textbf{0.658}} & \multicolumn{1}{c|}{{\color[HTML]{333333} \textbf{0.721}}} & {\color[HTML]{333333} \textbf{0.158}} & {\color[HTML]{333333} \textbf{0.553}} \\ \cline{2-11} 
                                            & \textbf{3-SSIM}       & \multicolumn{1}{c|}{{\color[HTML]{333333} \textbf{0.373}}} & \multicolumn{1}{c|}{{\color[HTML]{333333} \textbf{0.737}}} & \multicolumn{1}{c|}{{\color[HTML]{333333} \textbf{0.469}}} & {\color[HTML]{333333} \textbf{0.577}} & \multicolumn{1}{c|}{{\color[HTML]{333333} \textbf{0.818}}} & {\color[HTML]{333333} \textbf{0.585}} & \multicolumn{1}{c|}{{\color[HTML]{333333} \textbf{0.744}}} & {\color[HTML]{333333} \textbf{0.113}} & {\color[HTML]{333333} \textbf{0.518}} \\ \cline{2-11} 
                                            & \textbf{VQM}          & \multicolumn{1}{c|}{{\color[HTML]{333333} \textbf{0.540}}} & \multicolumn{1}{c|}{{\color[HTML]{333333} \textbf{0.734}}} & \multicolumn{1}{c|}{{\color[HTML]{333333} \textbf{0.548}}} & {\color[HTML]{333333} \textbf{0.639}} & \multicolumn{1}{c|}{{\color[HTML]{3166FF} \textbf{0.848}}} & {\color[HTML]{333333} \textbf{0.573}} & \multicolumn{1}{c|}{{\color[HTML]{333333} \textbf{0.737}}} & {\color[HTML]{333333} \textbf{0.264}} & {\color[HTML]{333333} \textbf{0.457}} \\ \cline{2-11} 
\multirow{-9}{*}{\textbf{Projection-based}} & \textbf{VMAF}         & \multicolumn{1}{c|}{{\color[HTML]{333333} \textbf{0.476}}} & \multicolumn{1}{c|}{{\color[HTML]{333333} \textbf{0.727}}} & \multicolumn{1}{c|}{{\color[HTML]{333333} \textbf{0.484}}} & {\color[HTML]{333333} \textbf{0.601}} & \multicolumn{1}{c|}{{\color[HTML]{333333} \textbf{0.846}}} & {\color[HTML]{333333} \textbf{0.650}} & \multicolumn{1}{c|}{{\color[HTML]{333333} \textbf{0.757}}} & {\color[HTML]{333333} \textbf{0.245}} & {\color[HTML]{333333} \textbf{0.527}} \\ \hline
                                            & \textbf{MESH}         & \multicolumn{1}{c|}{{\color[HTML]{333333} \textbf{0.134}}} & \multicolumn{1}{c|}{{\color[HTML]{333333} \textbf{0.124}}} & \multicolumn{1}{c|}{{\color[HTML]{333333} \textbf{0.142}}} & {\color[HTML]{333333} \textbf{0.179}} & \multicolumn{1}{c|}{{\color[HTML]{333333} \textbf{-}}}     & {\color[HTML]{333333} \textbf{-}}     & \multicolumn{1}{c|}{{\color[HTML]{333333} \textbf{0.241}}} & {\color[HTML]{333333} \textbf{0.086}} & {\color[HTML]{333333} \textbf{0.058}} \\ \cline{2-11} 
                                            & \textbf{GL}           & \multicolumn{1}{c|}{{\color[HTML]{333333} \textbf{0.015}}} & \multicolumn{1}{c|}{{\color[HTML]{333333} \textbf{0.110}}} & \multicolumn{1}{c|}{{\color[HTML]{333333} \textbf{0.011}}} & {\color[HTML]{333333} \textbf{0.137}} & \multicolumn{1}{c|}{{\color[HTML]{333333} \textbf{-}}}     & {\color[HTML]{333333} \textbf{-}}     & \multicolumn{1}{c|}{{\color[HTML]{333333} \textbf{0.223}}} & {\color[HTML]{333333} \textbf{0.076}} & {\color[HTML]{333333} \textbf{0.078}} \\ \cline{2-11} 
                                            & \textbf{TPDM}         & \multicolumn{1}{c|}{{\color[HTML]{FE0000} \textbf{0.766}}} & \multicolumn{1}{c|}{{\color[HTML]{333333} \textbf{0.599}}} & \multicolumn{1}{c|}{{\color[HTML]{333333} \textbf{0.649}}} & {\color[HTML]{FE0000} \textbf{0.801}} & \multicolumn{1}{c|}{{\color[HTML]{333333} \textbf{-}}}     & {\color[HTML]{333333} \textbf{-}}     & \multicolumn{1}{c|}{{\color[HTML]{333333} \textbf{0.611}}} & {\color[HTML]{333333} \textbf{0.282}} & {\color[HTML]{333333} \textbf{0.101}} \\ \cline{2-11} 
                                            & \textbf{MSDM2}        & \multicolumn{1}{c|}{{\color[HTML]{333333} \textbf{0.360}}} & \multicolumn{1}{c|}{{\color[HTML]{333333} \textbf{0.495}}} & \multicolumn{1}{c|}{{\color[HTML]{333333} \textbf{0.049}}} & {\color[HTML]{333333} \textbf{0.641}} & \multicolumn{1}{c|}{{\color[HTML]{333333} \textbf{-}}}     & {\color[HTML]{333333} \textbf{-}}     & \multicolumn{1}{c|}{{\color[HTML]{333333} \textbf{0.510}}} & {\color[HTML]{333333} \textbf{0.169}} & {\color[HTML]{333333} \textbf{0.053}} \\ \cline{2-11} 
\multirow{-5}{*}{\textbf{Model-based}}      & \textbf{GeodesicPSIM} & \multicolumn{1}{c|}{{\color[HTML]{3166FF} \textbf{0.578}}} & \multicolumn{1}{c|}{{\color[HTML]{333333} \textbf{0.642}}} & \multicolumn{1}{c|}{{\color[HTML]{333333} \textbf{0.631}}} & {\color[HTML]{32CB00} \textbf{0.778}} & \multicolumn{1}{c|}{{\color[HTML]{FE0000} \textbf{0.902}}} & {\color[HTML]{FE0000} \textbf{0.889}} & \multicolumn{1}{c|}{{\color[HTML]{333333} \textbf{0.697}}} & {\color[HTML]{3166FF} \textbf{0.592}} & {\color[HTML]{32CB00} \textbf{0.757}} \\ \hline
                                            & \textbf{D1}           & \multicolumn{1}{c|}{{\color[HTML]{333333} \textbf{0.436}}} & \multicolumn{1}{c|}{{\color[HTML]{32CB00} \textbf{0.754}}} & \multicolumn{1}{c|}{{\color[HTML]{333333} \textbf{0.420}}} & {\color[HTML]{333333} \textbf{0.656}} & \multicolumn{1}{c|}{{\color[HTML]{333333} \textbf{-}}}     & {\color[HTML]{333333} \textbf{-}}     & \multicolumn{1}{c|}{{\color[HTML]{32CB00} \textbf{0.793}}} & {\color[HTML]{333333} \textbf{0.305}} & {\color[HTML]{333333} \textbf{0.132}} \\ \cline{2-11} 
                                            & \textbf{D2}           & \multicolumn{1}{c|}{{\color[HTML]{333333} \textbf{0.419}}} & \multicolumn{1}{c|}{{\color[HTML]{3166FF} \textbf{0.752}}} & \multicolumn{1}{c|}{{\color[HTML]{333333} \textbf{0.409}}} & {\color[HTML]{333333} \textbf{0.651}} & \multicolumn{1}{c|}{{\color[HTML]{333333} \textbf{-}}}     & {\color[HTML]{333333} \textbf{-}}     & \multicolumn{1}{c|}{{\color[HTML]{3166FF} \textbf{0.791}}} & {\color[HTML]{333333} \textbf{0.304}} & {\color[HTML]{333333} \textbf{0.129}} \\ \cline{2-11} 
                                            & \textbf{YUV\_PSNR}    & \multicolumn{1}{c|}{{\color[HTML]{333333} \textbf{0.244}}} & \multicolumn{1}{c|}{{\color[HTML]{333333} \textbf{0.596}}} & \multicolumn{1}{c|}{{\color[HTML]{333333} \textbf{0.569}}} & {\color[HTML]{333333} \textbf{0.455}} & \multicolumn{1}{c|}{{\color[HTML]{333333} \textbf{0.711}}} & {\color[HTML]{333333} \textbf{0.661}} & \multicolumn{1}{c|}{{\color[HTML]{333333} \textbf{0.673}}} & {\color[HTML]{333333} \textbf{0.522}} & {\color[HTML]{333333} \textbf{0.671}} \\ \cline{2-11} 
                                            & \textbf{PCQM\_PSNR}   & \multicolumn{1}{c|}{{\color[HTML]{333333} \textbf{0.326}}} & \multicolumn{1}{c|}{{\color[HTML]{333333} \textbf{0.701}}} & \multicolumn{1}{c|}{{\color[HTML]{3166FF} \textbf{0.671}}} & {\color[HTML]{333333} \textbf{0.523}} & \multicolumn{1}{c|}{{\color[HTML]{333333} \textbf{0.814}}} & {\color[HTML]{333333} \textbf{0.611}} & \multicolumn{1}{c|}{{\color[HTML]{333333} \textbf{0.730}}} & {\color[HTML]{333333} \textbf{0.415}} & {\color[HTML]{333333} \textbf{0.576}} \\ \cline{2-11} 
                                            & \textbf{GraphSIM}     & \multicolumn{1}{c|}{{\color[HTML]{333333} \textbf{0.391}}} & \multicolumn{1}{c|}{{\color[HTML]{333333} \textbf{0.733}}} & \multicolumn{1}{c|}{{\color[HTML]{FE0000} \textbf{0.777}}} & {\color[HTML]{333333} \textbf{0.652}} & \multicolumn{1}{c|}{{\color[HTML]{32CB00} \textbf{0.880}}} & {\color[HTML]{32CB00} \textbf{0.829}} & \multicolumn{1}{c|}{{\color[HTML]{333333} \textbf{0.766}}} & {\color[HTML]{FE0000} \textbf{0.699}} & {\color[HTML]{3166FF} \textbf{0.755}} \\ \cline{2-11} 
\multirow{-6}{*}{\textbf{Point-based}}      & \textbf{FMQM}         & \multicolumn{1}{c|}{{\color[HTML]{32CB00} \textbf{0.693}}} & \multicolumn{1}{c|}{{\color[HTML]{FE0000} \textbf{0.802}}} & \multicolumn{1}{c|}{{\color[HTML]{32CB00} \textbf{0.737}}} & {\color[HTML]{3166FF} \textbf{0.758}} & \multicolumn{1}{c|}{{\color[HTML]{3166FF} \textbf{0.848}}} & {\color[HTML]{3166FF} \textbf{0.807}} & \multicolumn{1}{c|}{{\color[HTML]{FE0000} \textbf{0.803}}} & {\color[HTML]{32CB00} \textbf{0.688}} & {\color[HTML]{FE0000} \textbf{0.760}} \\ \hline
\end{tabular}
}
\end{table*}

\subsubsection{Per-sample performance and challenging samples}
To further understand the limitations of TMQA metrics, we conduct a per-sample correlation analysis to identify particularly challenging mesh models.  
\begin{figure}[htbp]   
    \centering
    \subfloat[Per-model Correlation\label{perModel-1:sub1}]{
        \centering
        \includegraphics[width=0.45\linewidth]{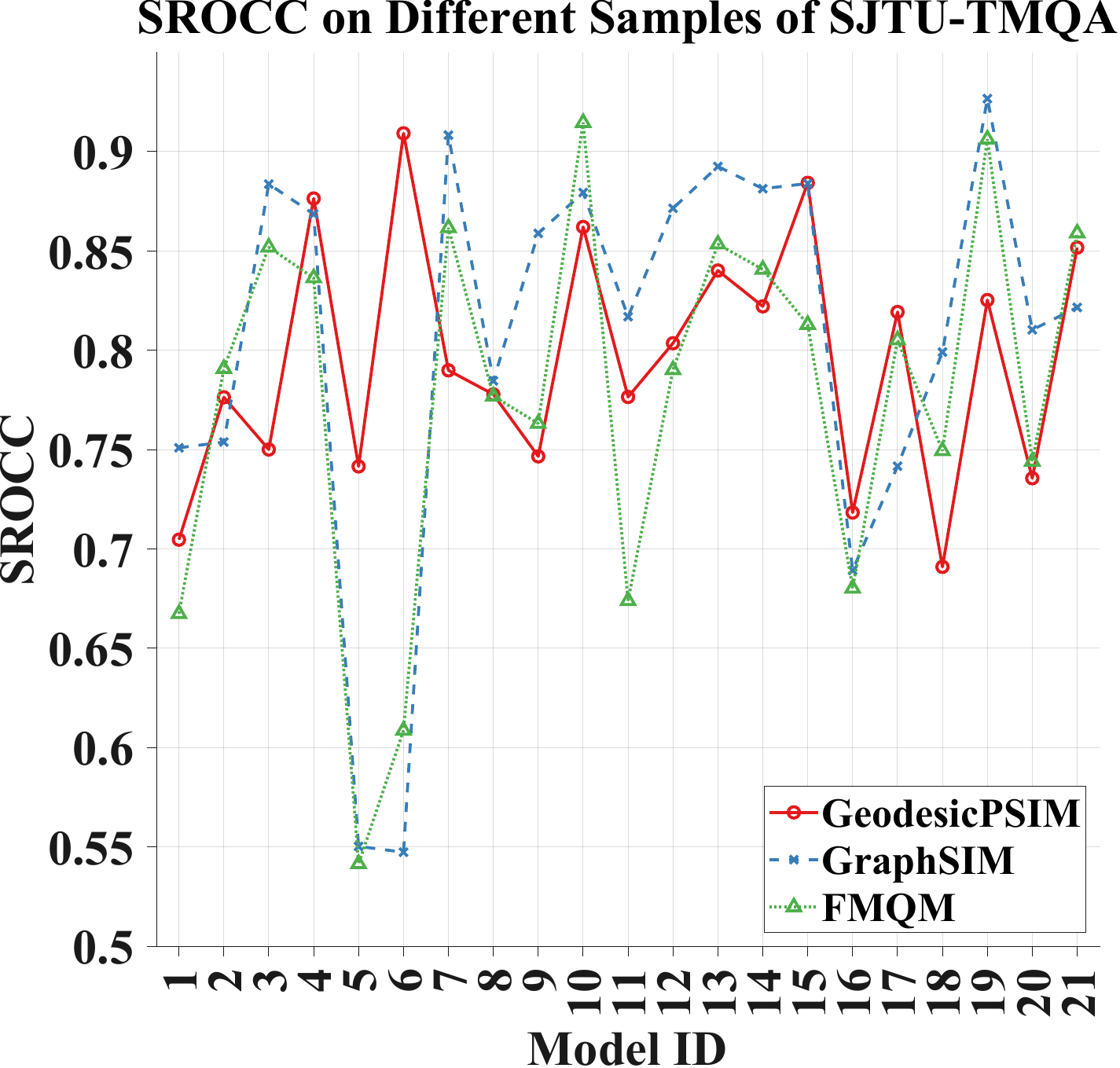}
    }
    \hfill
    \subfloat[MOS Distribution\label{perModel-1:sub2}]{
        \centering
        \includegraphics[width=0.45\linewidth]{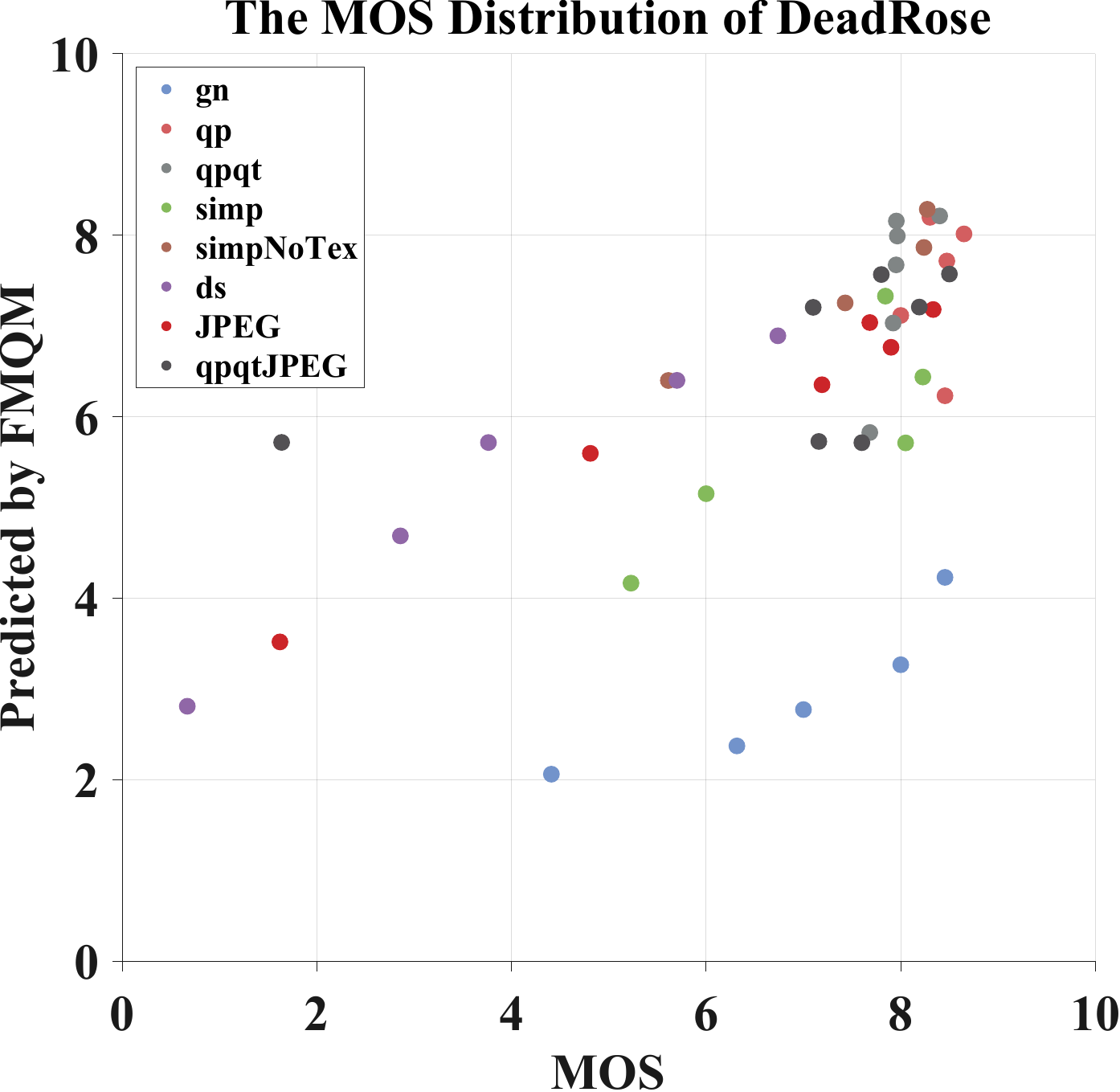}
    }
    \caption{Per-model correlation analysis}
    \label{fig:perModel-1}
\end{figure}

\begin{figure}[htbp]
    \centering
    \subfloat[FruitSet\label{perModel:sub1}]{
        \includegraphics[width=0.3\linewidth]{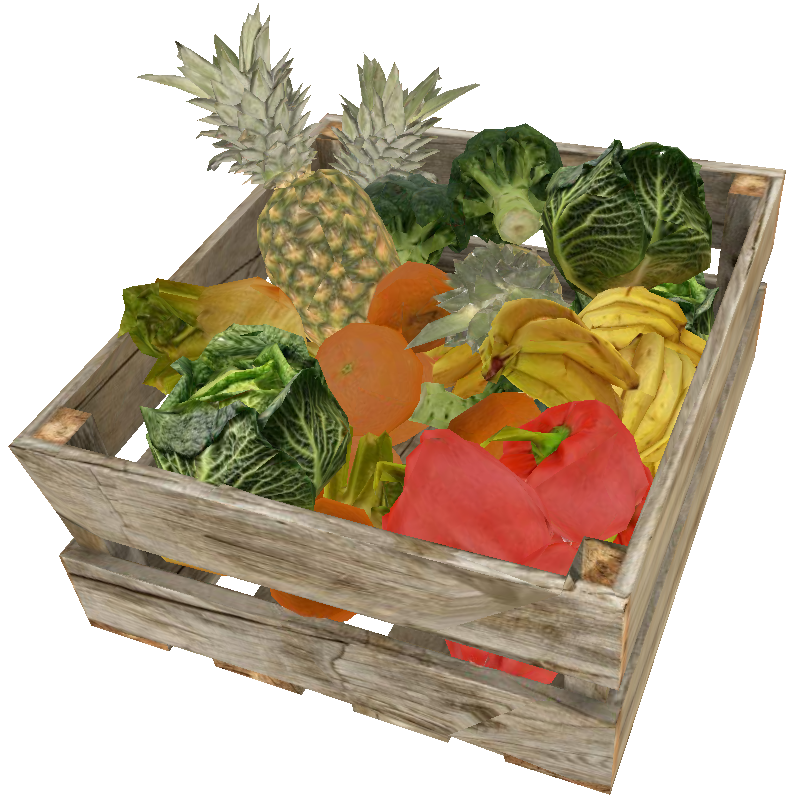}
    }
    \hfill
    \subfloat[ZakopaneChair\label{perModel:sub4}]{
        \includegraphics[width=0.3\linewidth]{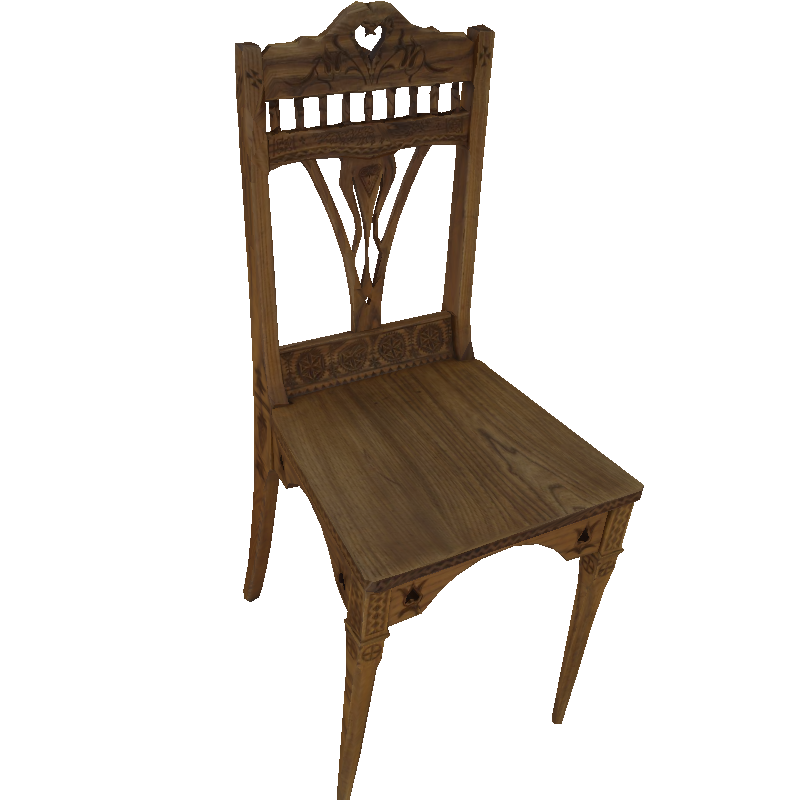}
    }
    \hfill
    \subfloat[DeadRose\label{perModel:sub2}]{
        \includegraphics[width=0.3\linewidth]{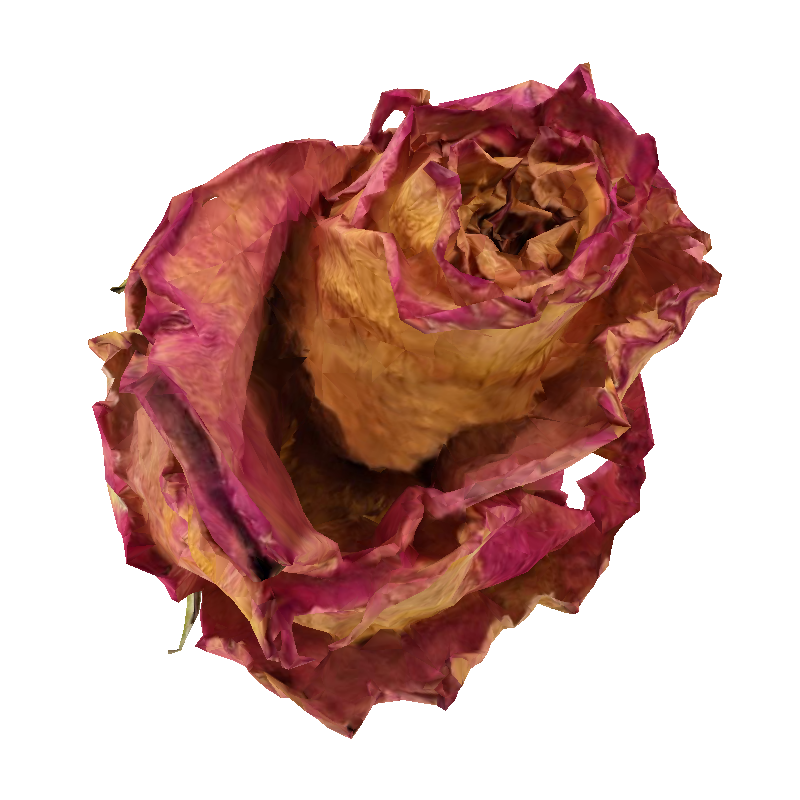}
    }
    \par\medskip
    \subfloat[DeadRose Geometry\label{perModel:sub3}]{
        \includegraphics[width=0.3\linewidth]{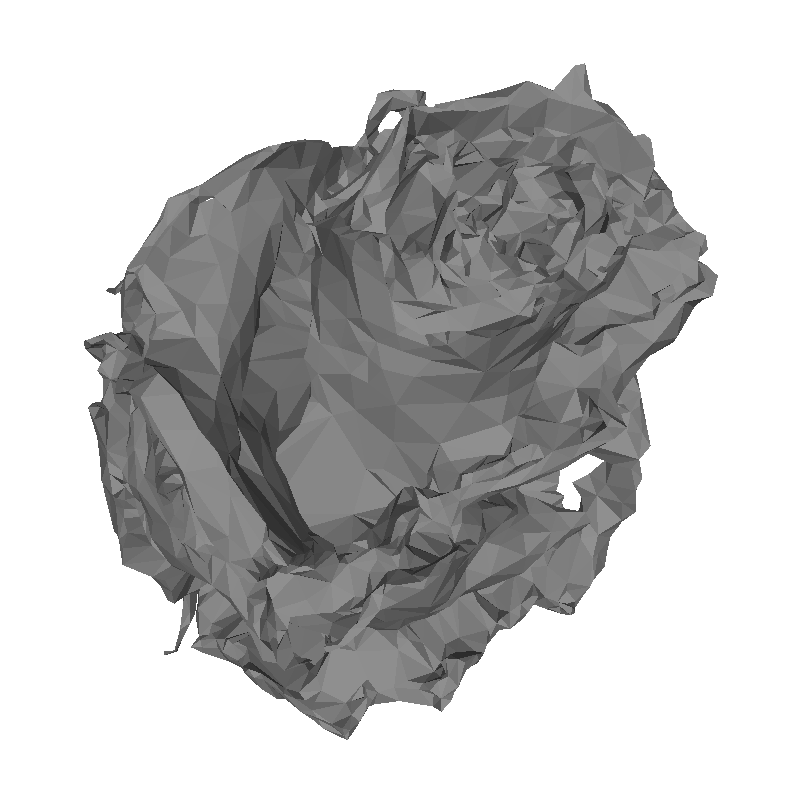}
    }
    \hfill
    \subfloat[Bread\label{perModel:sub5}]{
        \includegraphics[width=0.3\linewidth]{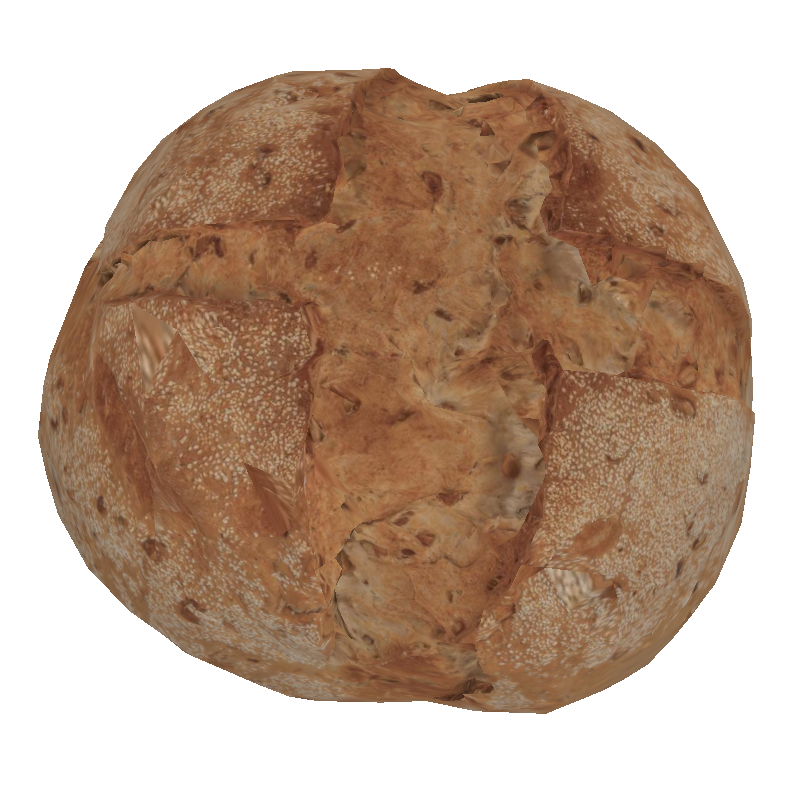}
    }
    \hfill
    \subfloat[Bread Texture\label{perModel:sub6}]{
        \includegraphics[width=0.3\linewidth]{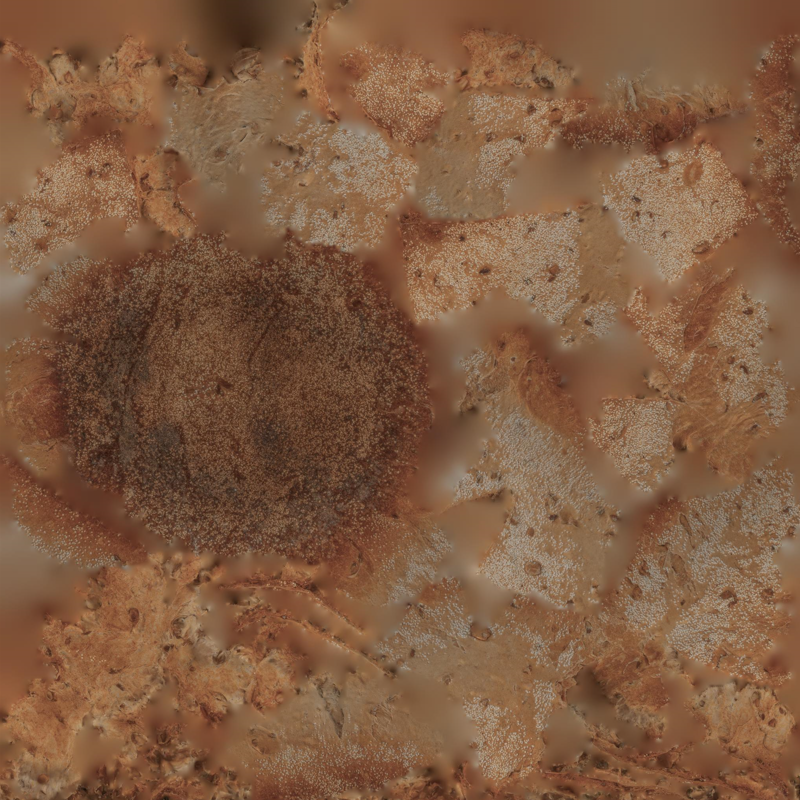}
    }
    \caption{The changling mesh samples on SJTU-TMQA}
    \label{fig:perModel}
\end{figure}
Fig.~\ref{perModel-1:sub1} shows the correlation scores of the top three metrics—FMQM, GeodesicPSIM, GraphSIM—computed per reference model on the SJTU-TMQA dataset.
Several models stand out as especially challenging, including model 5 (DeadRose), model 6 (Bread), model 8 (ZakopaneChair), and model 16 (FruitSet), as highlighted in Fig.~\ref{fig:perModel}. 
These samples typically fall into two categories: those with highly intricate geometry, such as the densely layered petals in DeadRose (Fig.~\ref{perModel:sub3}) and the fine-grained foliage in FruitSet; and those with complex and visually cluttered textures, such as the irregular wooden surfaces in ZakopaneChair and the mold-covered roughness in Bread (Fig.~\ref{perModel:sub6}). 
These structural and textural complexities pose significant challenges for automatic quality prediction.

A closer look at the subjective score distribution of DeadRose (Fig.~\ref{fig:perModel-1}(b)) shows that most MOS values fall in the high-quality range (7–9). This suggests that complex geometry and rich textures may induce masking effects, reducing perceptual sensitivity to distortions and thus weakening the correlation between objective and subjective scores.
These findings underscore the need for TMQA metrics to account for perceptual masking, especially in visually complex regions.

\subsubsection{Robutness of FMQM}
\label{sec:robutness}
In this section, we evaluate the robustness of FMQM with respect to changes in key hyperparameters, including the number of sampled points, surface and color patch sizes, sample variance $\sigma_\text{base}$, and the tangent/normal thresholds and radii for color patch selection. All experiments are conducted on the SJTU-TMQA dataset.

$\bullet $ Sample point number

\begin{figure}[htbp]   
    \centering
    \subfloat[Total sample number\label{hyper:sub1}]{
        \includegraphics[width=0.45\linewidth]{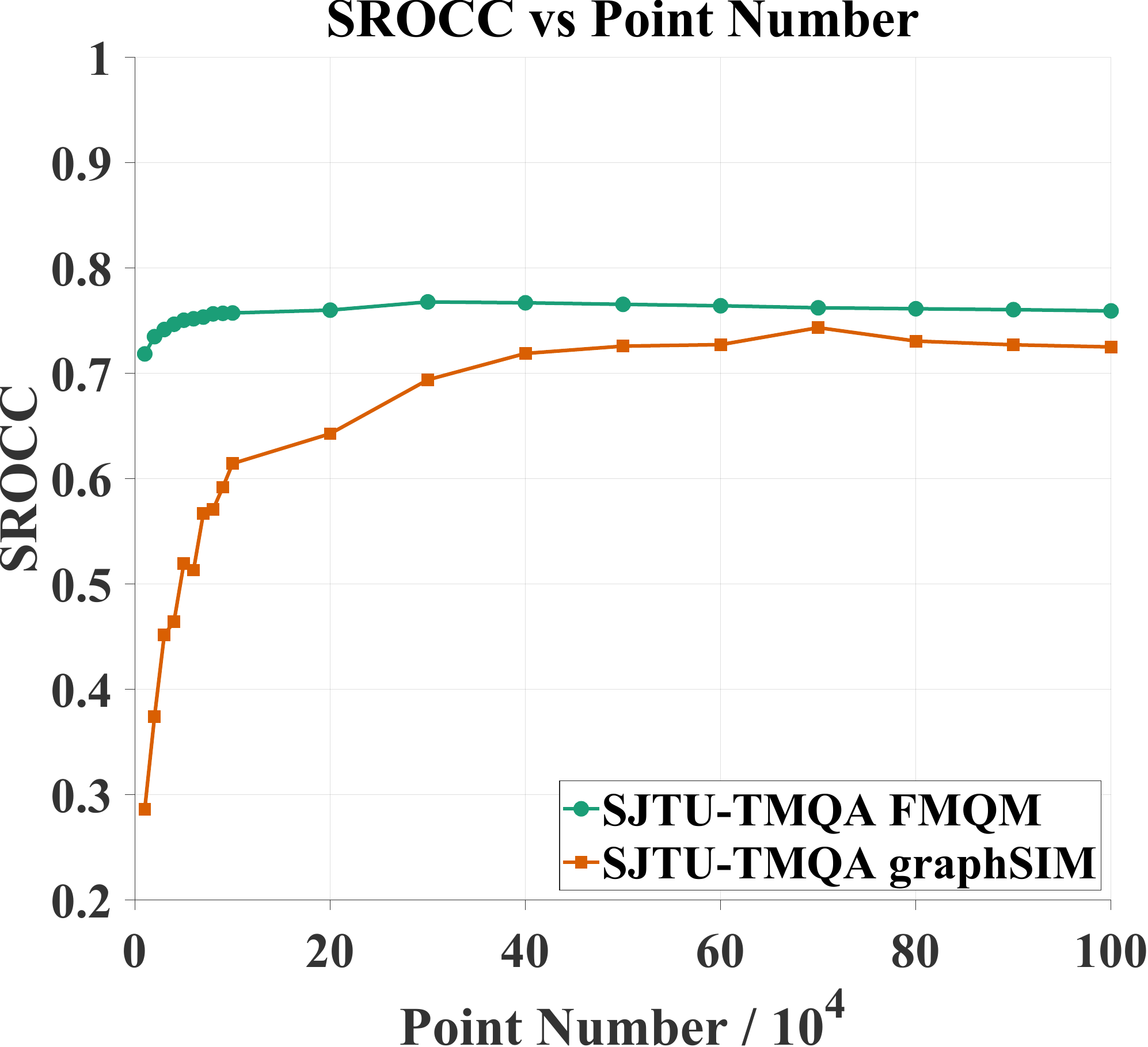}  
    }
    \hfill
    \subfloat[Surface patch number\label{hyper:sub2}]{
        \includegraphics[width=0.45\linewidth]{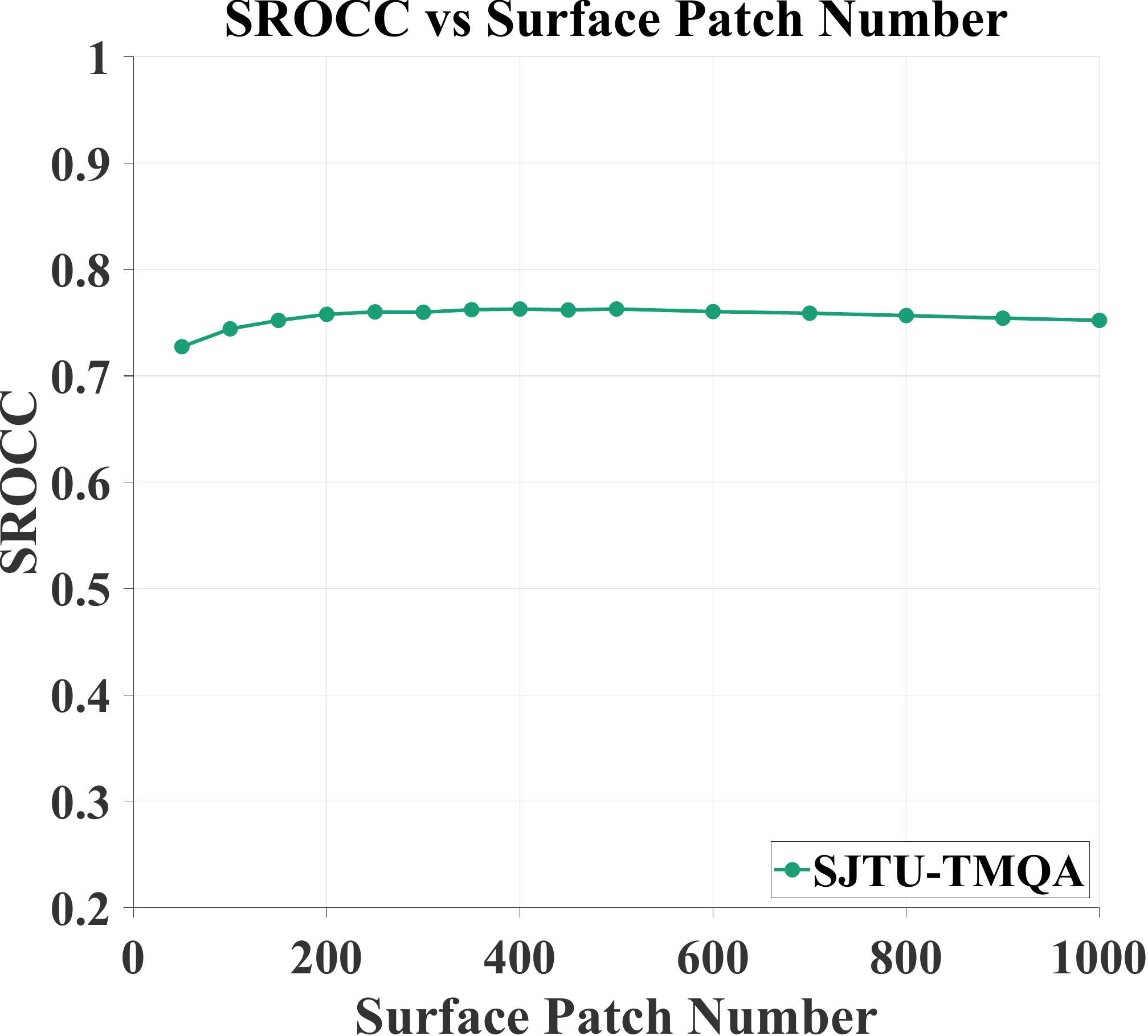}  
    }
    \par\medskip
    \subfloat[Sample offset variance\label{hyper:sub3}]{
        \includegraphics[width=0.45\linewidth]{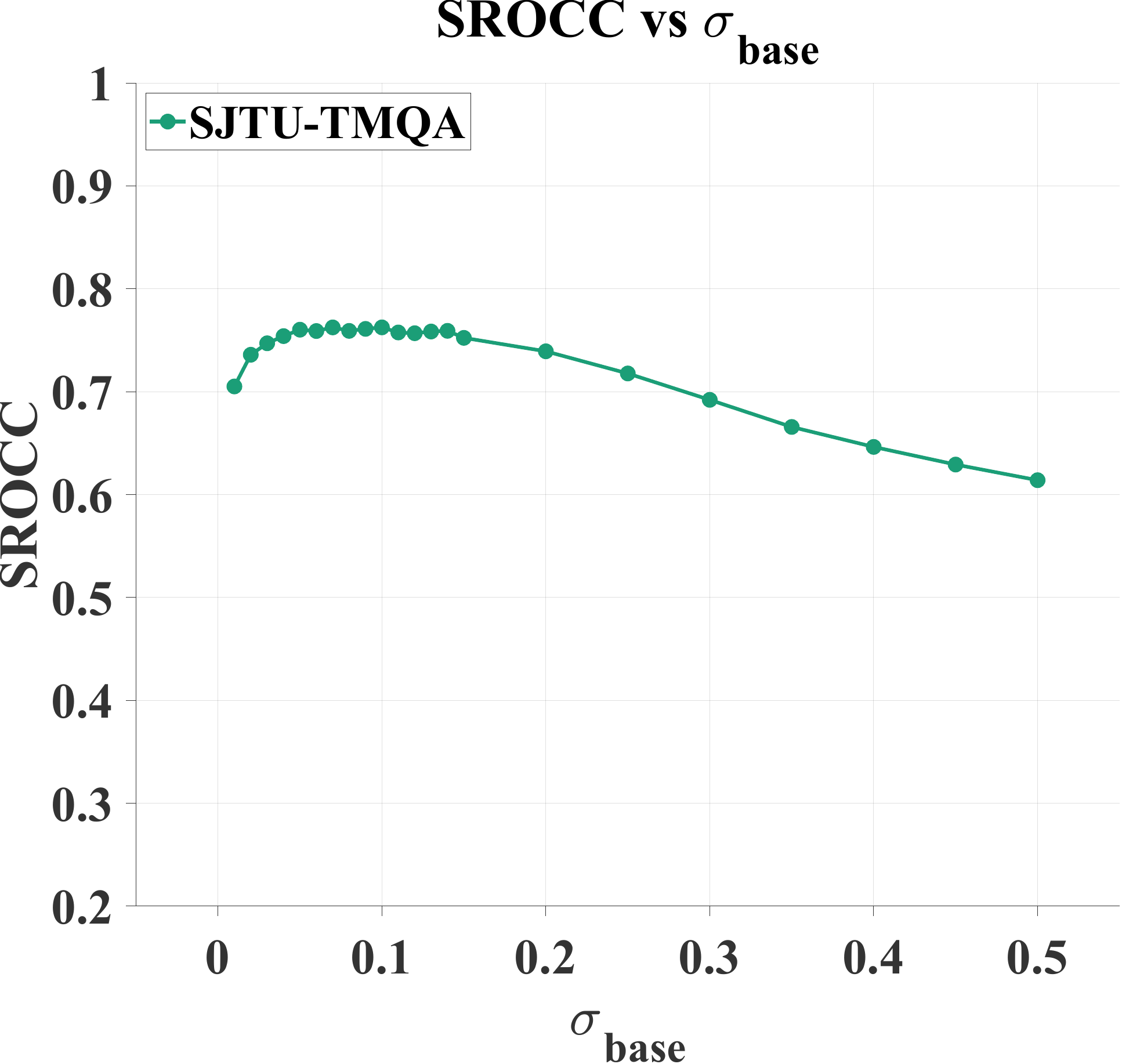}  
    }
    \hfill
    \subfloat[Color patch selection threshold\label{hyper:sub4}]{
        \includegraphics[width=0.45\linewidth]{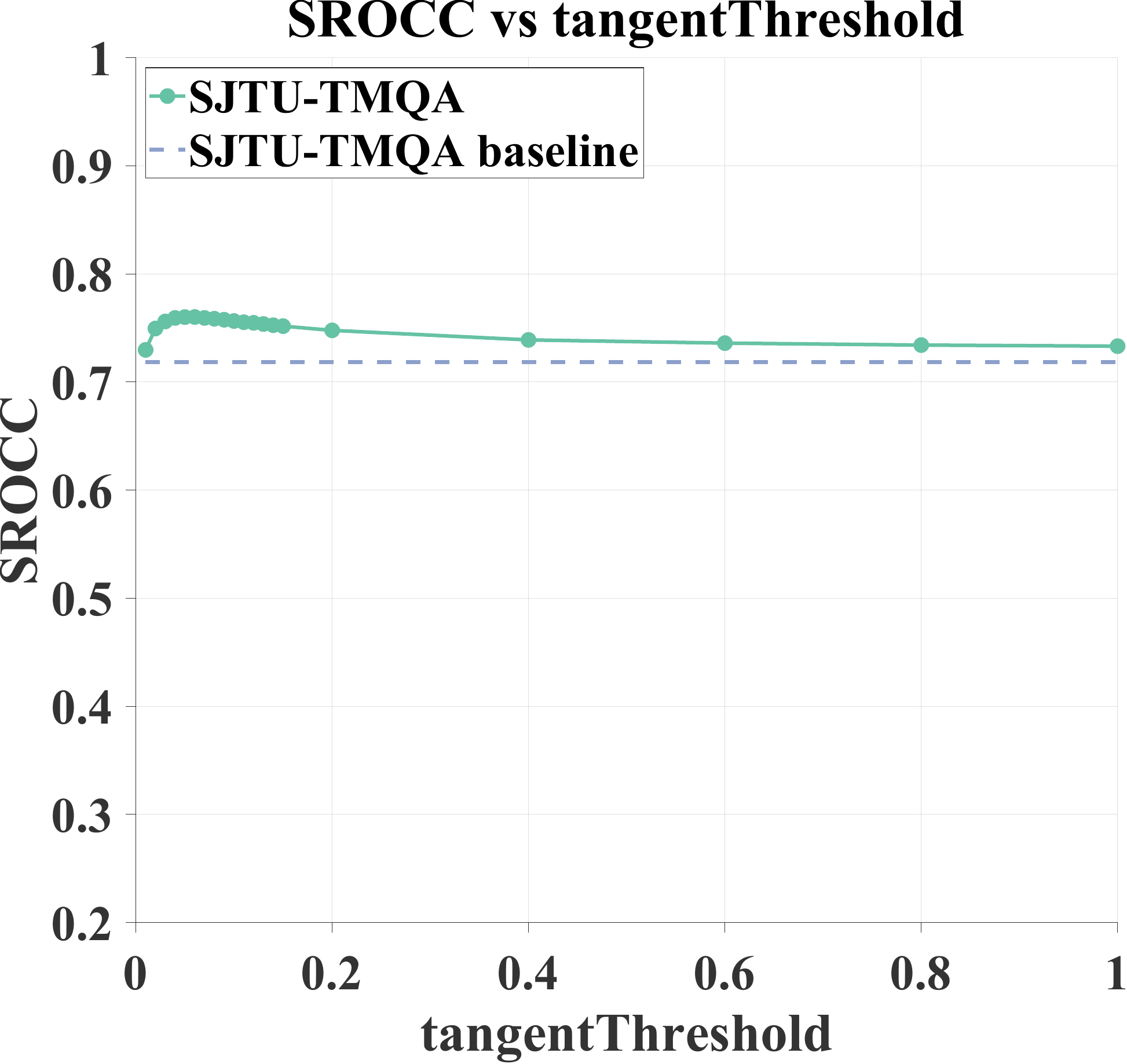}  
    }
    \par\medskip
    \subfloat[Color patch selection radius\label{hyper:sub5}]{
        \includegraphics[width=0.45\linewidth]{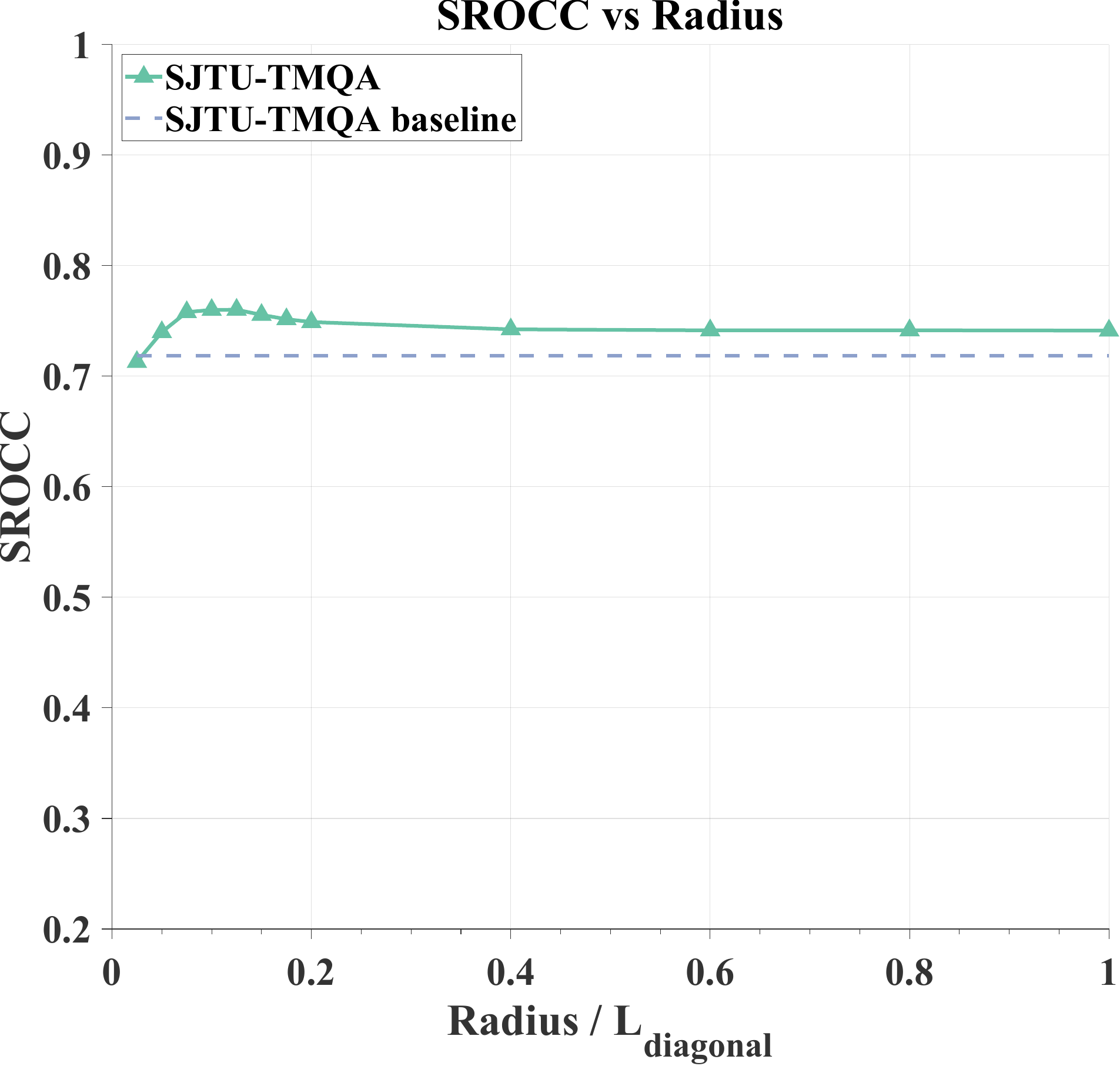}  
    }
    \hfill
    \subfloat[Color patch number\label{hyper:sub6}]{
        \includegraphics[width=0.45\linewidth]{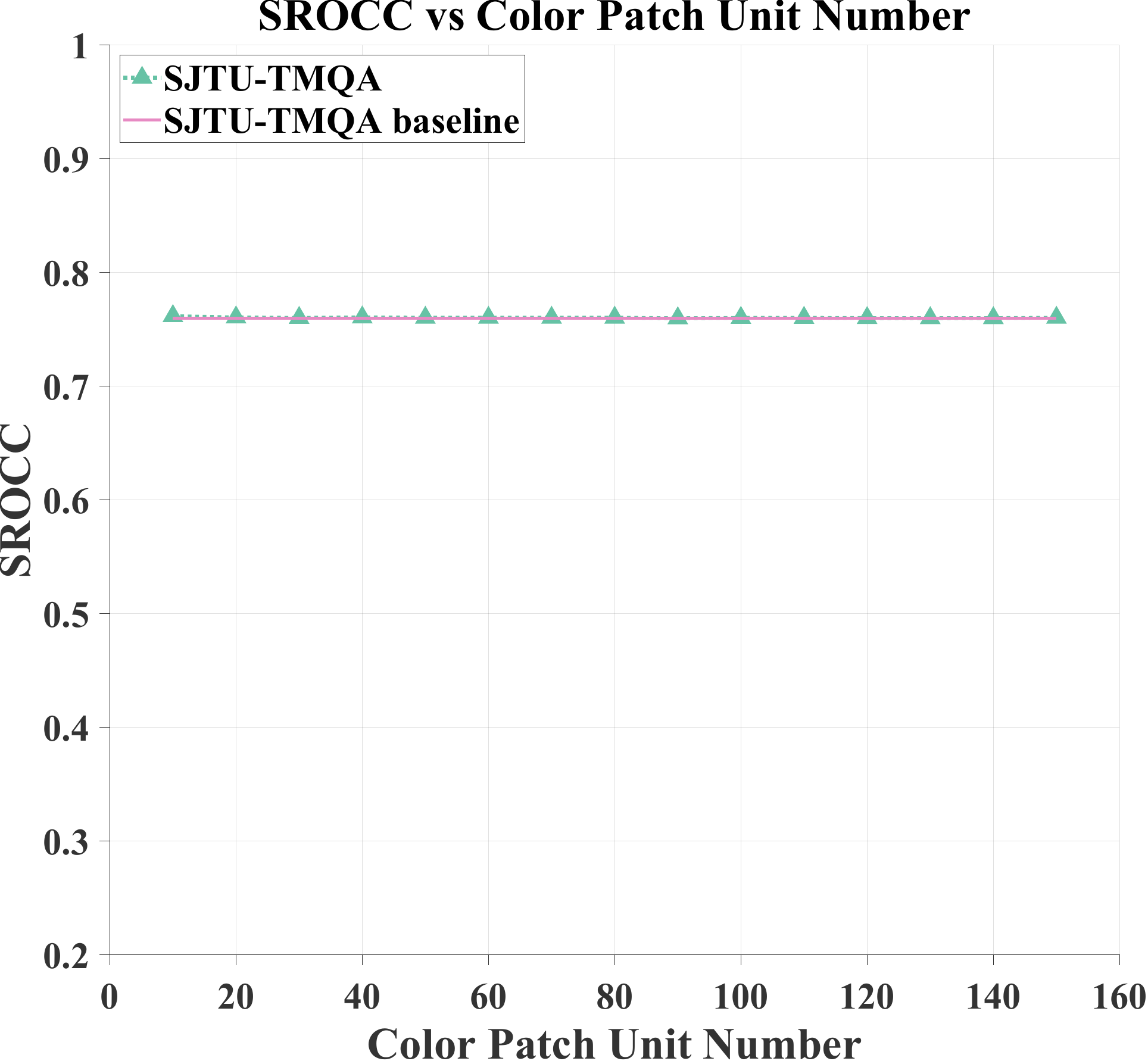}  
    }
    \caption{Metric performance VS hyperparameters}  
    \label{fig:hyper}  
\end{figure}

For point-based metrics, the number of sampled points significantly influences both storage cost and computational complexity. We compare FMQM with GraphSIM—the other top-performing metric—to assess sensitivity to this factor.
As shown in Fig.~\ref{hyper:sub1}:

i) FMQM exhibits consistently better and more stable performance than GraphSIM when the number of sampled points is high. This robustness stems from FMQM’s sampling strategy: the local patches used for analysis remain unchanged, and fewer points simply resemble lower-resolution sampling within the same patch. In contrast, GraphSIM relies on keypoint detection; thus, variations in sampling affect keypoint positions and local patch structures, introducing fluctuations in performance.

ii) When the number of sampled points decreases below 300,000, GraphSIM’s performance declines sharply, while FMQM remains stable. FMQM only starts to degrade when sampling drops below 30,000 points. This highlights the efficiency of FMQM’s field-based representation in capturing geometric and color structure, compared to GraphSIM’s graph-based approach.

$\bullet $ Surface patch number

We examined the impact of varying the number of surface patches on FMQM's performance. As the number of patches increases, each patch becomes smaller, effectively altering the LoD in mesh observations. Fig.~\ref{hyper:sub2} illustrates that both excessively large and excessively small patch counts lead to performance degradation. This suggests that an optimal patch scale exists, balancing local detail and global context. Future work could explore integrating multiple patch scales to better align with the multi-scale processing characteristics of the HVS.

$\bullet$ Sample variance $\sigma_\text{base}$

As discussed in Section~\ref{sec:feature}, FMQM restricts sampling to near-surface regions to ensure both representational fidelity and perceptual relevance.  
Fig.~\ref{hyper:sub3} shows that the performance drops significantly as $\sigma_\text{base}$ increases, confirming that sampling points too far from the surface introduces noise and ambiguity unrelated to the target geometry.  
Interestingly, an overly small $\sigma_\text{base}$ also leads to a slight decrease in performance. This is because excessively narrow sampling limits the spatial diversity of the field values, resulting in degraded feature representation and reduced sensitivity to geometric and textural variations.  
Thus, an appropriate sampling variance is essential to balance representational richness and noise robustness.

$\bullet $ Color patch unit threshold and radius

To validate the effectiveness of the tangent and normal color patch unit defined in Section~\ref{sec:ncfFeature}, we evaluated the impact of varying the angle thresholds (\text{tanThre} and \text{norThre}) and the spatial \text{radius} constraint, as shown in Fig.~\ref{hyper:sub4} and~\ref{hyper:sub5}.  
The baseline corresponds to using all samples within a surface patch without applying feature extraction units.  

As observed, applying either only the \text{radius} constraint (\text{tanThre} = 1 in Fig.~\ref{hyper:sub4}) or only the angular constraint (\text{radius} = $L_\text{diagonal}$ in Fig.~\ref{hyper:sub5}) yields approximately 0.02 higher SROCC than the baseline, demonstrating the effectiveness of localized feature aggregation.  
Further decreasing \text{tanThre} improves performance by an additional 0.04 on SJTU-TMQA (optimal at \text{tanThre} = 0.02). However, excessively tightening the thresholds leads to rapid performance degradation, as too few samples remain to form meaningful feature units.  
A similar trend is observed with respect to the \text{radius} parameter, highlighting the importance of balancing locality and sample sufficiency when defining feature extraction units.

$\bullet $ Color patch unit number

As discussed in Section~\ref{sec:colorssim}, only a subset of color patch units, denoted by $N_{\text{color\_patch}}$, is randomly selected for computing \text{colorSSIM}.  
Fig.~\ref{hyper:sub6} shows the performance as a function of $N_{\text{color\_patch}}$.  
It can be observed that FMQM maintains nearly identical performance when $N_{\text{color\_patch}}$ varies from 10 to 150 compared to using all available units.  
This demonstrates that substantial computational savings can be achieved without compromising accuracy.

\subsubsection{Computation Cost}
In this section, the computation cost of three types of TMQA methods are compared.
\begin{table*}[]
\centering
\caption{Computation time(s) of different kinds of TMQA methods on three datasets}
\label{tab:computationTime}
\resizebox{0.8\textwidth}{!}{%
\begin{tabular}{|l|cccc|cccc|cccc|}
\hline
\textbf{}             & \multicolumn{4}{c|}{\textbf{TSMD}}                                                                                                       & \multicolumn{4}{c|}{\textbf{SJTUMQA}}                                                                                                   & \multicolumn{4}{c|}{\textbf{YANA}}                                                                                                     \\ \hline
\textbf{}             & \multicolumn{1}{c|}{\textbf{Prep.}}   & \multicolumn{1}{c|}{\textbf{Conv.}}   & \multicolumn{1}{c|}{\textbf{Metric}}  & \textbf{Overall} & \multicolumn{1}{c|}{\textbf{Prep.}}   & \multicolumn{1}{c|}{\textbf{Conv.}}  & \multicolumn{1}{c|}{\textbf{Metric}}  & \textbf{Overall} & \multicolumn{1}{c|}{\textbf{Prep.}}   & \multicolumn{1}{c|}{\textbf{Conv.}}  & \multicolumn{1}{c|}{\textbf{Metric}} & \textbf{Overall} \\ \hline
\textbf{GEO\_PSNR}    & \multicolumn{1}{c|}{\textbf{-}}       & \multicolumn{1}{c|}{\textbf{9.849}}   & \multicolumn{1}{c|}{\textbf{0.055}}   & \textbf{9.904}   & \multicolumn{1}{c|}{\textbf{-}}       & \multicolumn{1}{c|}{\textbf{6.776}}  & \multicolumn{1}{c|}{\textbf{0.054}}   & \textbf{6.830}   & \multicolumn{1}{c|}{\textbf{-}}       & \multicolumn{1}{c|}{\textbf{8.580}}  & \multicolumn{1}{c|}{\textbf{0.051}}  & \textbf{8.631}   \\ \hline
\textbf{RGB\_PSNR}    & \multicolumn{1}{c|}{\textbf{-}}       & \multicolumn{1}{c|}{\textbf{9.849}}   & \multicolumn{1}{c|}{\textbf{0.103}}   & \textbf{9.952}   & \multicolumn{1}{c|}{\textbf{-}}       & \multicolumn{1}{c|}{\textbf{6.776}}  & \multicolumn{1}{c|}{\textbf{0.098}}   & \textbf{6.874}   & \multicolumn{1}{c|}{\textbf{-}}       & \multicolumn{1}{c|}{\textbf{8.580}}  & \multicolumn{1}{c|}{\textbf{0.096}}  & \textbf{8.676}   \\ \hline
\textbf{YUV\_PSNR}    & \multicolumn{1}{c|}{\textbf{-}}       & \multicolumn{1}{c|}{\textbf{9.849}}   & \multicolumn{1}{c|}{\textbf{0.112}}   & \textbf{9.961}   & \multicolumn{1}{c|}{\textbf{-}}       & \multicolumn{1}{c|}{\textbf{6.776}}  & \multicolumn{1}{c|}{\textbf{0.105}}   & \textbf{6.881}   & \multicolumn{1}{c|}{\textbf{-}}       & \multicolumn{1}{c|}{\textbf{8.580}}  & \multicolumn{1}{c|}{\textbf{0.095}}  & \textbf{8.675}   \\ \hline
\textbf{PSNR}         & \multicolumn{1}{c|}{\textbf{-}}       & \multicolumn{1}{c|}{\textbf{141.350}} & \multicolumn{1}{c|}{\textbf{1.353}}   & \textbf{142.703} & \multicolumn{1}{c|}{\textbf{-}}       & \multicolumn{1}{c|}{\textbf{64.237}} & \multicolumn{1}{c|}{\textbf{8.978}}   & \textbf{73.215}  & \multicolumn{1}{c|}{\textbf{-}}       & \multicolumn{1}{c|}{\textbf{50.022}} & \multicolumn{1}{c|}{\textbf{0.476}}  & \textbf{50.498}  \\ \hline
\textbf{SSIM}         & \multicolumn{1}{c|}{\textbf{-}}       & \multicolumn{1}{c|}{\textbf{141.350}} & \multicolumn{1}{c|}{\textbf{1.387}}   & \textbf{142.738} & \multicolumn{1}{c|}{\textbf{-}}       & \multicolumn{1}{c|}{\textbf{64.237}} & \multicolumn{1}{c|}{\textbf{8.924}}   & \textbf{73.161}  & \multicolumn{1}{c|}{\textbf{-}}       & \multicolumn{1}{c|}{\textbf{50.022}} & \multicolumn{1}{c|}{\textbf{0.490}}  & \textbf{50.512}  \\ \hline
\textbf{MS-SSIM}      & \multicolumn{1}{c|}{\textbf{-}}       & \multicolumn{1}{c|}{\textbf{141.350}} & \multicolumn{1}{c|}{\textbf{1.530}}   & \textbf{142.880} & \multicolumn{1}{c|}{\textbf{-}}       & \multicolumn{1}{c|}{\textbf{64.237}} & \multicolumn{1}{c|}{\textbf{9.106}}   & \textbf{73.342}  & \multicolumn{1}{c|}{\textbf{-}}       & \multicolumn{1}{c|}{\textbf{50.022}} & \multicolumn{1}{c|}{\textbf{0.480}}  & \textbf{50.502}  \\ \hline
\textbf{3-SSIM}       & \multicolumn{1}{c|}{\textbf{-}}       & \multicolumn{1}{c|}{\textbf{141.350}} & \multicolumn{1}{c|}{\textbf{11.820}}  & \textbf{153.170} & \multicolumn{1}{c|}{\textbf{-}}       & \multicolumn{1}{c|}{\textbf{64.237}} & \multicolumn{1}{c|}{\textbf{12.448}}  & \textbf{76.685}  & \multicolumn{1}{c|}{\textbf{-}}       & \multicolumn{1}{c|}{\textbf{50.022}} & \multicolumn{1}{c|}{\textbf{1.164}}  & \textbf{51.186}  \\ \hline
\textbf{VQM}          & \multicolumn{1}{c|}{\textbf{-}}       & \multicolumn{1}{c|}{\textbf{141.350}} & \multicolumn{1}{c|}{\textbf{4.103}}   & \textbf{145.454} & \multicolumn{1}{c|}{\textbf{-}}       & \multicolumn{1}{c|}{\textbf{64.237}} & \multicolumn{1}{c|}{\textbf{11.533}}  & \textbf{75.769}  & \multicolumn{1}{c|}{\textbf{-}}       & \multicolumn{1}{c|}{\textbf{50.022}} & \multicolumn{1}{c|}{\textbf{0.775}}  & \textbf{50.798}  \\ \hline
\textbf{VMAF}         & \multicolumn{1}{c|}{\textbf{-}}       & \multicolumn{1}{c|}{\textbf{141.350}} & \multicolumn{1}{c|}{\textbf{19.831}}  & \textbf{161.181} & \multicolumn{1}{c|}{\textbf{-}}       & \multicolumn{1}{c|}{\textbf{64.237}} & \multicolumn{1}{c|}{\textbf{21.171}}  & \textbf{85.408}  & \multicolumn{1}{c|}{\textbf{-}}       & \multicolumn{1}{c|}{\textbf{50.022}} & \multicolumn{1}{c|}{\textbf{1.826}}  & \textbf{51.848}  \\ \hline
\textbf{MESH}         & \multicolumn{1}{c|}{\textbf{-}}       & \multicolumn{1}{c|}{\textbf{-}}       & \multicolumn{1}{c|}{\textbf{0.628}}   & \textbf{0.628}   & \multicolumn{1}{c|}{\textbf{-}}       & \multicolumn{1}{c|}{\textbf{-}}      & \multicolumn{1}{c|}{\textbf{0.028}}   & \textbf{0.028}   & \multicolumn{1}{c|}{\textbf{-}}       & \multicolumn{1}{c|}{\textbf{-}}      & \multicolumn{1}{c|}{\textbf{0.698}}  & \textbf{0.698}   \\ \hline
\textbf{GL}           & \multicolumn{1}{c|}{\textbf{-}}       & \multicolumn{1}{c|}{\textbf{-}}       & \multicolumn{1}{c|}{\textbf{2.520}}   & \textbf{2.520}   & \multicolumn{1}{c|}{\textbf{-}}       & \multicolumn{1}{c|}{\textbf{-}}      & \multicolumn{1}{c|}{\textbf{0.112}}   & \textbf{0.112}   & \multicolumn{1}{c|}{\textbf{-}}       & \multicolumn{1}{c|}{\textbf{-}}      & \multicolumn{1}{c|}{\textbf{2.844}}  & \textbf{2.844}   \\ \hline
\textbf{TPDM}         & \multicolumn{1}{c|}{\textbf{-}}       & \multicolumn{1}{c|}{\textbf{-}}       & \multicolumn{1}{c|}{\textbf{27.276}}  & \textbf{27.276}  & \multicolumn{1}{c|}{\textbf{-}}       & \multicolumn{1}{c|}{\textbf{-}}      & \multicolumn{1}{c|}{\textbf{0.668}}   & \textbf{0.668}   & \multicolumn{1}{c|}{\textbf{-}}       & \multicolumn{1}{c|}{\textbf{-}}      & \multicolumn{1}{c|}{\textbf{14.301}} & \textbf{14.301}  \\ \hline
\textbf{MSDM2}        & \multicolumn{1}{c|}{\textbf{-}}       & \multicolumn{1}{c|}{\textbf{-}}       & \multicolumn{1}{c|}{\textbf{-}}       & \textbf{-}       & \multicolumn{1}{c|}{\textbf{-}}       & \multicolumn{1}{c|}{\textbf{-}}      & \multicolumn{1}{c|}{\textbf{3.028}}   & \textbf{3.028}   & \multicolumn{1}{c|}{\textbf{-}}       & \multicolumn{1}{c|}{\textbf{-}}      & \multicolumn{1}{c|}{\textbf{96.264}} & \textbf{96.264}  \\ \hline
\textbf{GeodesicPSIM} & \multicolumn{1}{c|}{\textbf{-}}       & \multicolumn{1}{c|}{\textbf{-}}       & \multicolumn{1}{c|}{\textbf{74.172}}  & \textbf{74.172}  & \multicolumn{1}{c|}{\textbf{-}}       & \multicolumn{1}{c|}{\textbf{-}}      & \multicolumn{1}{c|}{\textbf{3.443}}   & \textbf{3.443}   & \multicolumn{1}{c|}{\textbf{-}}       & \multicolumn{1}{c|}{\textbf{-}}      & \multicolumn{1}{c|}{\textbf{58.191}} & \textbf{58.191}  \\ \hline
\textbf{D1+D2}        & \multicolumn{1}{c|}{\textbf{-}}       & \multicolumn{1}{c|}{\textbf{5.715}}   & \multicolumn{1}{c|}{\textbf{30.071}}  & \textbf{35.786}  & \multicolumn{1}{c|}{\textbf{-}}       & \multicolumn{1}{c|}{\textbf{1.507}}  & \multicolumn{1}{c|}{\textbf{4.090}}   & \textbf{5.597}   & \multicolumn{1}{c|}{\textbf{-}}       & \multicolumn{1}{c|}{\textbf{4.713}}  & \multicolumn{1}{c|}{\textbf{9.423}}  & \textbf{14.136}  \\ \hline
\textbf{PCQM}         & \multicolumn{1}{c|}{\textbf{-}}       & \multicolumn{1}{c|}{\textbf{5.715}}   & \multicolumn{1}{c|}{\textbf{26.961}}  & \textbf{32.676}  & \multicolumn{1}{c|}{\textbf{-}}       & \multicolumn{1}{c|}{\textbf{1.507}}  & \multicolumn{1}{c|}{\textbf{4.368}}   & \textbf{5.875}   & \multicolumn{1}{c|}{\textbf{-}}       & \multicolumn{1}{c|}{\textbf{4.713}}  & \multicolumn{1}{c|}{\textbf{8.637}}  & \textbf{13.350}  \\ \hline
\textbf{GraphSim}     & \multicolumn{1}{c|}{\textbf{176.122}} & \multicolumn{1}{c|}{\textbf{5.715}}   & \multicolumn{1}{c|}{\textbf{147.894}} & \textbf{153.609} & \multicolumn{1}{c|}{\textbf{238.865}} & \multicolumn{1}{c|}{\textbf{1.507}}  & \multicolumn{1}{c|}{\textbf{152.237}} & \textbf{153.744} & \multicolumn{1}{c|}{\textbf{196.007}} & \multicolumn{1}{c|}{\textbf{4.713}}  & \multicolumn{1}{c|}{\textbf{16.499}} & \textbf{21.213}  \\ \hline
\textbf{FMQM}         & \multicolumn{1}{c|}{\textbf{69.737}}  & \multicolumn{1}{c|}{\textbf{10.638}}  & \multicolumn{1}{c|}{\textbf{3.049}}   & \textbf{13.687}  & \multicolumn{1}{c|}{\textbf{1.072}}   & \multicolumn{1}{c|}{\textbf{2.847}}  & \multicolumn{1}{c|}{\textbf{3.463}}   & \textbf{6.310}   & \multicolumn{1}{c|}{\textbf{67.083}}  & \multicolumn{1}{c|}{\textbf{11.117}} & \multicolumn{1}{c|}{\textbf{2.712}}  & \textbf{13.829}  \\ \hline
\end{tabular}%
}
\end{table*}
\begin{table*}[]
\centering
\caption{Per-feature SROCC of FMQM on TSMD, SJTU-TMQA, and YANA dataset.
}
\label{tab:per-feature}
\resizebox{0.8\textwidth}{!}{%
\begin{tabular}{|l|c|ccccccccc|c|}
\hline
\textbf{}                  & \textbf{TSMD}                         & \multicolumn{9}{c|}{\textbf{SJTU-TMQA}}                                                                                                                                                                                                                                                                                                                                                                                                                                                                                                       & \textbf{YANA}                         \\ \hline
\textbf{}                  & \textbf{Mix}                          & \multicolumn{4}{c|}{\textbf{Geometry}}                                                                                                                                                                                                            & \multicolumn{2}{c|}{\textbf{Texture}}                                                                                   & \multicolumn{2}{c|}{\textbf{Mix}}                                                                                       & \textbf{Overall}                      & \textbf{Mix}                          \\ \hline
\textbf{}                  & \textbf{}                             & \multicolumn{1}{c|}{\textbf{GN}}                           & \multicolumn{1}{c|}{\textbf{QP}}                           & \multicolumn{1}{c|}{\textbf{SOT}}                          & \multicolumn{1}{c|}{\textbf{SWT}}                          & \multicolumn{1}{c|}{\textbf{DS}}                           & \multicolumn{1}{c|}{\textbf{TMC}}                          & \multicolumn{1}{c|}{\textbf{MQ}}                           & \multicolumn{1}{c|}{\textbf{GTC}}                          & \textbf{}                             & \textbf{}                             \\ \hline
\textbf{geoSSIM}           & {\color[HTML]{32CB00} \textbf{0.779}} & \multicolumn{1}{c|}{{\color[HTML]{000000} \textbf{0.547}}} & \multicolumn{1}{c|}{{\color[HTML]{000000} \textbf{0.630}}} & \multicolumn{1}{c|}{{\color[HTML]{000000} \textbf{0.527}}} & \multicolumn{1}{c|}{{\color[HTML]{000000} \textbf{0.681}}} & \multicolumn{1}{c|}{{\color[HTML]{000000} \textbf{-}}}     & \multicolumn{1}{c|}{{\color[HTML]{000000} \textbf{-}}}     & \multicolumn{1}{c|}{{\color[HTML]{000000} \textbf{0.632}}} & \multicolumn{1}{c|}{{\color[HTML]{000000} \textbf{0.268}}} & {\color[HTML]{000000} \textbf{0.114}} & {\color[HTML]{000000} \textbf{0.353}} \\ \hline
\textbf{geoGradientSSIM}   & {\color[HTML]{3166FF} \textbf{0.674}} & \multicolumn{1}{c|}{{\color[HTML]{000000} \textbf{0.534}}} & \multicolumn{1}{c|}{{\color[HTML]{000000} \textbf{0.571}}} & \multicolumn{1}{c|}{{\color[HTML]{000000} \textbf{0.119}}} & \multicolumn{1}{c|}{{\color[HTML]{000000} \textbf{0.560}}} & \multicolumn{1}{c|}{{\color[HTML]{000000} \textbf{-}}}     & \multicolumn{1}{c|}{{\color[HTML]{000000} \textbf{-}}}     & \multicolumn{1}{c|}{{\color[HTML]{000000} \textbf{0.586}}} & \multicolumn{1}{c|}{{\color[HTML]{000000} \textbf{0.183}}} & {\color[HTML]{000000} \textbf{0.093}} & {\color[HTML]{000000} \textbf{0.363}} \\ \hline
\textbf{colorSSIM}         & {\color[HTML]{000000} \textbf{0.670}} & \multicolumn{1}{c|}{{\color[HTML]{32CB00} \textbf{0.677}}} & \multicolumn{1}{c|}{{\color[HTML]{32CB00} \textbf{0.797}}} & \multicolumn{1}{c|}{{\color[HTML]{FE0000} \textbf{0.755}}} & \multicolumn{1}{c|}{{\color[HTML]{32CB00} \textbf{0.726}}} & \multicolumn{1}{c|}{{\color[HTML]{3166FF} \textbf{0.803}}} & \multicolumn{1}{c|}{{\color[HTML]{32CB00} \textbf{0.798}}} & \multicolumn{1}{c|}{{\color[HTML]{3166FF} \textbf{0.797}}} & \multicolumn{1}{c|}{{\color[HTML]{FE0000} \textbf{0.727}}} & {\color[HTML]{FE0000} \textbf{0.795}} & {\color[HTML]{32CB00} \textbf{0.571}} \\ \hline
\textbf{colorGradientSSIM} & {\color[HTML]{FE0000} \textbf{0.811}} & \multicolumn{1}{c|}{{\color[HTML]{3166FF} \textbf{0.570}}} & \multicolumn{1}{c|}{{\color[HTML]{3166FF} \textbf{0.794}}} & \multicolumn{1}{c|}{{\color[HTML]{32CB00} \textbf{0.743}}} & \multicolumn{1}{c|}{{\color[HTML]{3166FF} \textbf{0.715}}} & \multicolumn{1}{c|}{{\color[HTML]{32CB00} \textbf{0.837}}} & \multicolumn{1}{c|}{{\color[HTML]{3166FF} \textbf{0.765}}} & \multicolumn{1}{c|}{{\color[HTML]{32CB00} \textbf{0.799}}} & \multicolumn{1}{c|}{{\color[HTML]{3166FF} \textbf{0.619}}} & {\color[HTML]{3166FF} \textbf{0.690}} & {\color[HTML]{3166FF} \textbf{0.565}} \\ \hline
\textbf{FMQM}              & {\color[HTML]{FE0000} \textbf{0.811}} & \multicolumn{1}{c|}{{\color[HTML]{FE0000} \textbf{0.693}}} & \multicolumn{1}{c|}{{\color[HTML]{FE0000} \textbf{0.802}}} & \multicolumn{1}{c|}{{\color[HTML]{3166FF} \textbf{0.737}}} & \multicolumn{1}{c|}{{\color[HTML]{FE0000} \textbf{0.758}}} & \multicolumn{1}{c|}{{\color[HTML]{FE0000} \textbf{0.848}}} & \multicolumn{1}{c|}{{\color[HTML]{FE0000} \textbf{0.807}}} & \multicolumn{1}{c|}{{\color[HTML]{FE0000} \textbf{0.803}}} & \multicolumn{1}{c|}{{\color[HTML]{32CB00} \textbf{0.688}}} & {\color[HTML]{32CB00} \textbf{0.760}} & {\color[HTML]{FE0000} \textbf{0.666}} \\ \hline
\end{tabular}%
}
\end{table*}
The hardware platform features an 11th Gen Intel(R) Core(TM) i7-11800H CPU (2.3 GHz, 16 cores) with 16GB RAM. GraphSIM is implemented in MATLAB, FMQM in Python, and all other metrics, including rendering and point cloud sampling, in C++.

Computation times, detailed in Table~\ref{tab:computationTime}, are divided into three components: preprocessing time, media-type conversion time, and metric computation time. Preprocessing—including keypoint extraction for GraphSIM and local patch construction and sampling for FMQM—is performed solely on reference meshes and can be completed offline; thus, it is excluded from the total computation time. Media-type conversion refers to rendering for projection-based methods, point cloud generation for point-based methods, and field value computation for FMQM.
The following observations can be made:
We can see that:

i) In projection-based methods, rendering accounts for the majority of the computation time, while metric evaluation itself is lightweight. Video-based methods are slower than image-based ones due to the need for more rendered views to cover a full rotation, and differences in rendering resolution across datasets further contribute to time variability.
The rendering time also varies significantly between datasets, reflecting its sensitivity to mesh complexity such as vertex count and geometric detail.
Consequently, video-based methods are the slowest overall, while image-based methods remain relatively efficient.

ii) Model-based methods are also sensitive to mesh complexity due to neighbor search operations. Simpler geometry-only metrics (e.g., MESH, GL) are the fastest overall, while metrics involving topological structures are slower, especially on complex datasets like TSMD and YANA.

iii) Although point-based methods eliminate the need for explicit topology processing, they rely on computationally intensive neighbor searches within unstructured point clouds, resulting in generally higher computation times compared to image-based and simple model-based methods.
Notably, GraphSIM exhibits significant variance in computation time, as a few samples with exceptionally dense point clouds (e.g., over 4 million points) disproportionately inflate the average time.

iv) FMQM's major cost lies in local patch construction and sample generation, which can be fully preprocessed. Consequently, FMQM achieves fast and stable runtime across all datasets, demonstrating strong potential for real-time TMQA applications.

\subsection{Ablation Study}
\subsubsection{Effectiveness of four features}
The performances of FMQM and its four individual features across various datasets are summarized in Table~\ref{tab:per-feature}.  
It is important to note that for pure texture distortions, where the mesh geometry remains unchanged, \(\mathrm{geoSSIM}\) and \(\mathrm{geoGraSSIM}\) are not applicable because the underlying geometric structures between the reference and distorted meshes are identical.

For distortions involving only geometric changes, \(\mathrm{geoSSIM}\) and \(\mathrm{geoGraSSIM}\) generally perform reasonably well, with the notable exception of the SOT distortion.  
SOT is generated using the QEM~\cite{qem}, which minimizes geometric error by balancing SDF values symmetrically around the surface.  
As a result, the geometric field differences captured by \(\mathrm{geoSSIM}\) and \(\mathrm{geoGraSSIM}\) are significantly weakened, leading to poor performance under such distortions.  
Furthermore, when texture degradation is introduced, the performance of geometric features deteriorates rapidly.

In contrast, color-based features (\(\mathrm{colorSSIM}\) and \(\mathrm{colorGraSSIM}\)) consistently achieve strong performance across all types of distortions.  
This is because even purely geometric deformations typically induce visible texture mapping changes, making color features highly sensitive and effective.  
These results highlight that color information often contributes more critically than pure geometry in the TMQA task.

While combining all four features generally leads to the best overall performance, in some cases, individual color features outperform the full FMQM (e.g., \(\mathrm{colorSSIM}\) surpasses FMQM on SJTU-TMQA).  
This is mainly because FMQM currently assigns equal weight to geometric and color features; when geometric features perform poorly on certain distortions, they can drag down the overall performance.  
Future work could address this limitation by adopting HVS-inspired weighting strategies, allowing color and geometry contributions to be adaptively balanced based on perceptual importance.

\section{Conclusion}
\label{sec:conclusion}
In this paper, we proposed the Field Mesh Quality Metric (FMQM), the first TMQA metric based on 3D field representations.  
By leveraging SDF and a newly proposed novel color field NCF, FMQM assesses geometric and textural degradations through distribution comparisons of field values and gradients at corresponding spatial locations.  
This field-based formulation effectively addresses key challenges in TMQA, such as handling topology differences, texture utilization, and view selection biases.  
Extensive experiments conducted on three TMQA datasets demonstrate that FMQM achieves robust and superior performance compared to state-of-the-art metrics, while maintaining a low computational cost that favors practical deployment.  
Our work not only advances research in the TMQA domain, but also introduces a novel mesh representation approach for designing subsequent TMQA metrics. In our future work, we aim to explore additional features based on field representation to further enhance metric performance.

\bibliographystyle{IEEEtran}
\input{main.bbl}

\clearpage
\appendix
\section*{Supplementary Materials for Textured Mesh Quality Assessment using Geometry and Color Field Similarity}

\author{Kaifa Yang,
        Qi Yang,~\IEEEmembership{Member,~IEEE},
        Zhu Li,~\IEEEmembership{Senior Member,~IEEE},
        Yiling Xu,~\IEEEmembership{Member,~IEEE}
}

\subsection{NCF Visualizations on Various Mesh Samples}
\begin{figure*}[p]
  \centering
  \setlength{\tabcolsep}{1pt}
  \renewcommand{\arraystretch}{1.0}
  \begin{tabular}{ccccc}
    \includegraphics[width=0.19\textwidth]{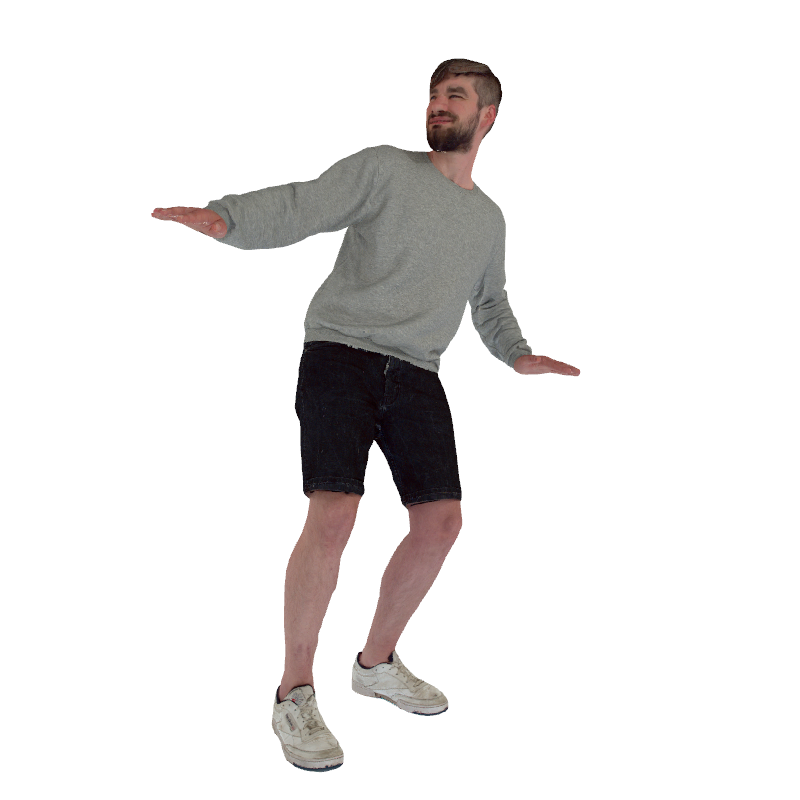} &
    \includegraphics[width=0.19\textwidth]{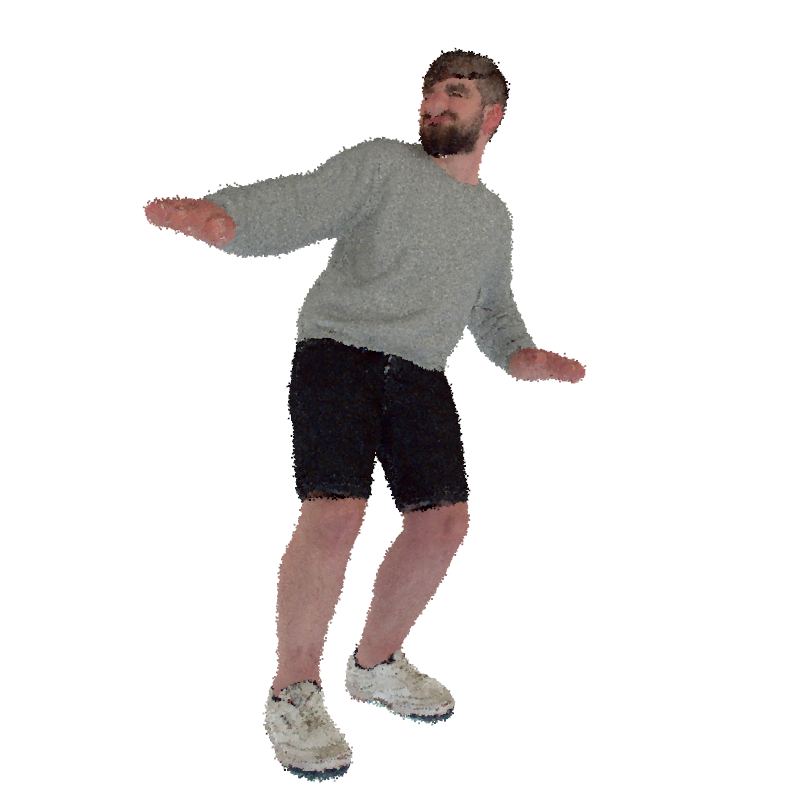} &
    \includegraphics[width=0.19\textwidth]{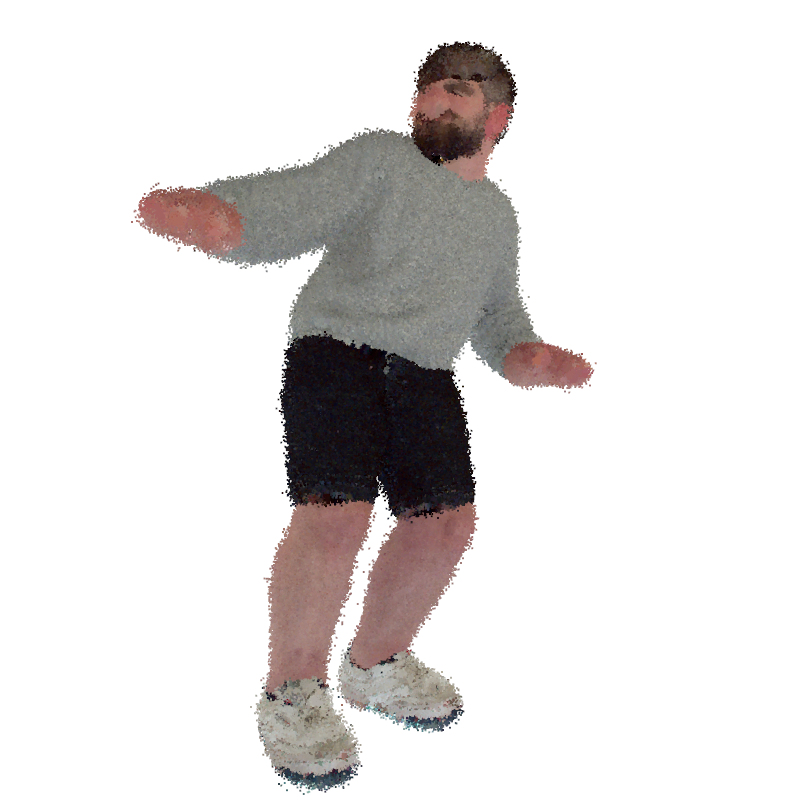} &
    \includegraphics[width=0.19\textwidth]{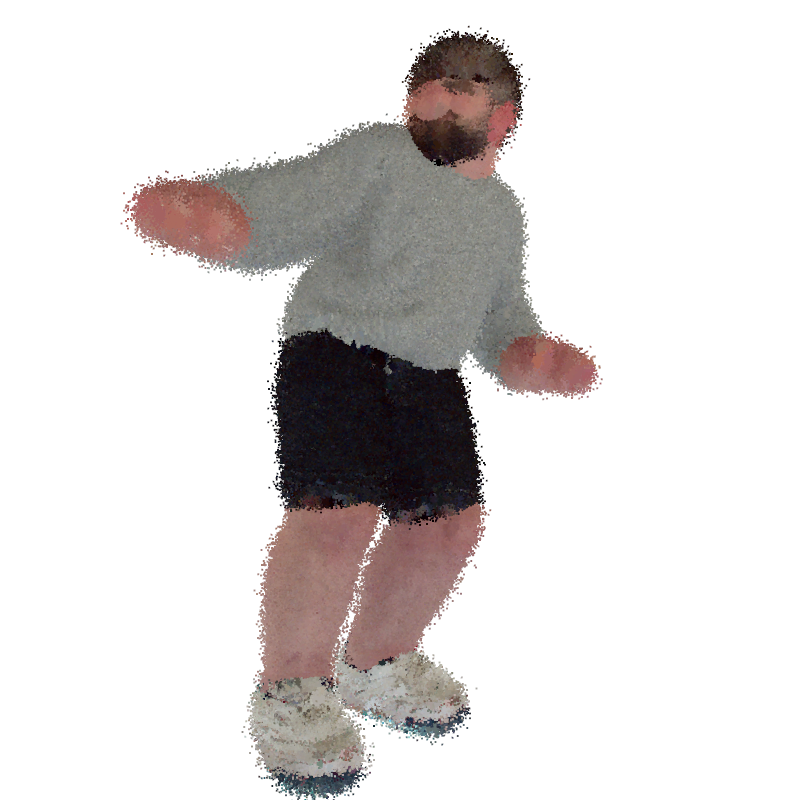} &
    \includegraphics[width=0.19\textwidth]{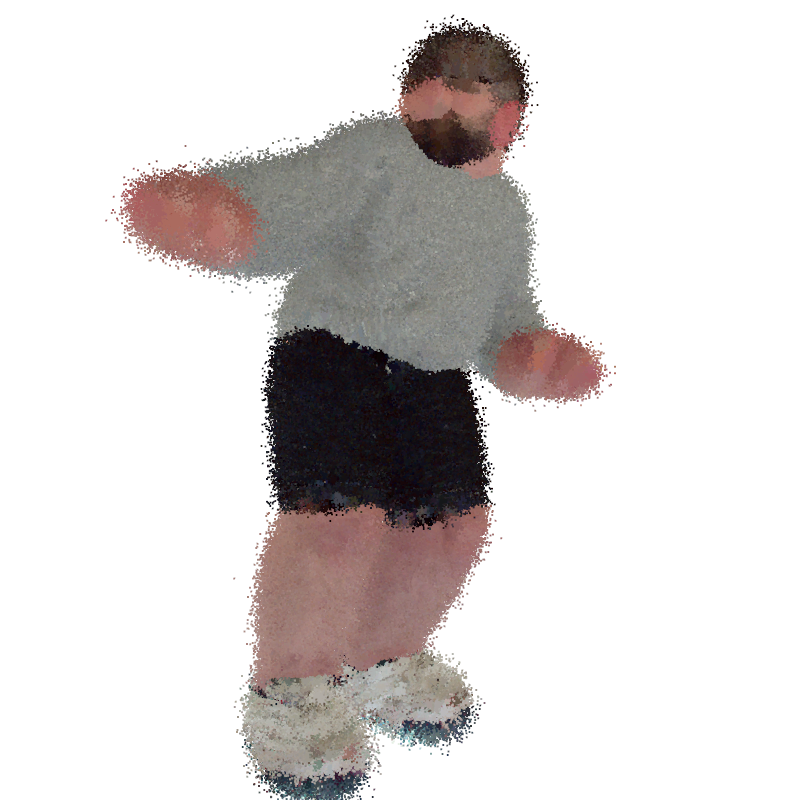} \\
    
    \includegraphics[width=0.19\textwidth]{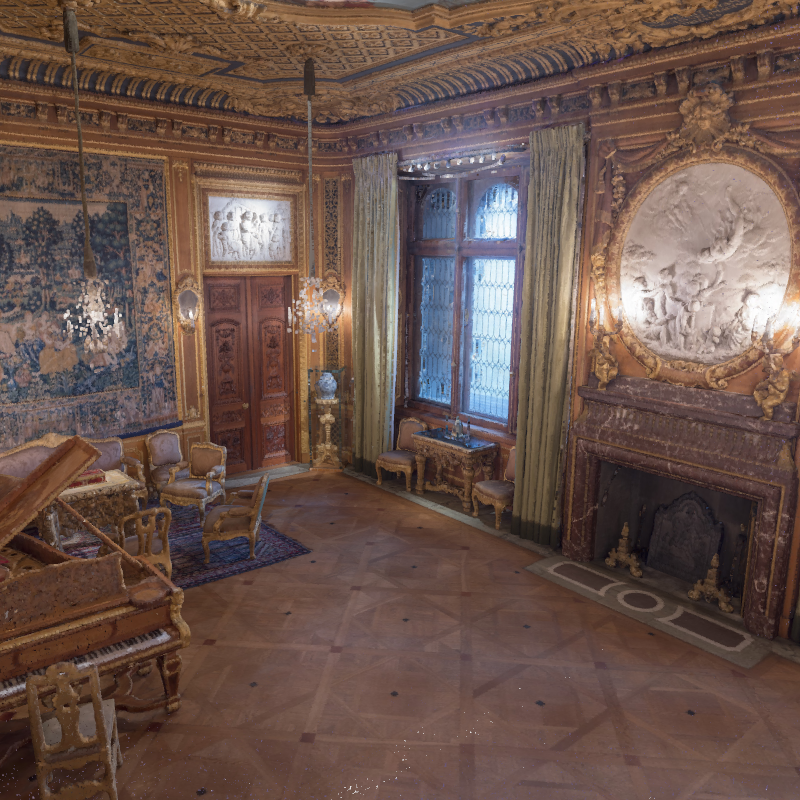} &
    \includegraphics[width=0.19\textwidth]{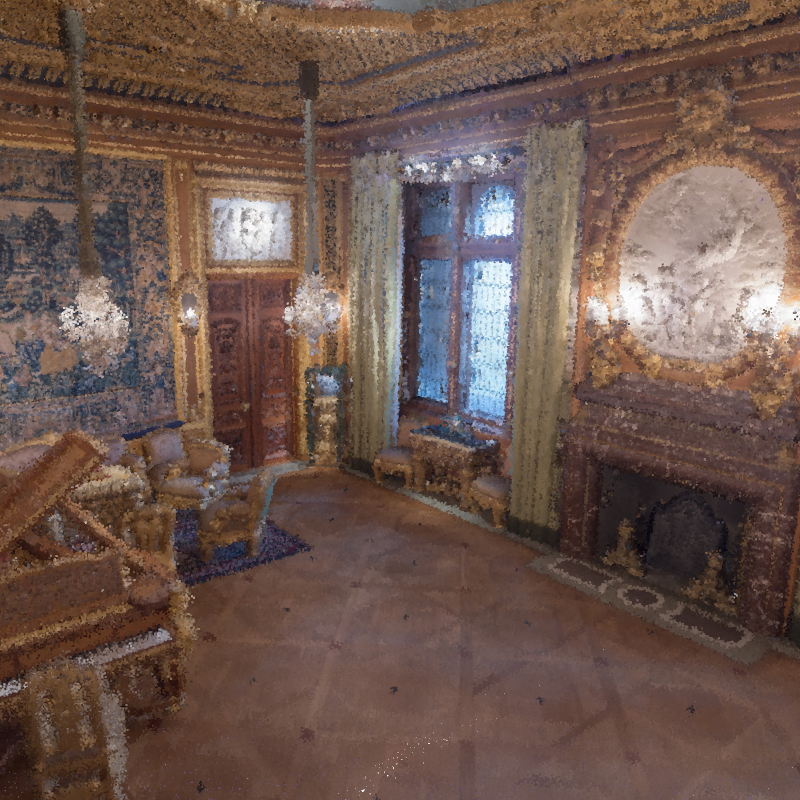} &
    \includegraphics[width=0.19\textwidth]{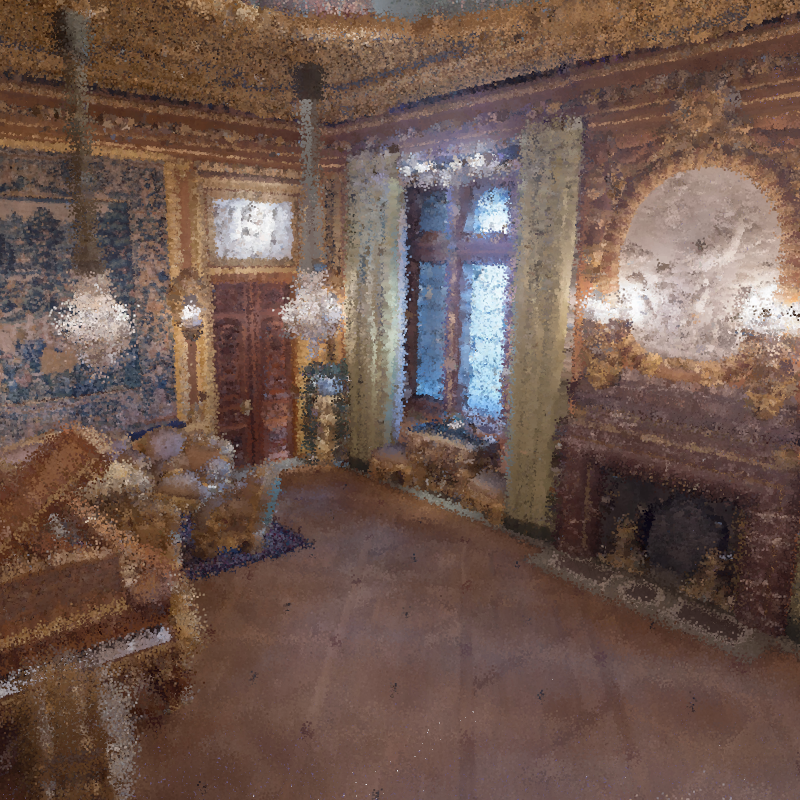} &
    \includegraphics[width=0.19\textwidth]{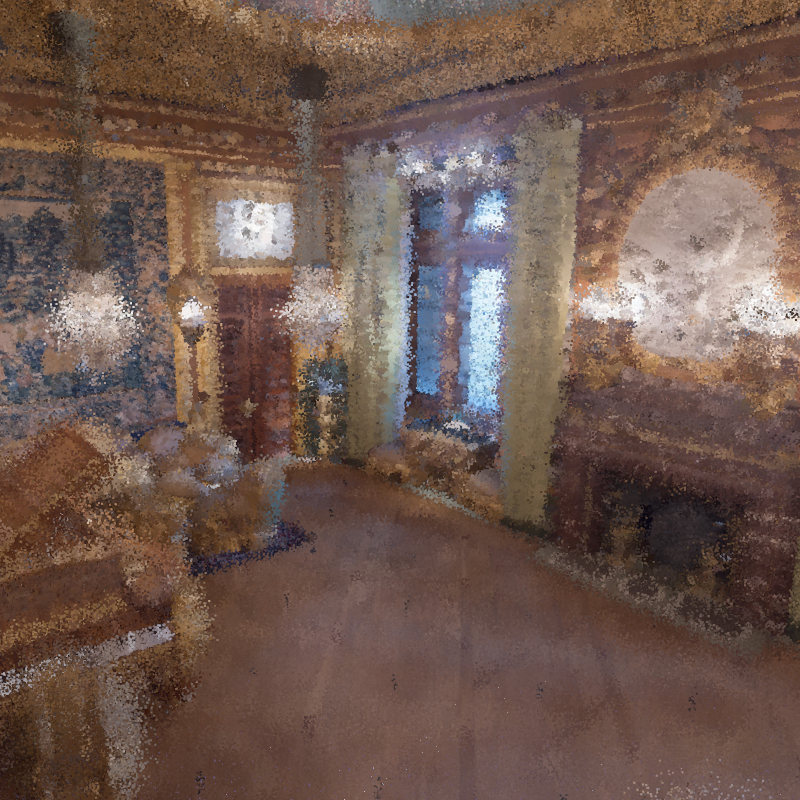} &
    \includegraphics[width=0.19\textwidth]{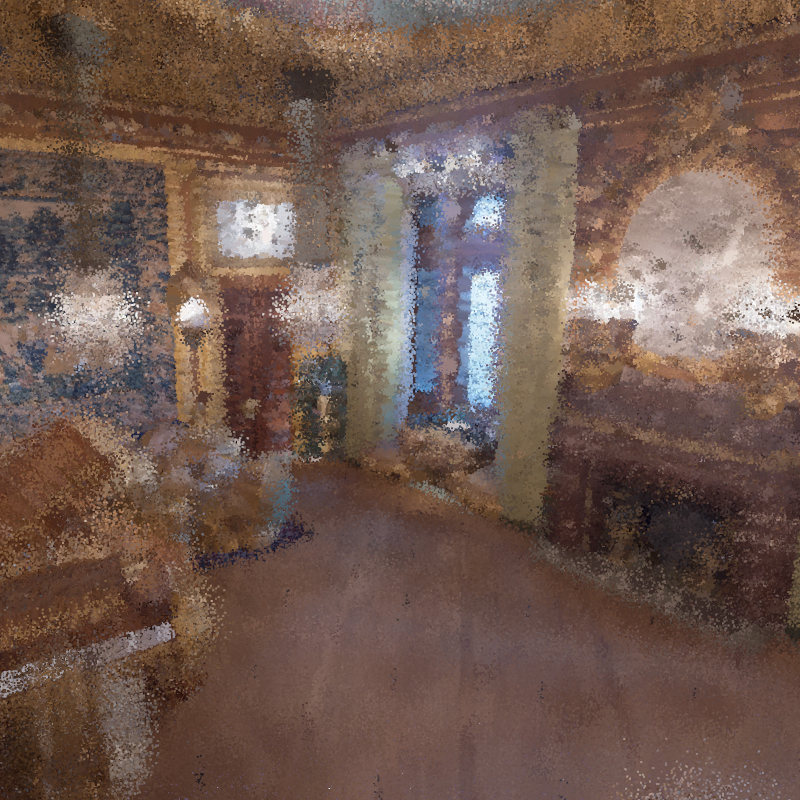} \\

    \includegraphics[width=0.19\textwidth]{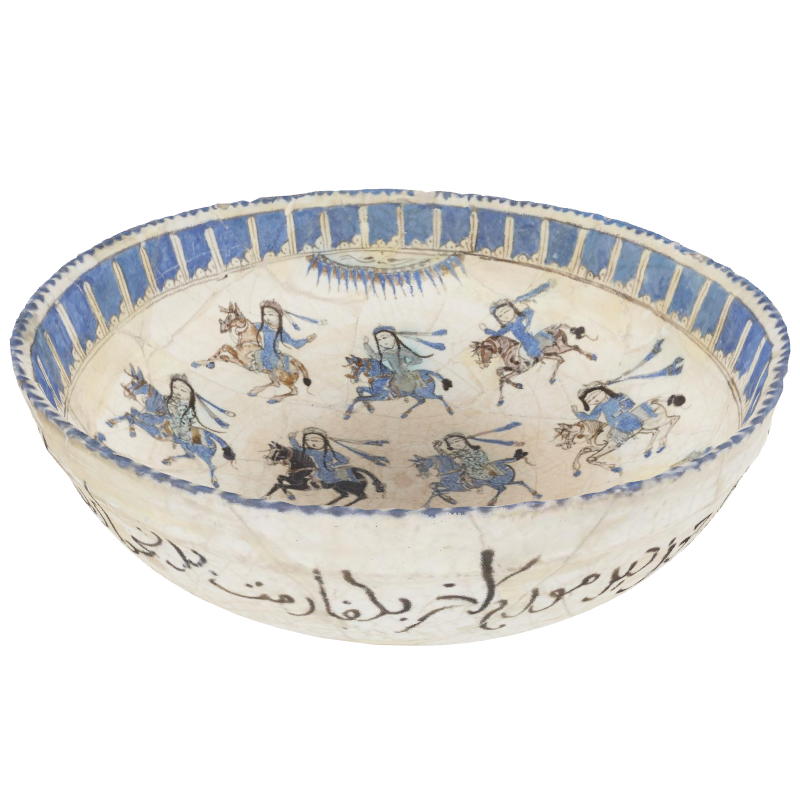} &
    \includegraphics[width=0.19\textwidth]{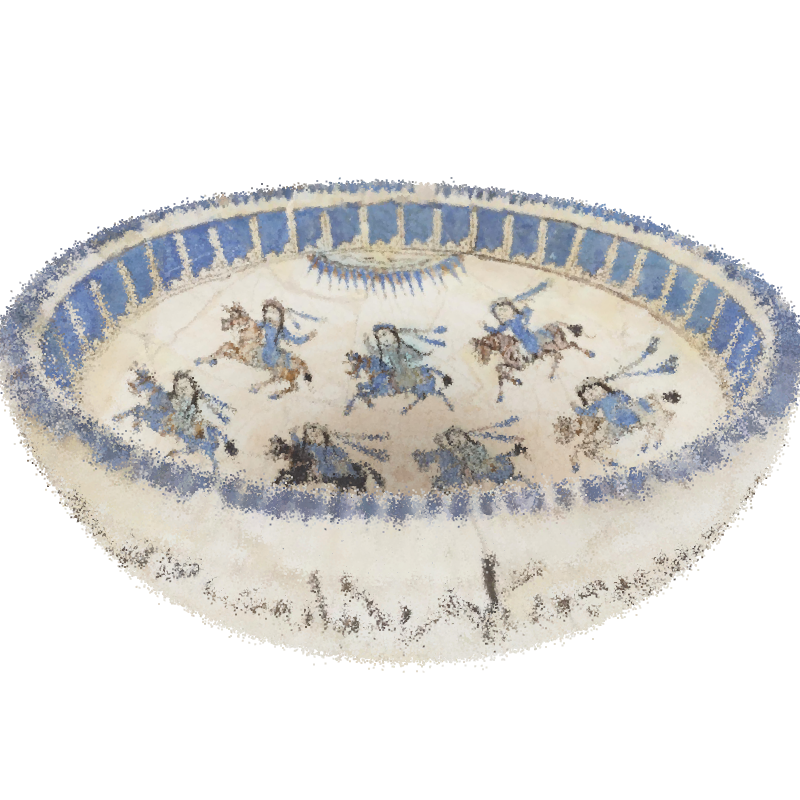} &
    \includegraphics[width=0.19\textwidth]{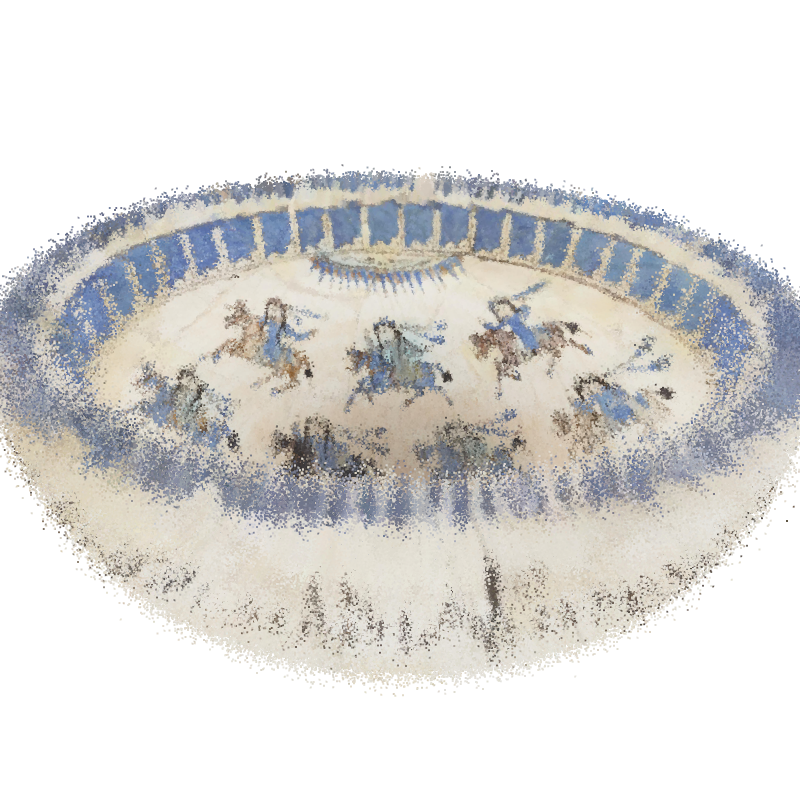} &
    \includegraphics[width=0.19\textwidth]{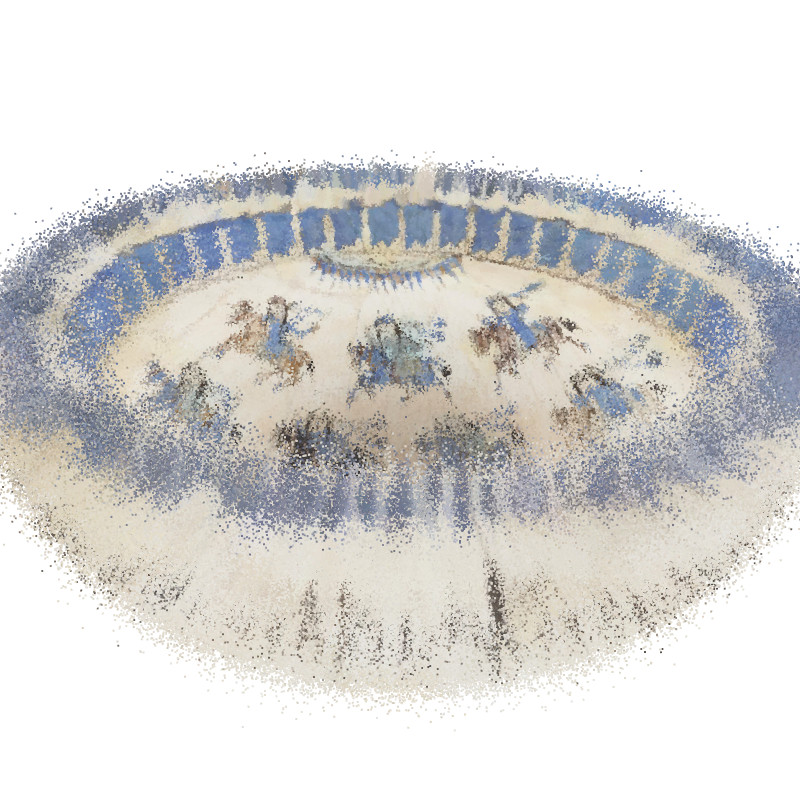} &
    \includegraphics[width=0.19\textwidth]{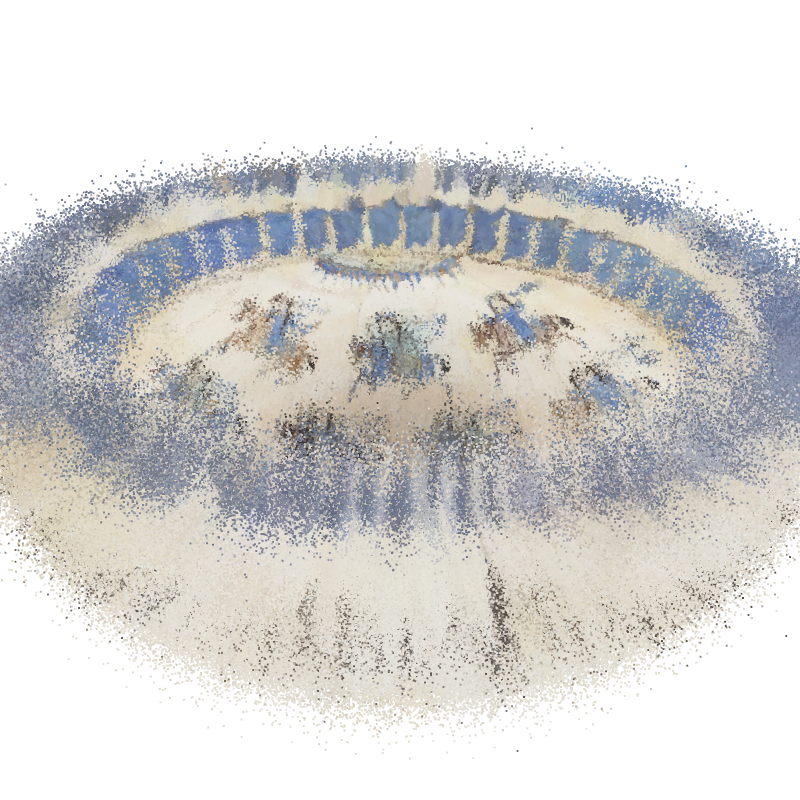} \\

    \includegraphics[width=0.19\textwidth]{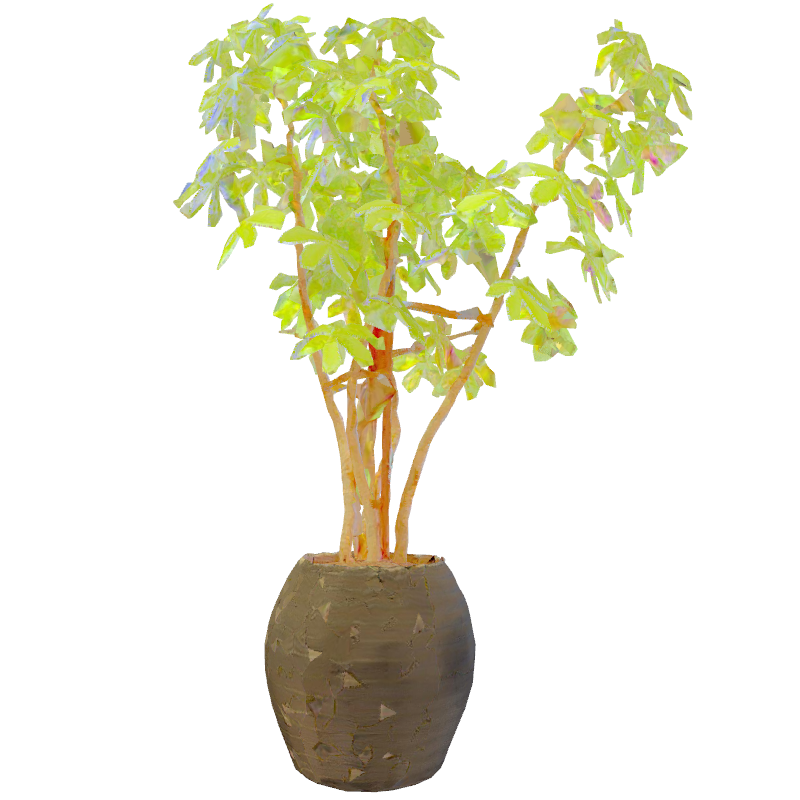} &
    \includegraphics[width=0.19\textwidth]{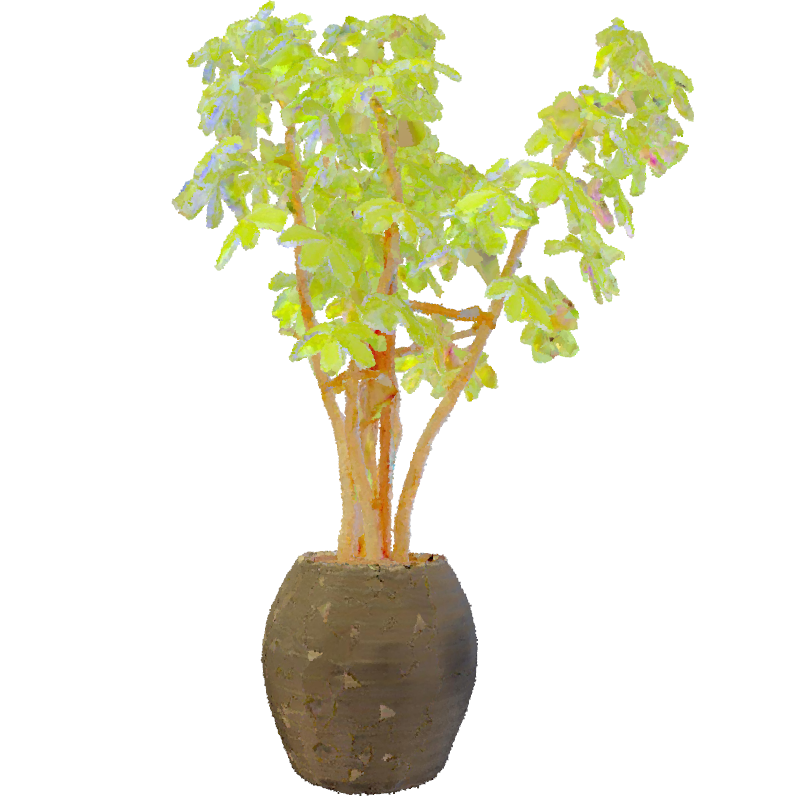} &
    \includegraphics[width=0.19\textwidth]{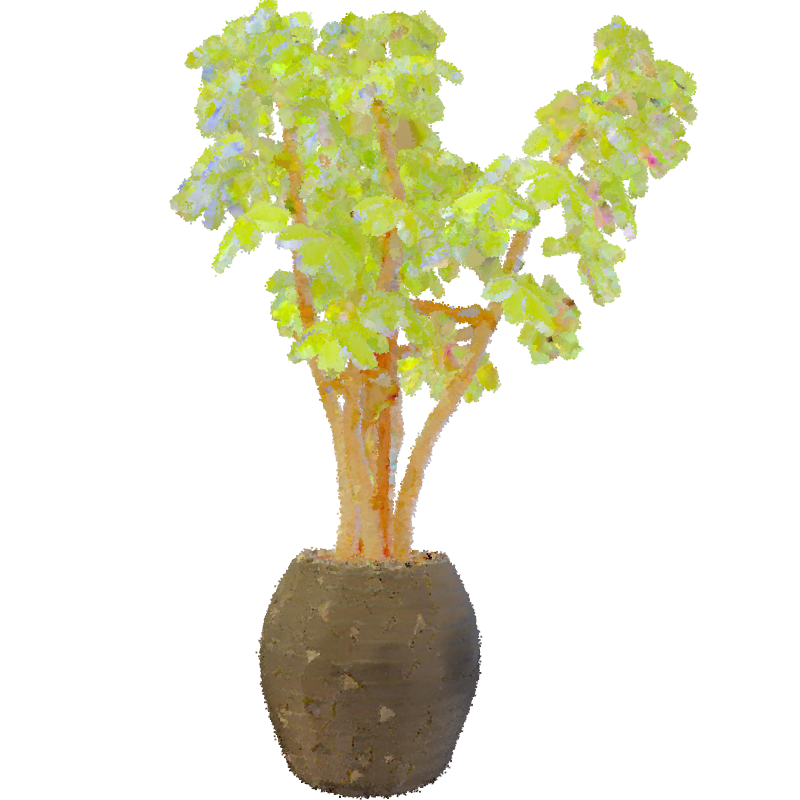} &
    \includegraphics[width=0.19\textwidth]{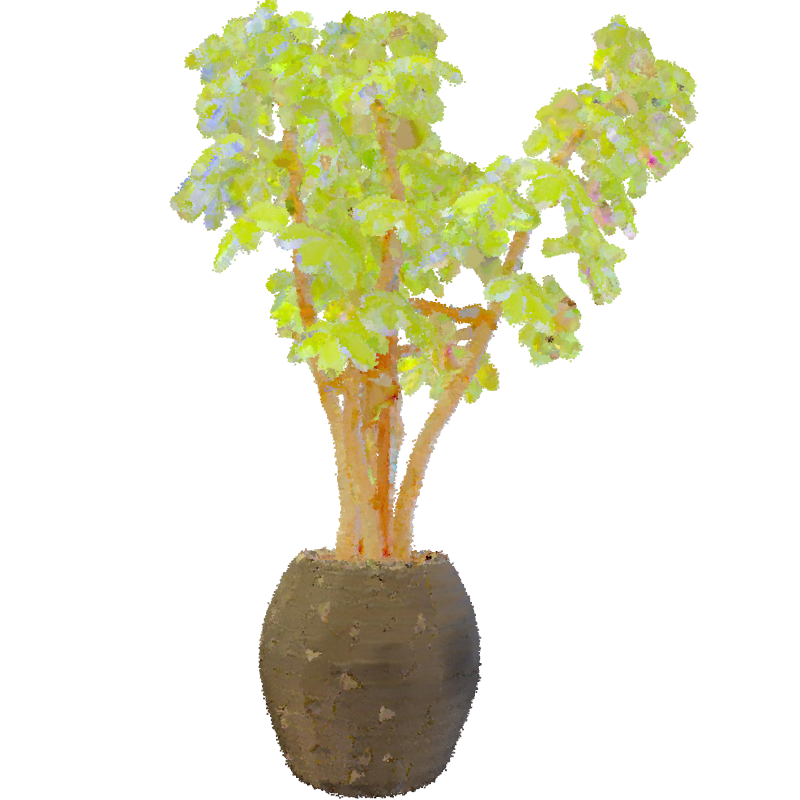} &
    \includegraphics[width=0.19\textwidth]{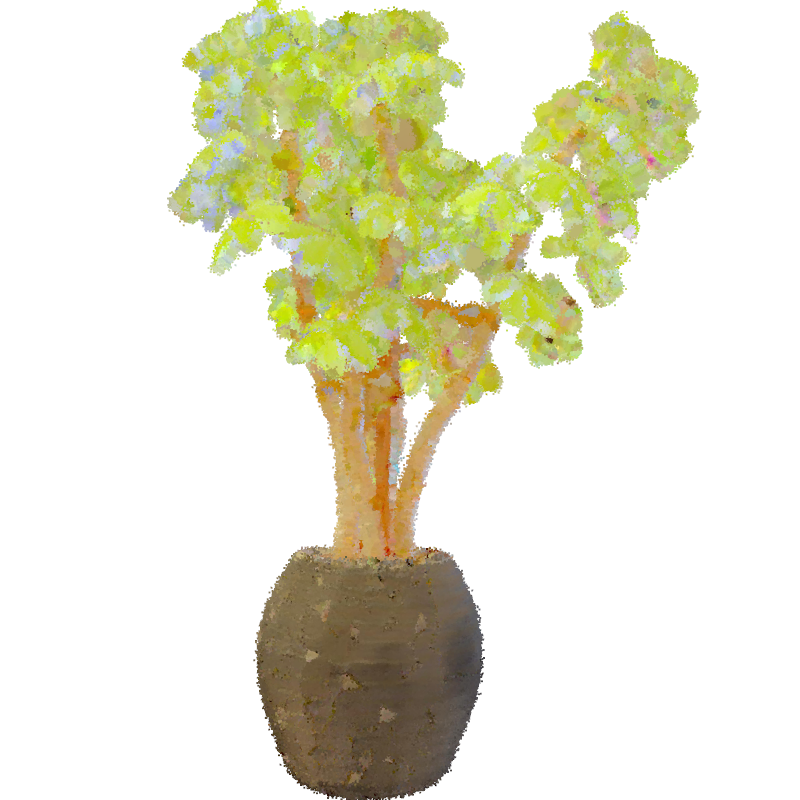} \\

    \includegraphics[width=0.19\textwidth]{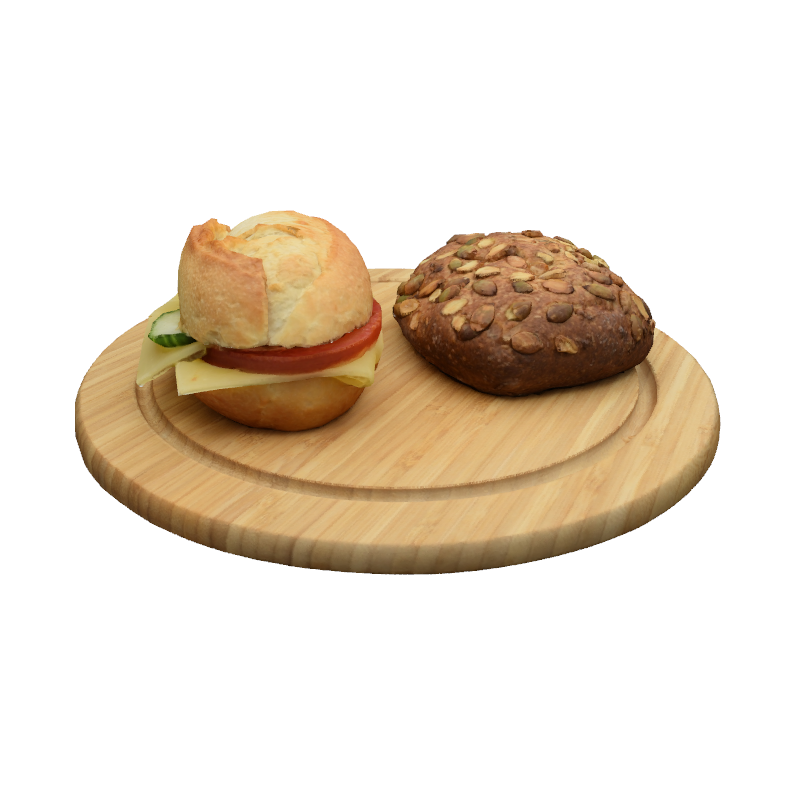} &
    \includegraphics[width=0.19\textwidth]{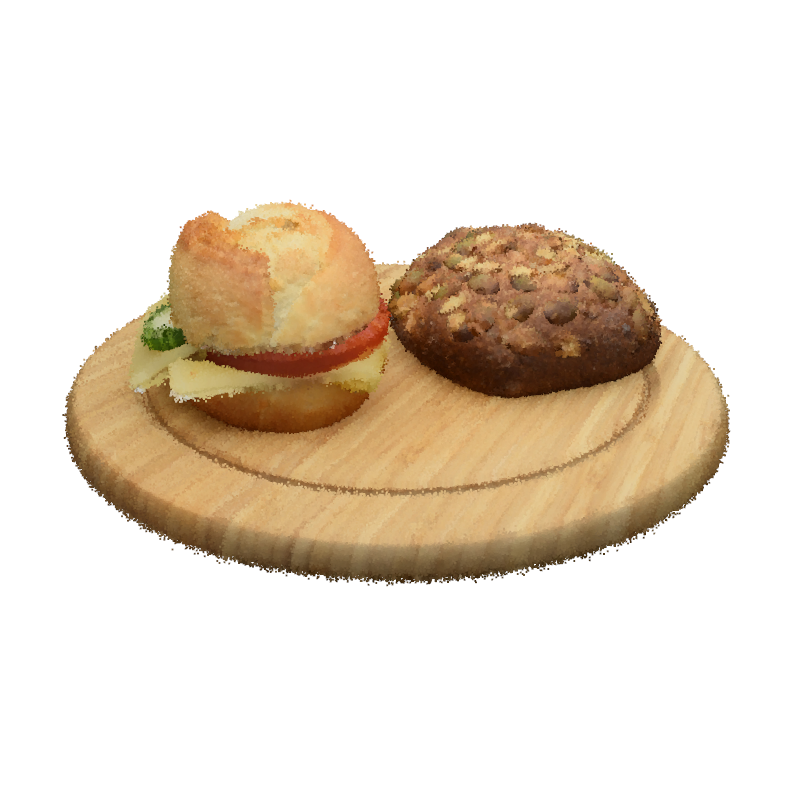} &
    \includegraphics[width=0.19\textwidth]{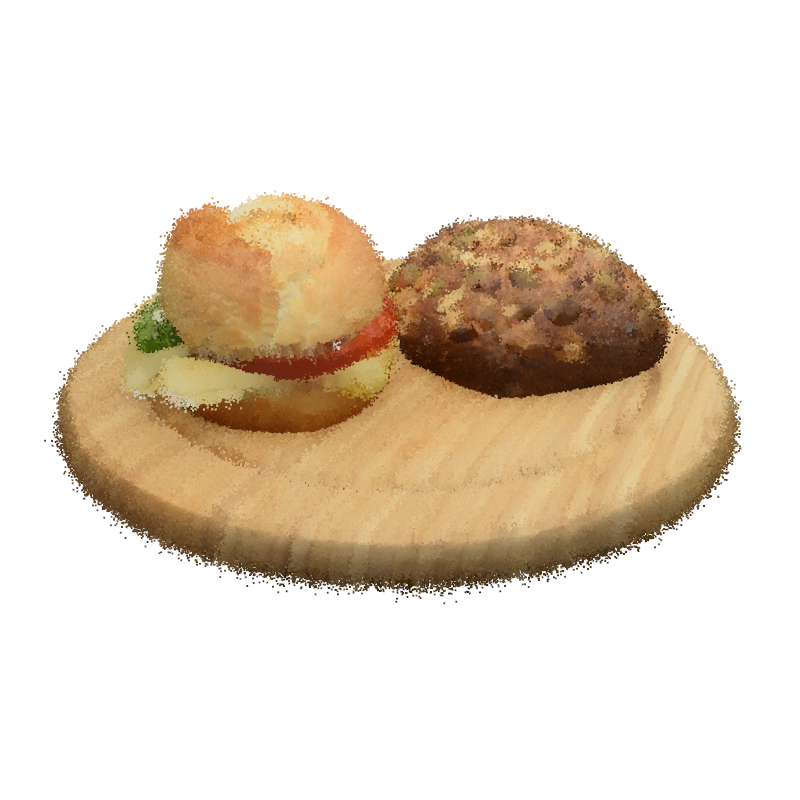} &
    \includegraphics[width=0.19\textwidth]{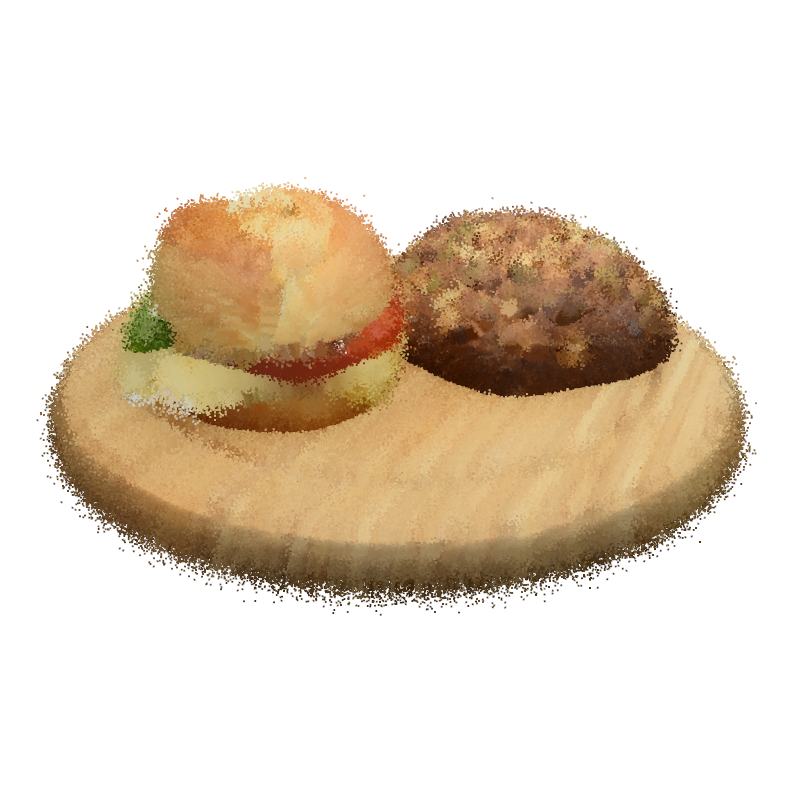} &
    \includegraphics[width=0.19\textwidth]{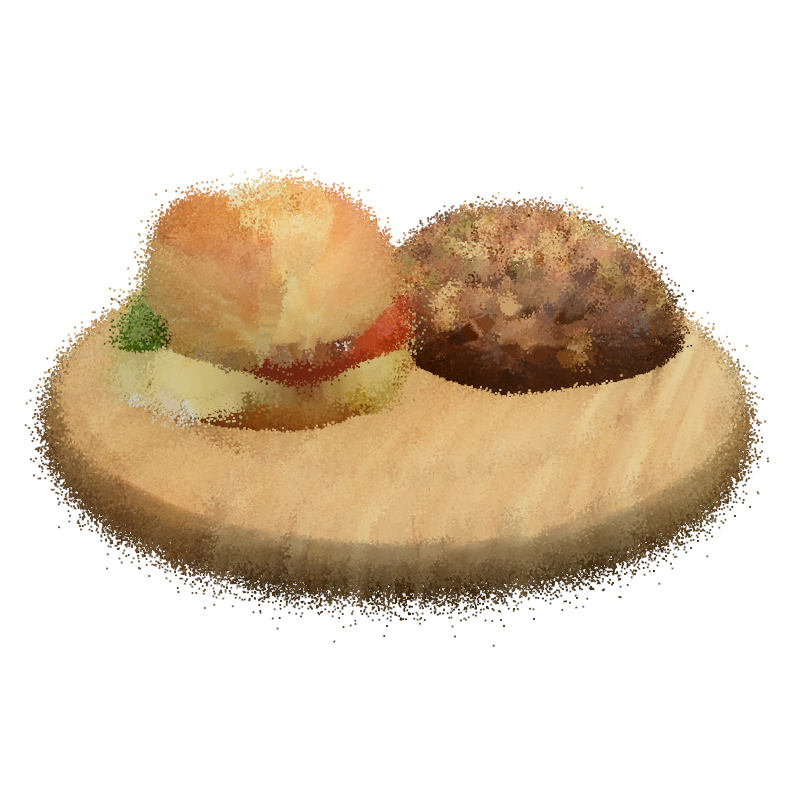} \\

    \includegraphics[width=0.19\textwidth]{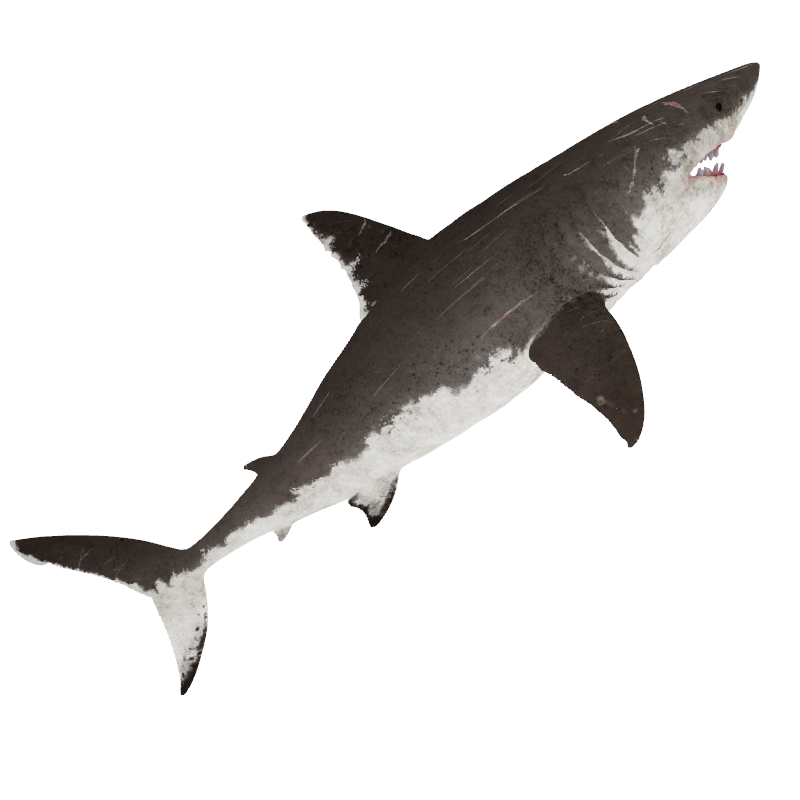} &
    \includegraphics[width=0.19\textwidth]{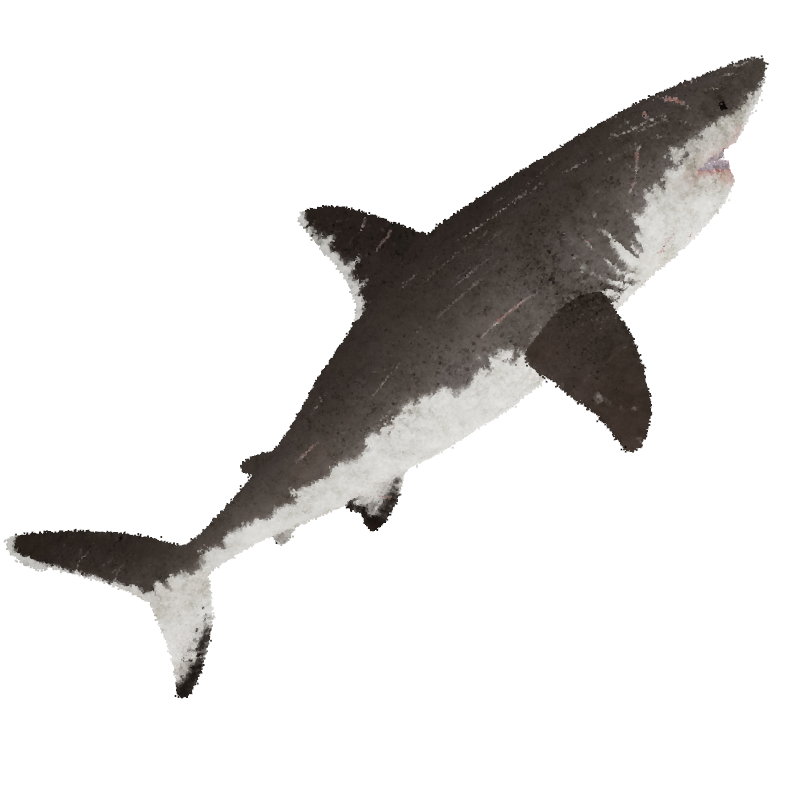} &
    \includegraphics[width=0.19\textwidth]{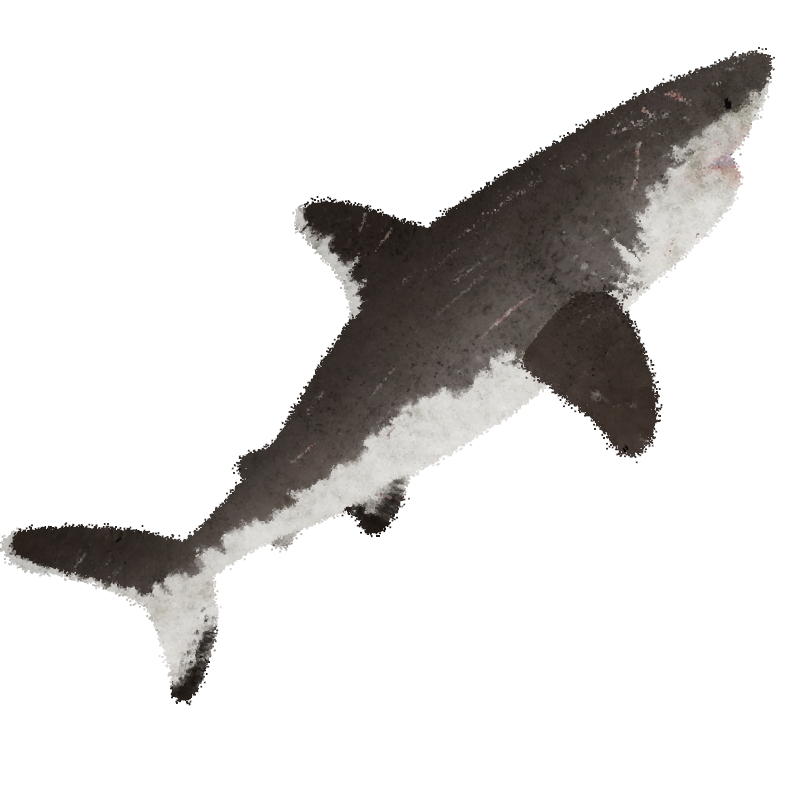} &
    \includegraphics[width=0.19\textwidth]{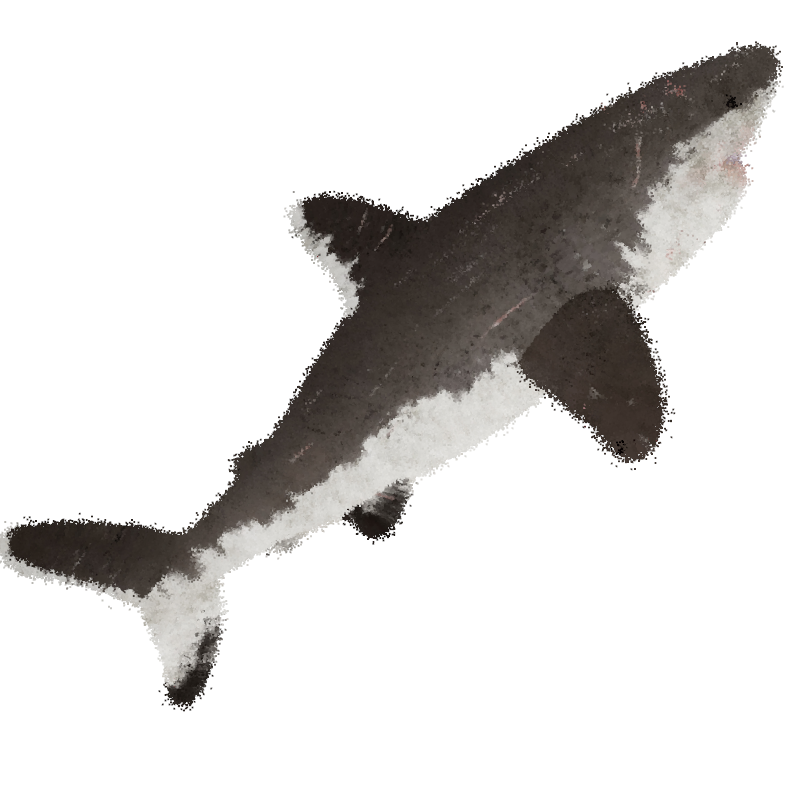} &
    \includegraphics[width=0.19\textwidth]{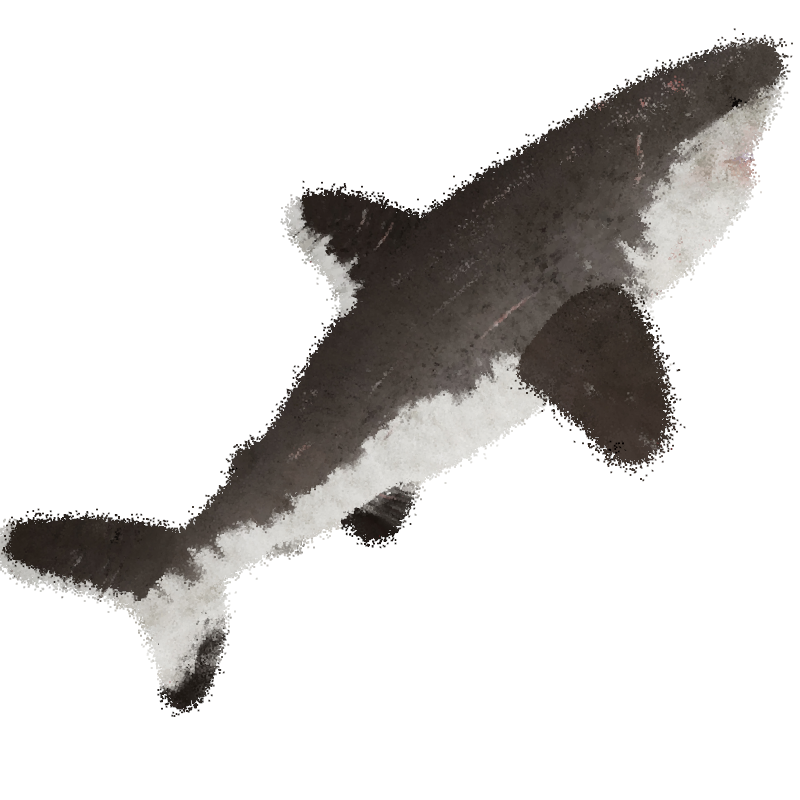} \\
  \end{tabular}
  \caption{NCF visualizations across multiple mesh samples and spatial locations. For each row, a mesh sample is shown, and the five images represent sampling points at increasing distances along the surface normal direction.}
  \label{fig:ncf_diffusion}
\end{figure*}

To provide an intuitive understanding of the spatial behavior of the nearest surface point color field (NCF), we visualize its values across several mesh samples in Fig.~\ref{fig:ncf_diffusion}. Specifically, we select six representative models from our datasets: \texttt{mitch\_fr00001} and \texttt{the\_great\_drawing\_room} from TSMD~\cite{tsmd}, \texttt{wareBowl} and \texttt{plant} from SJTU-MQA~\cite{sjtumqa}, and \texttt{bread} and \texttt{shark} from YANA~\cite{yana2023}. For each sample, we generate five sets of sampling points progressively displaced along the surface normal direction. Each point is assigned a color according to its NCF value evaluated via Equation~(7) in the main paper, i.e., by tracing to its nearest surface point and acquiring the corresponding texture color.

These visualizations clearly reflect Property~(1) and Property(3) of the NCF, as defined in Section~IV.B of the main paper. First, the colors remain stable along the normal direction, demonstrating the normal-aligned diffusion behavior of NCF, where texture information is extended outward in a perceptually consistent manner. Second, despite the displacement, the overall structure of the texture remains coherent—features such as object contours, text, or patterns are preserved across sampling layers. This confirms that NCF maintains strong structural consistency even under normal deviation. These characteristics highlight the perceptual alignment and robustness of NCF in capturing local texture appearance.

\subsection{NCF Feature Extraction Unit Visualization}
\begin{figure*}[p]
    \centering
    \includegraphics[width=0.3\textwidth]{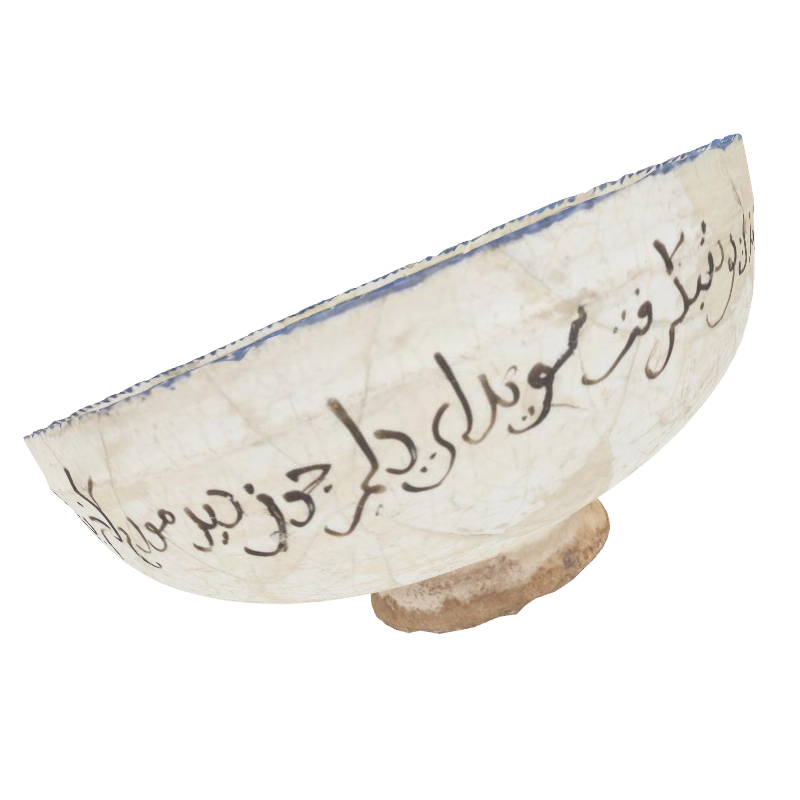}
    \hfill
    \includegraphics[width=0.3\textwidth]{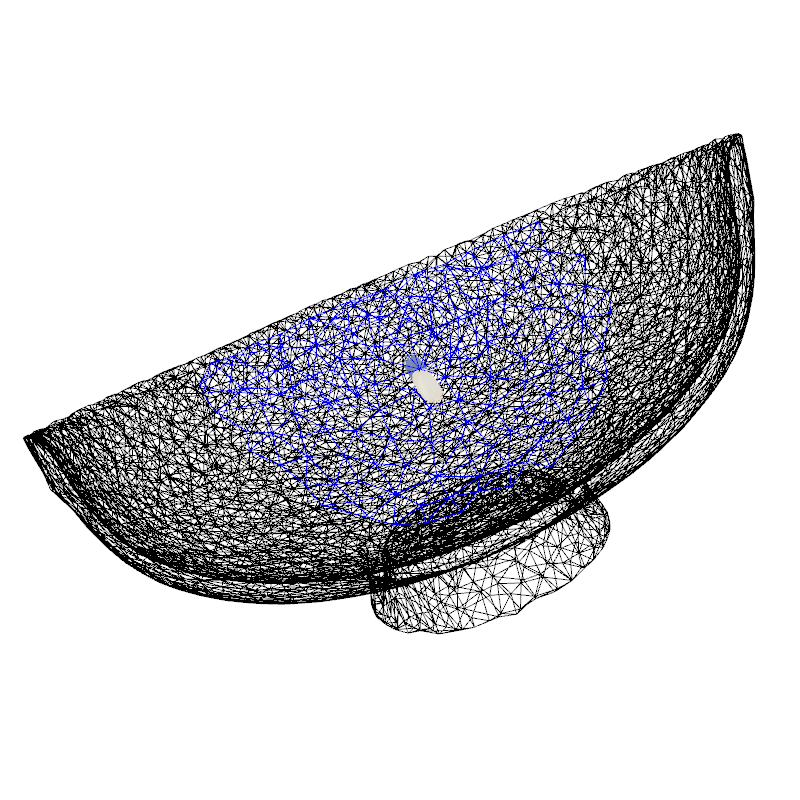}
    \hfill
    \includegraphics[width=0.3\textwidth]{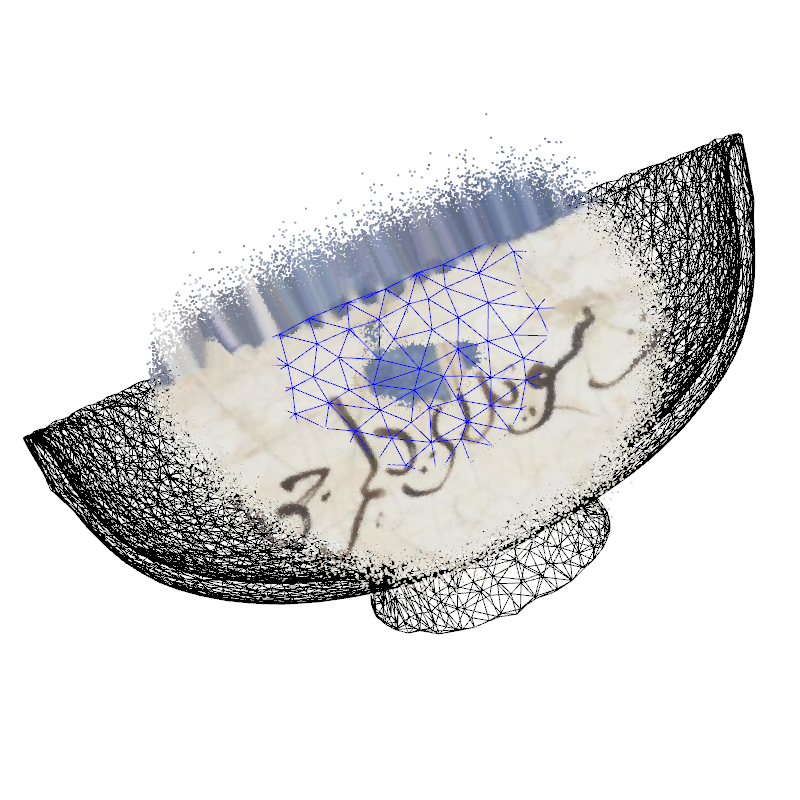}

    \vspace{2pt}
    \includegraphics[width=0.3\textwidth]{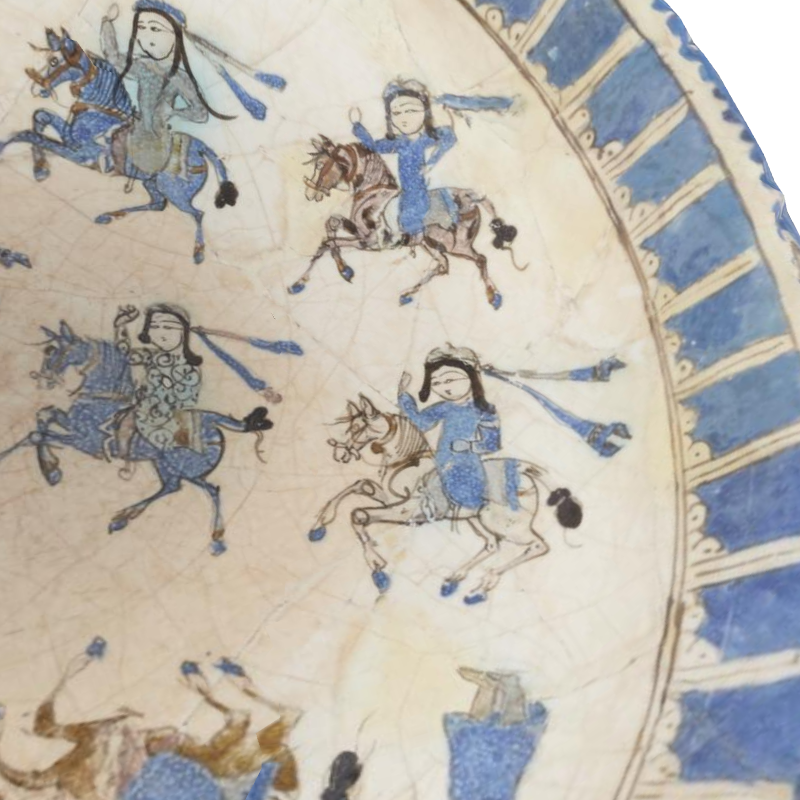}
    \hfill
    \includegraphics[width=0.3\textwidth]{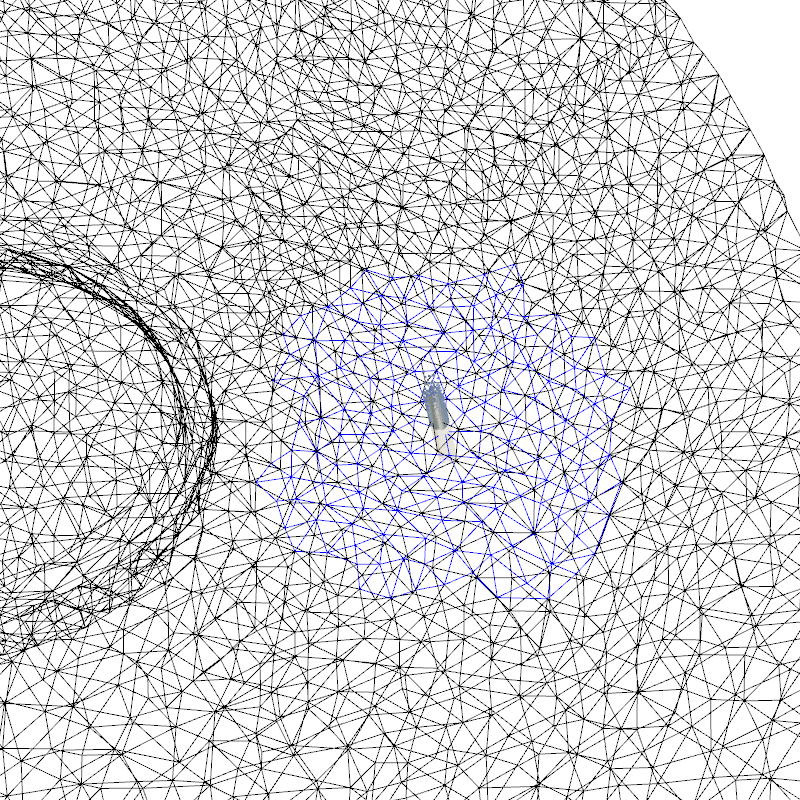}
    \hfill
    \includegraphics[width=0.3\textwidth]{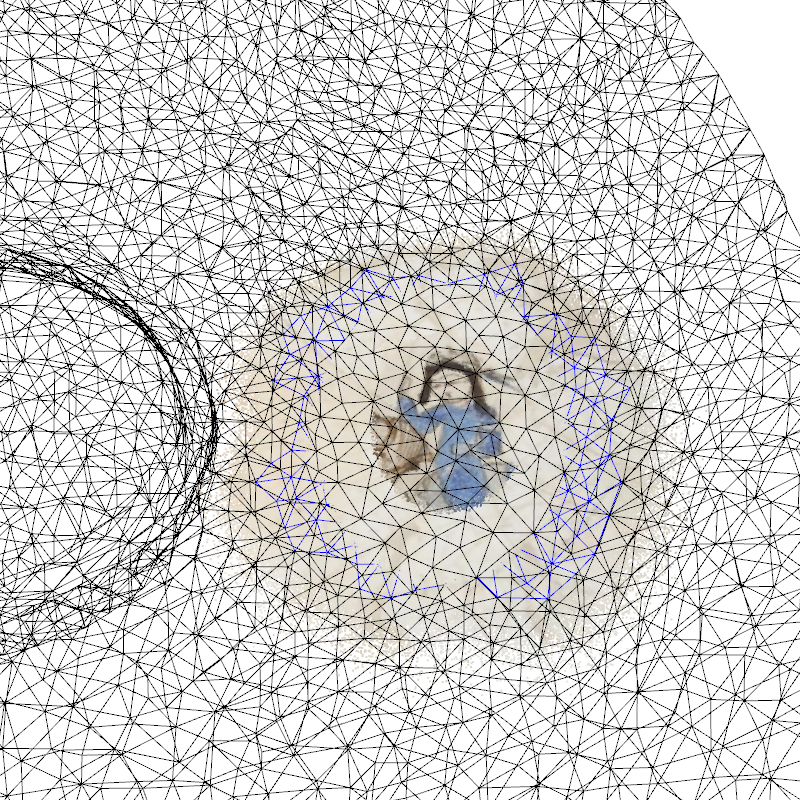}

    \vspace{2pt}
    \includegraphics[width=0.3\textwidth]{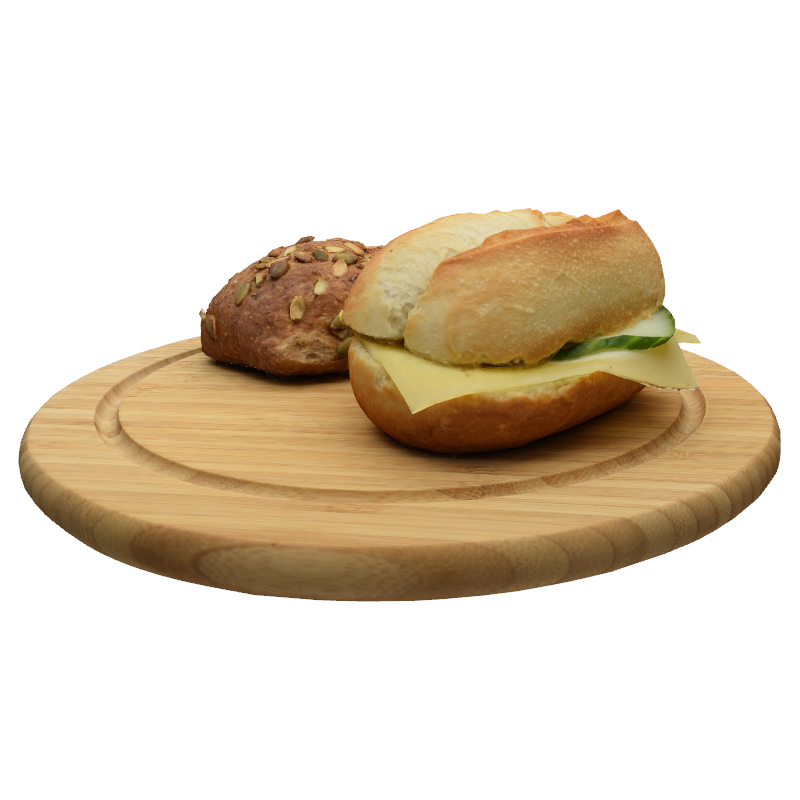}
    \hfill
    \includegraphics[width=0.3\textwidth]{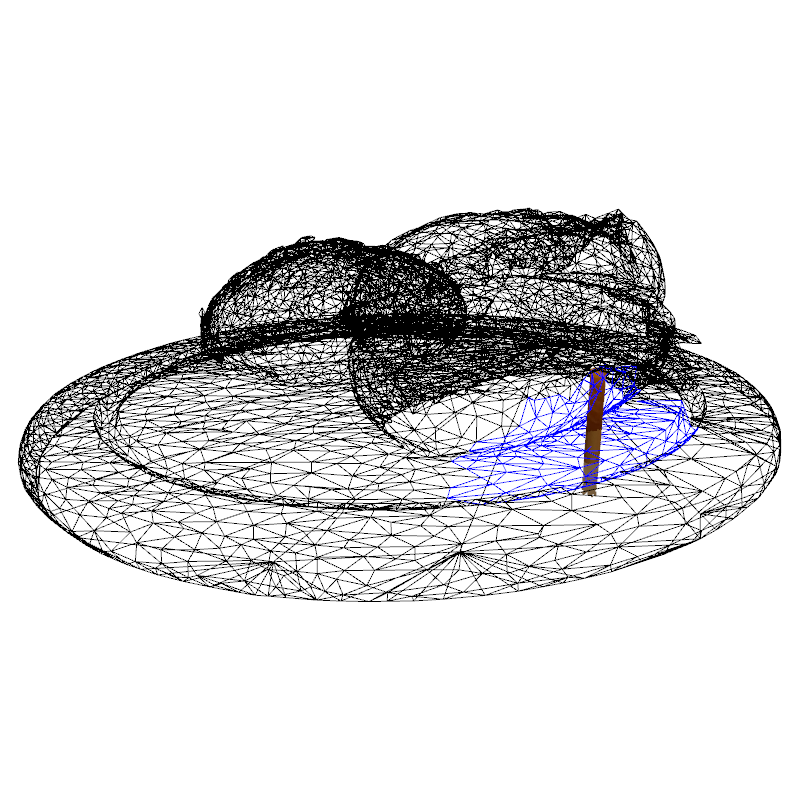}
    \hfill
    \includegraphics[width=0.3\textwidth]{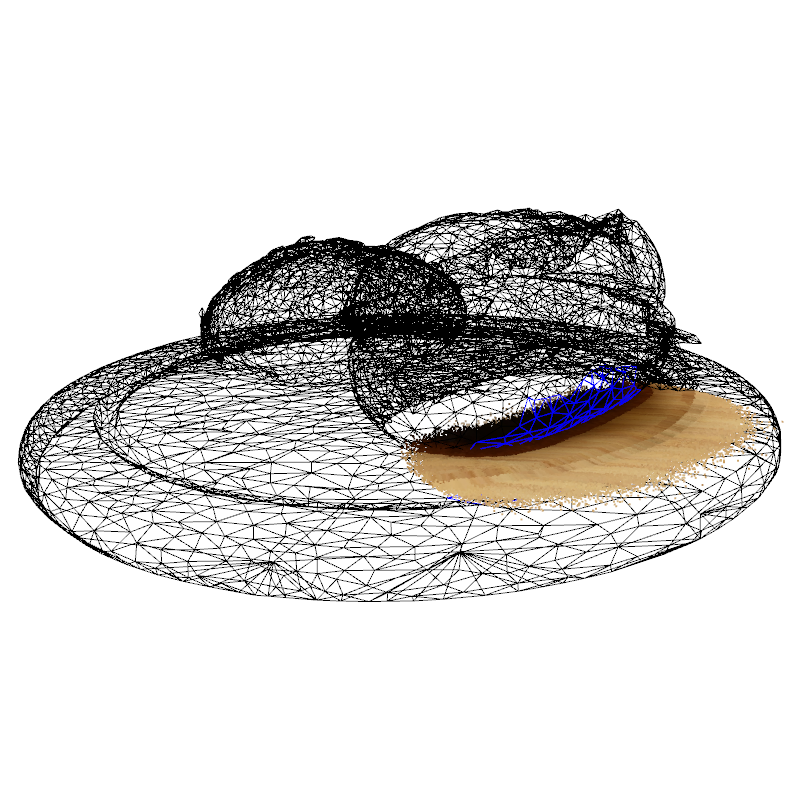}

    \vspace{2pt}
    \includegraphics[width=0.3\textwidth]{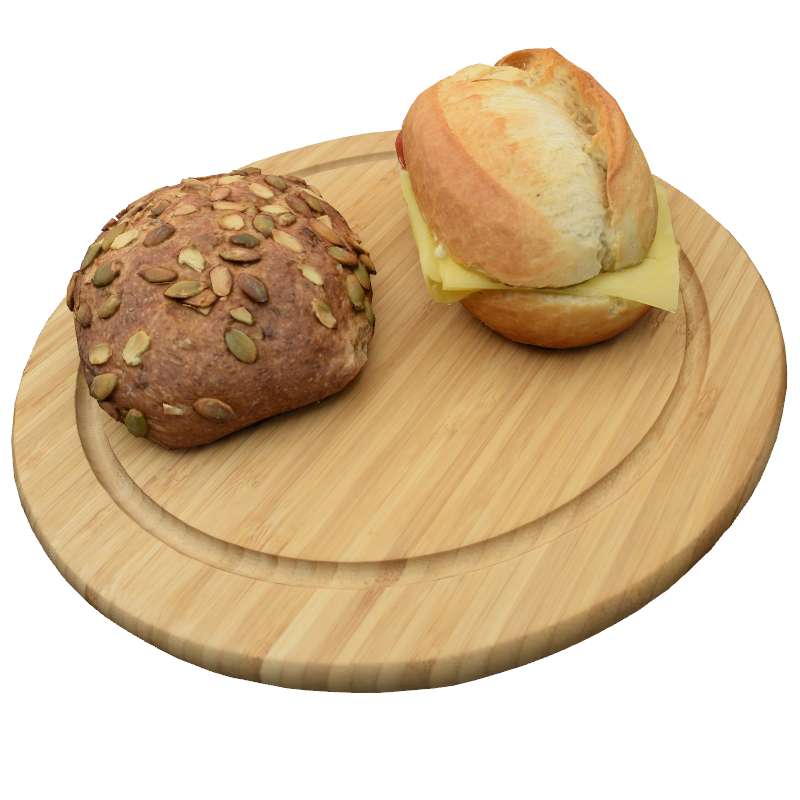}
    \hfill
    \includegraphics[width=0.3\textwidth]{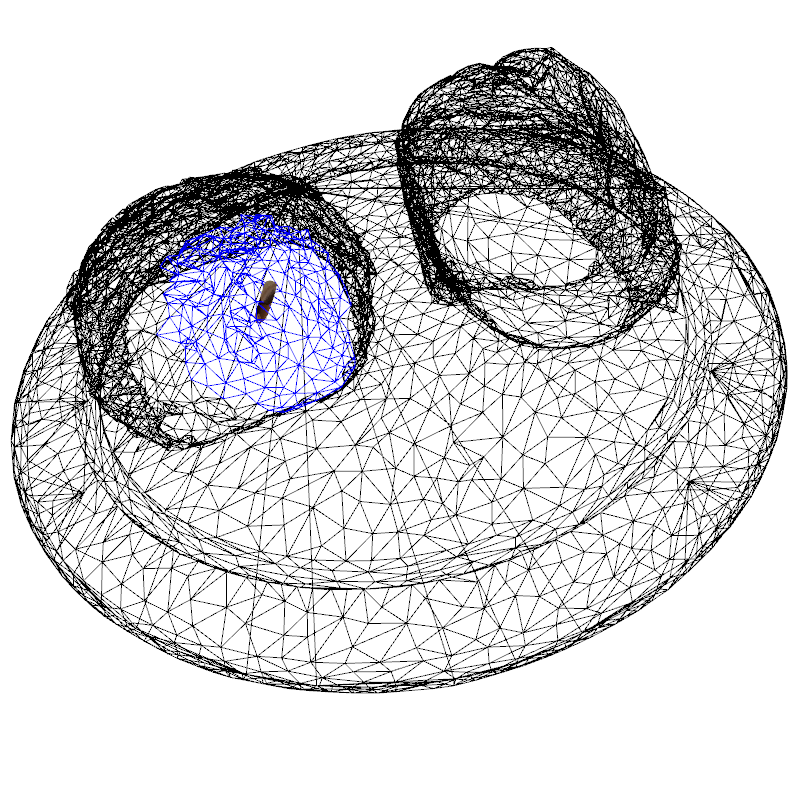}
    \hfill
    \includegraphics[width=0.3\textwidth]{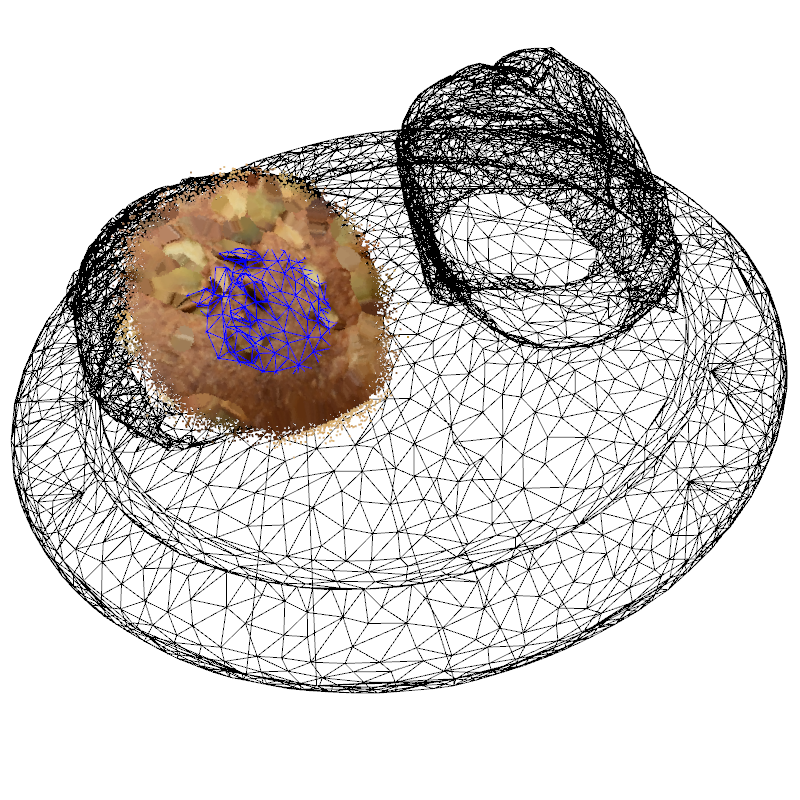}

    \caption{Visualization of NCF feature extraction units across different neighborhoods. Each row corresponds to a local region on the mesh, showing the original view (left), the normal-direction patch (middle), and the tangent-direction patch (right).}
    \label{fig:ncf_patch_units}
\end{figure*}

To further illustrate the local structure captured by our patch-based feature extraction modules, we visualize the normal- and tangent-direction color patch units centered at selected sampling points, as shown in Fig.~\ref{fig:ncf_patch_units}. The first two rows correspond to two sample points on the \texttt{wareBowl} mesh, while the third and fourth rows show two sample points on the \texttt{bread} mesh. Each row contains three images: the original mesh view, the normal-direction patch unit (computed via Equation~(10)), and the tangent-direction patch unit (computed via Equation~(11)).

While Fig.~\ref{fig:ncf_diffusion} presents global diffusion behavior across multiple sampling layers, it may be less intuitive to discern how meaningful features are extracted from a cluster of NCF-valued points. In contrast, Fig.~\ref{fig:ncf_patch_units} shows that once the normal- and tangent-direction patch units are extracted, the meaning of NCF values and their relation to rendering results become much clearer.

As shown in the second column of Fig.~\ref{fig:ncf_patch_units}, the NCF values along the sampling point and its nearest surface point are identical, which corresponds to Property~(1) and Property~(2) of NCF. Regarding the tangent-direction patch units (third column), they closely resemble rendering results. As the tangent patch points extend farther from the center, the color distribution gradually deviates from that of the original surface. This phenomenon reflects Property~(3), since the surface normal at the intersection points increasingly deviates from the viewing direction (i.e., the normal at the patch center). To address this effect, we apply a radius constraint when selecting tangent-direction patch points.

The above observations also exemplify Property~(4) of NCF: the generated color fields resemble localized renderings without requiring an explicit viewpoint. Instead, the viewing direction is implicitly defined by the surface normal at the patch center.

\subsection{FMQM Sampling Procedure}
\begin{figure*}[tp]
  \centering
  \subfloat[Original mesh]{
    \includegraphics[width=0.3\textwidth]{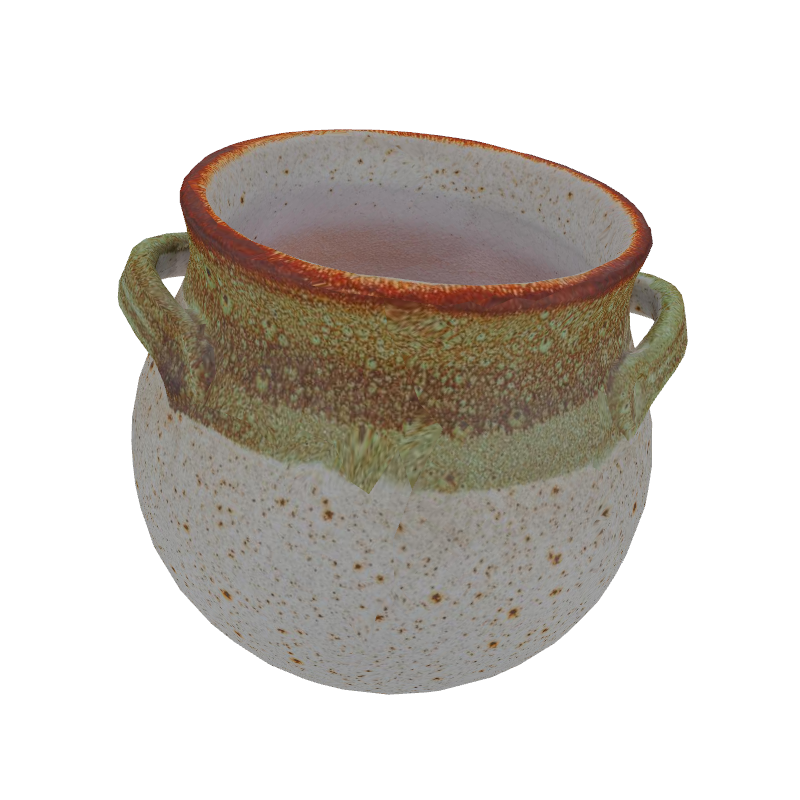}
  }
  \hfill
  \subfloat[FPS selected points]{
    \includegraphics[width=0.3\textwidth]{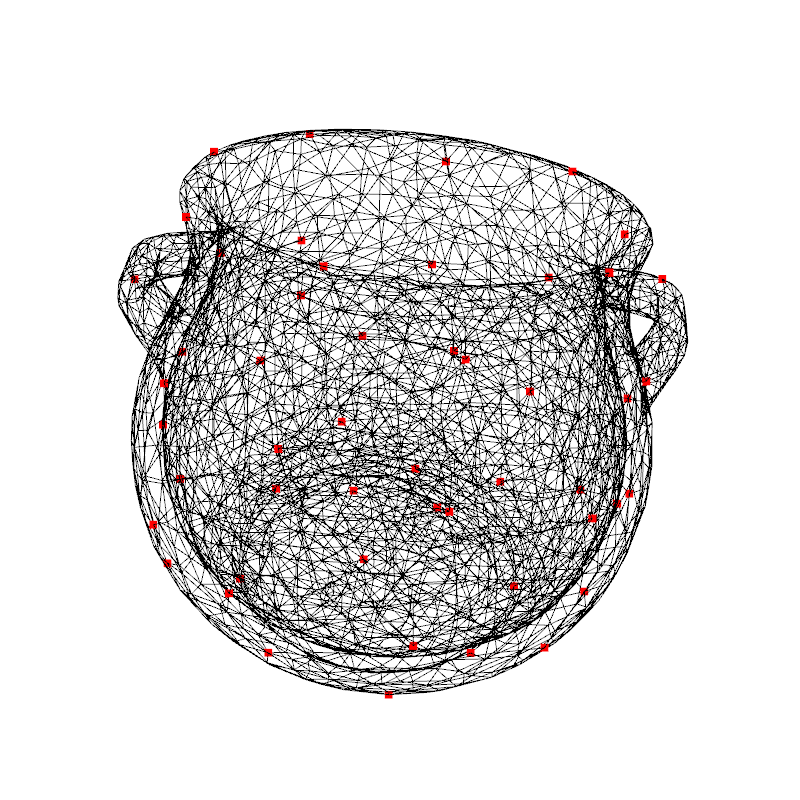}
  }
  \hfill
  \subfloat[Constructed local patch]{
    \includegraphics[width=0.3\textwidth]{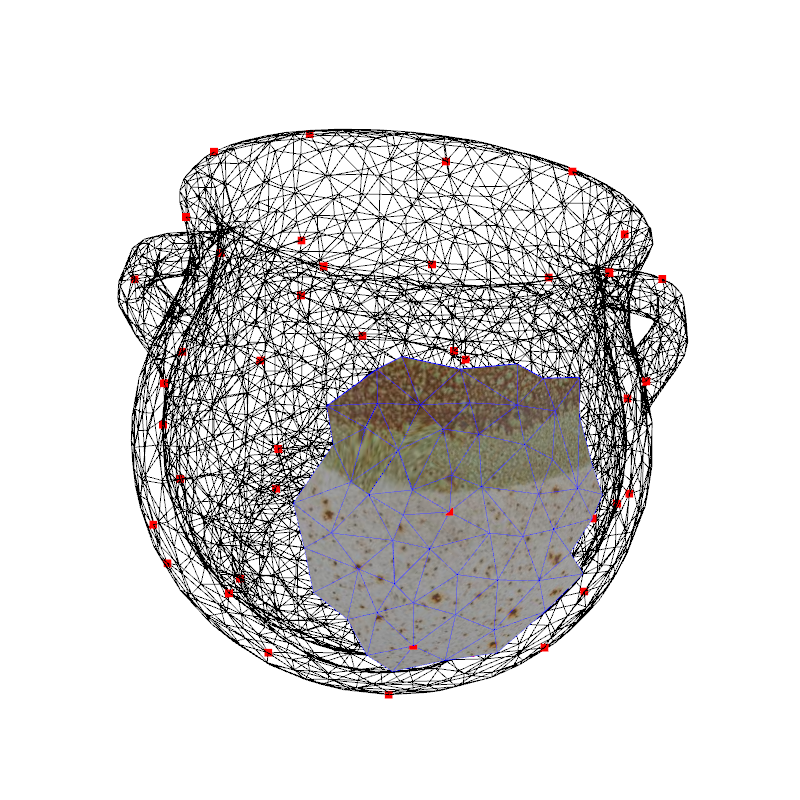}
  }

  \vspace{1em}

  \subfloat[Local patch sampled points]{
    \includegraphics[width=0.3\textwidth]{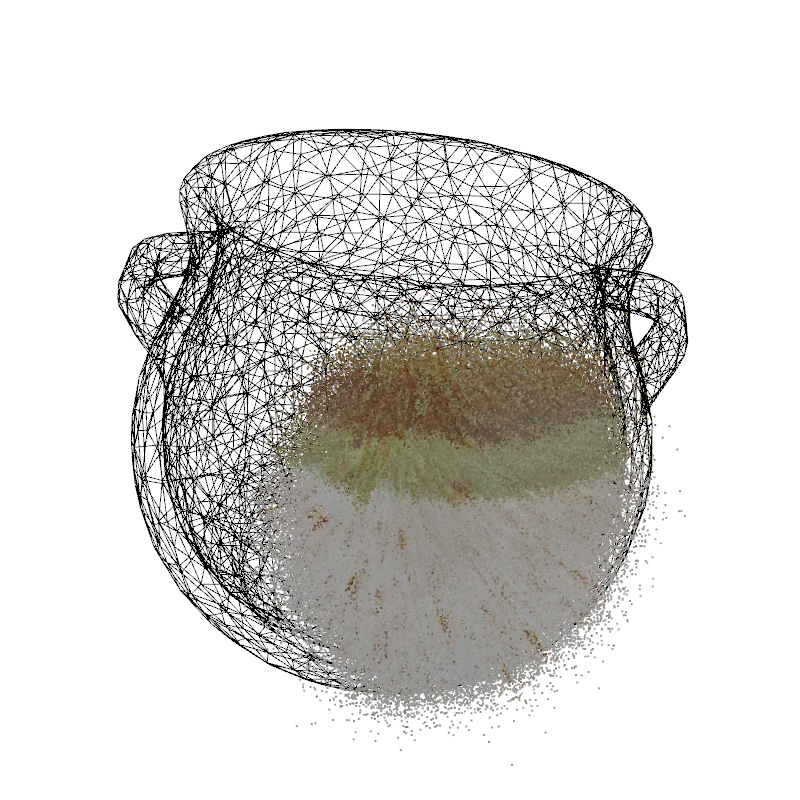}
  }
  \hfill
  \subfloat[Normal color patch unit]{
    \includegraphics[width=0.3\textwidth]{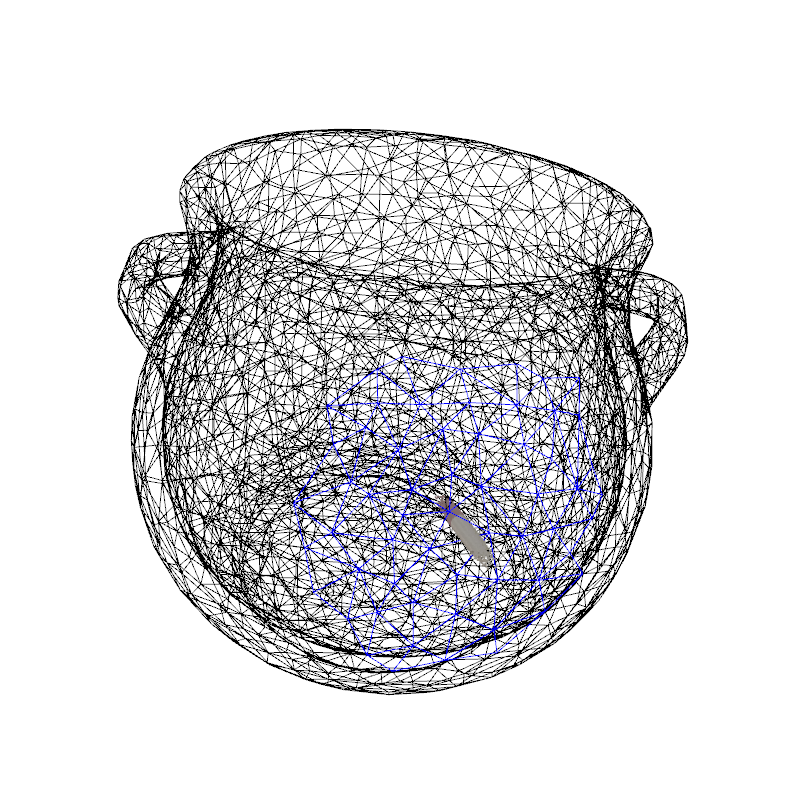}
  }
  \hfill
  \subfloat[Tangent color patch unit]{
    \includegraphics[width=0.3\textwidth]{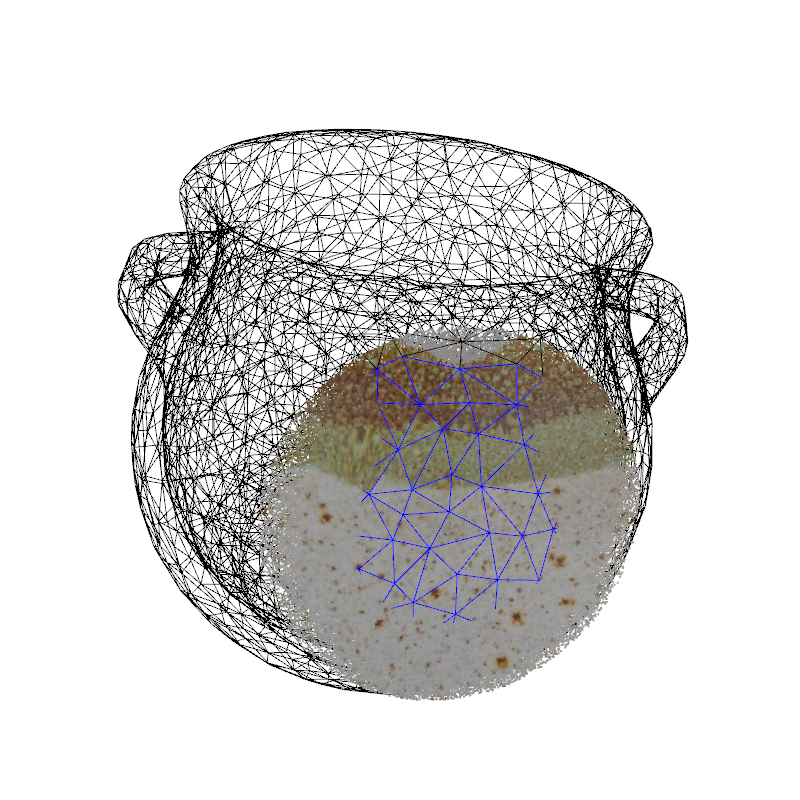}
  }

  \caption{3D illustration of the FMQM pipeline, showing the progression from the original mesh through patch construction, sample point generation, and field-based representation.}
  \label{fig:fmqm_pipeline}
\end{figure*}

Fig.~\ref{fig:fmqm_pipeline} illustrates the 3D sampling procedure of our proposed FMQM framework. We use the \textit{vaso} mesh from SJTU-MQA~\cite{sjtumqa} as an example to visualize the key stages.

Starting from the original mesh geometry (a), we apply farthest point sampling (FPS) on the mesh surface to select representative patch centers (b). For each selected center, a local surface patch is dynamically constructed by expanding the neighborhood until a sufficient surface area is covered (c). This ensures each patch captures meaningful local structure for robust analysis.

Next, we perform near-surface sampling on the constructed patch using a normal-offset strategy (d). Sample points are drawn from triangle barycentric coordinates and perturbed along the local surface normals, enabling dense spatial probing around the surface.

Finally, we visualize two structured patch units: the normal-aligned (e) and tangent-aligned (f) color patches. The normal-aligned patch follows the surface normal direction and mimics how colors are integrated along viewing rays in rendering, capturing distortion along depth. The tangent-aligned patch spans a local surface neighborhood, resembling a projected image patch, and is effective in capturing texture coherence. Together, these two units provide complementary and interpretable cues for feature extraction in FMQM.

\subsection{Ablations on Color Space and Pooling Strategy}
We conduct ablation experiments to validate two design choices in FMQM: the color feature representation space and the feature pooling strategy. For the color-related terms (\(\mathrm{colorSSIM}\) and \(\mathrm{colorGraSSIM}\)), our default configuration uses the RGB color space. As an alternative, we evaluate these metrics in the Lab space, which is considered more aligned with human visual perception.
\begin{table}[htbp]
\centering
\caption{Performance comparison across color spaces and pooling methods}
\label{tab:pooling_lab}
\resizebox{\columnwidth}{!}{%
\begin{tabular}{|l|ccc|ccc|ccc|}
\hline
\multicolumn{1}{|c|}{\textbf{}} & \multicolumn{3}{c|}{\textbf{TSMD}}                                                         & \multicolumn{3}{c|}{\textbf{SJTU-TMQA}}                                                    & \multicolumn{3}{c|}{\textbf{YANA}}                                                         \\ \hline
\multicolumn{1}{|c|}{\textbf{}} & \multicolumn{1}{c|}{\textbf{PLCC}}  & \multicolumn{1}{c|}{\textbf{SROCC}} & \textbf{RMSE}  & \multicolumn{1}{c|}{\textbf{PLCC}}  & \multicolumn{1}{c|}{\textbf{SROCC}} & \textbf{RMSE}  & \multicolumn{1}{c|}{\textbf{PLCC}}  & \multicolumn{1}{c|}{\textbf{SROCC}} & \textbf{RMSE}  \\ \hline
\textbf{ColorSSIM\_RGB}         & \multicolumn{1}{c|}{\textbf{0.672}} & \multicolumn{1}{c|}{\textbf{0.670}} & \textbf{0.862} & \multicolumn{1}{c|}{\textbf{0.765}} & \multicolumn{1}{c|}{\textbf{0.795}} & \textbf{1.540} & \multicolumn{1}{c|}{\textbf{0.577}} & \multicolumn{1}{c|}{\textbf{0.571}} & \textbf{0.797} \\ \hline
\textbf{ColorSSIM\_Lab}         & \multicolumn{1}{c|}{\textbf{0.682}} & \multicolumn{1}{c|}{\textbf{0.678}} & \textbf{0.851} & \multicolumn{1}{c|}{\textbf{0.774}} & \multicolumn{1}{c|}{\textbf{0.807}} & \textbf{1.514} & \multicolumn{1}{c|}{\textbf{0.571}} & \multicolumn{1}{c|}{\textbf{0.566}} & \textbf{0.802} \\ \hline
\textbf{ColorGSSIM\_RGB}        & \multicolumn{1}{c|}{\textbf{0.817}} & \multicolumn{1}{c|}{\textbf{0.811}} & \textbf{0.671} & \multicolumn{1}{c|}{\textbf{0.658}} & \multicolumn{1}{c|}{\textbf{0.690}} & \textbf{1.801} & \multicolumn{1}{c|}{\textbf{0.573}} & \multicolumn{1}{c|}{\textbf{0.565}} & \textbf{0.800} \\ \hline
\textbf{ColorGSSIM\_Lab}        & \multicolumn{1}{c|}{\textbf{0.811}} & \multicolumn{1}{c|}{\textbf{0.803}} & \textbf{0.681} & \multicolumn{1}{c|}{\textbf{0.714}} & \multicolumn{1}{c|}{\textbf{0.740}} & \textbf{1.675} & \multicolumn{1}{c|}{\textbf{0.575}} & \multicolumn{1}{c|}{\textbf{0.565}} & \textbf{0.799} \\ \hline
\textbf{FMQM}                   & \multicolumn{1}{c|}{\textbf{0.810}} & \multicolumn{1}{c|}{\textbf{0.811}} & \textbf{0.683} & \multicolumn{1}{c|}{\textbf{0.728}} & \multicolumn{1}{c|}{\textbf{0.760}} & \textbf{1.639} & \multicolumn{1}{c|}{\textbf{0.674}} & \multicolumn{1}{c|}{\textbf{0.666}} & \textbf{0.721} \\ \hline
\textbf{FMQM\_Lab}              & \multicolumn{1}{c|}{\textbf{0.807}} & \multicolumn{1}{c|}{\textbf{0.808}} & \textbf{0.687} & \multicolumn{1}{c|}{\textbf{0.751}} & \multicolumn{1}{c|}{\textbf{0.776}} & \textbf{1.580} & \multicolumn{1}{c|}{\textbf{0.673}} & \multicolumn{1}{c|}{\textbf{0.665}} & \textbf{0.722} \\ \hline
\textbf{FMQM\_Ari}              & \multicolumn{1}{c|}{\textbf{0.822}} & \multicolumn{1}{c|}{\textbf{0.821}} & \textbf{0.663} & \multicolumn{1}{c|}{\textbf{0.721}} & \multicolumn{1}{c|}{\textbf{0.754}} & \textbf{1.658} & \multicolumn{1}{c|}{\textbf{0.675}} & \multicolumn{1}{c|}{\textbf{0.667}} & \textbf{0.720} \\ \hline
\end{tabular}%
}
\end{table}
Inspired by the human visual system's greater sensitivity to luminance over chrominance, we apply a weighted pooling strategy across the Lab channels using a 6:1:1 ratio for the L, a, and b channels, respectively. This weighting scheme is consistent with the perceptual prioritization found in YUV-based image quality metrics. As shown in Table~\ref{tab:pooling_lab}, using Lab space leads to comparable or even improved performance: SROCC drops marginally by 0.001–0.003 on TSMD and YANA, and increases by 0.016 on SJTU-TMQA. These results confirm that the Lab color space serves as a viable and effective alternative to RGB within our framework.
To support flexibility in practical applications, we include both RGB and Lab color space options in the released implementation of FMQM.

Regarding feature aggregation, FMQM adopts a geometric mean formulation:
\begin{equation}
\scalebox{0.8}{$
\mathrm{FMQM} = \sqrt[4]{\mathrm{geoSSIM} \cdot \mathrm{geoGraSSIM} \cdot \mathrm{colorSSIM} \cdot \mathrm{colorGraSSIM}}
$}
\end{equation}
We compare this with an arithmetic mean variant (\( \mathrm{FMQM}_{\mathrm{Ari}} \)). As shown in Table~\ref{tab:pooling_lab}, the performance differences are marginal. On TSMD, the arithmetic mean slightly outperforms the geometric mean by about 0.01 in both PLCC and SROCC. On SJTU-TMQA, the geometric mean performs marginally better (by approximately 0.007), and on YANA, both approaches achieve nearly identical results (with a difference of less than 0.001).

Despite the close performance, we opt for the geometric mean to mitigate the effect of scale imbalance across different feature components. This formulation helps stabilize the final score when individual feature magnitudes differ significantly, making it a more robust choice in practice.

\vfill

\end{document}